# A THEORY OF JET DEFINITION

*Fyodor V. Tkachov*

*Institute for Nuclear Research*
*of Russian Academy of Sciences*
*Moscow 117312 Russia*

A systematic framework for jet definition is developed from first principles of physical measurement, quantum field theory, and QCD.
A jet definition is found which:

- is theoretically optimal in regard of both minimization of detector errors and inversion of hadronization;
- is similar to a cone algorithm with dynamically negotiated jet shapes and positions found via shape observables that generalize the thrust to any number of axes;
- involves no ad hoc conventions;
- allows a fast computer implementation [hep-ph/9912415].

The framework offers an array of options for systematic construction of quasi-optimal observables for specific applications.

The second edition:

° clarifies, expands and solidifies the arguments behind the formal derivation;
° strengthens consistency of the derivation at the last step → fine-tunes the final criterion → the jet search algorithm is now much simpler, faster and robust [7];
° uncovers new options not available in conventional schemes.

**Eliminated**:

° The algorithmically cumbersome linear restriction on missing energy; now treated additively with a cumulative upper bound (6.21).

**Added:**

- a general theory of optimal observables (2.7–2.35);
- an analysis of inversion of hadronization (5.10);
- the relation to cone algorithms and thrust (8.11, 8.14);
- a model-independent tool to quantify hadronization (the $Y\text{–}E_{soft}$ distribution; 8.19);
- the option of multiple jet configurations (9).

Related materials (code etc.) are available at http://www.inr.ac.ru/~ftkachov/projects/jets/index.htm

## Introduction 1

Jet finding algorithms are a key tool in high-energy physics [1], and the problem of quantitative description of the structure of multi-hadron final states remains at the focus of physicists' attention (cf. e.g. [2]).

This paper continues the systematic investigation of quantitative description of multijet structure from first principles of physical measurements and quantum field theory undertaken in [3]–[5]. Our purpose here is to complete the analysis of [4] in regard of jet algorithms.[a] We are going to develop a systematic theory of jet definition and derive a jet finding criterion — the so-called *optimal jet definition (OJD)* — summarized in Sec. 7.16.

It is optimal in a well-defined sense — the sense which is ignored in the conventional deliberations about jet algorithms. Namely, the new principle on which the presented theory of jet definition is based is that the configuration of jets must inherit maximum physical information from the original event (Sec. 5.6).

Now the first difficulty (besides realizing its importance) is to give that axiom a systematic quantitative form. This is what the first part of the present work (Sections 2–4) deals with.

In the second part (Sections 5–7) we derive OJD which is summarized in Sec. 7.16.

The third part (Sections 8–11) investigates the definition.

A more detailed description of the content is given in Sec. 1.5.

The focus in this paper is on the analytical theory of the criterion and the underlying principles. Its software implementation is discussed in a separate publication [7]. A detailed numerical investigation of OJD requires a separate project.

Also beyond the scope of the present work are complete formal proofs of the background propositions of Section 3 (this especially concerns the arguments in Sec. 4.1): the purpose here is to uncover and clearly formulate the assumptions involved and to devise a formulaic way to talk about jet finding with hand-waving minimized. What may appear as fancy mathematical formulations is primarily intended as an invitation to mathematical physicists to fill in the remaining gaps.

Notations agree with [4] but the present paper is self-contained in this respect.

The theoretical attitude which permeates this work is that jets are not partons but a data processing tool motivated by the partonic structure of QCD dynamics at high energies [4]. Such a shift of emphasis allows one to remove artificial restrictions in the design of data processing algorithms.

In view of importance of the subject and the prevailing prejudice that the definition of jets is a matter of subjective preference (which, as the theory developed in the present paper shows, may not be quite true), I prefer to tax the reader's patience by explicitly stating trivial things (and even putting them in boxes) rather than leave an important axiom out of the picture.

[a] The $2 \to 1$ recombination version of the optimal jet definition was discussed in [6]; see Sec. 10.11 of the present paper. It is interesting to observe how the popularity of recombination schemes (which of course is due to their simplicity) led astray the study of jet algorithms within the framework of [4] which per se provides no motivation for considering $2 \to 1$ recombinations. This is not the first time that I realize, ex post facto examination, that the hardest part in the solving of seemingly intractable problems is invariably to escape the psychological traps created by quasi-solutions.

### Numbering and formatting 1.1

The two-level numbering conventions of [4] are adopted to facilitate searches of cross-referenced items: sub(sub)sections, equations, figures, tables, and textual propositions are numbered consecutively within sections. Section headings are set in bold type.

Sub- and subsubsection headings differ only by formatting (solid and dotted underlining, respectively).

*Underlined italic type* indicates an important term being defined. The underlining helps the eye to find definitions in the body of the text. The meaning of such terms in the context of our theory is usually narrowed compared with the conventional usage.

Double boxes enclose conceptually important propositions, which maybe numbered. 1.2

Simple solid boxes contain formulas and propositions which are part of the optimal jet definition (OJD) and related algorithmic options. 1.3

Dotted boxes denote important formulas and propositions. 1.4

- Bullets indicate further options, important asides, etc.

The reader is invited to begin to read this text by browsing through it using boxes and bullets as visual clues.

### Plan 1.5

The paper can be roughly divided into three parts.

The first part (Sections 2–4) is preparatory and devoted to a clarification of some general issues pertaining to data processing procedures of which jet algorithms are a part.

The second part (Sections 5–7) is devoted to the derivation of the optimal jet definition.

The third part (Sections 8–11) investigates the OJD.

Section 2 is devoted to a clarification of the relevant issues of mathematical statistics (this perhaps should have been done already in [4]). The reasoning is general and practically no specifics of high-energy physics is invoked. We introduce the notion of (quasi-) optimal observables for measurements of fundamental parameters such as $\alpha_S$ , $M_W$, etc. (Sec. 2.7). Such observables allow one in principle to reach the best possible precision for fundamental parameters. The resulting practical prescriptions (Sec. 2.25) improve upon the usual signal-vs.-noise considerations with an important new ingredient here being the notion of regularization of discontinuities (Sec. 2.52). The prescriptions of Section 2 motivate new ways for the use of jet algorithms some of which are described in Section 9.

Section 3 is essentially a clarification of the arguments of [4] in the light of the results of Section 1. It discusses the "kinematical" properties of observables (their so-called *C*-continuity [4]) which ensure their optimal sensitivity to errors and amenability to theoretical studies. *C*-continuity is described using a special distance among events (Sec. 3.23). The arguments here culminate in a quantitative description of the event's physical information content (Sec. 3.42) which serves as a formal starting point for a subsequent derivation of kinematical jet definition.

Section 4 investigates the specific structure of QCD probability densities. The purpose is to clarify the logical connection between the notions of *C*-continuity and IR safety (the former turns out to be a non-perturbative reformulation of the latter). This solidifies the conjecture of [8] concerning the possibility to confront perturbative calculations with hadronic data for IR safe observables (cf. Eq. 4.2). Then a formal description of hadronization is introduced (Eq. 4.12) to prepare ground for a subsequent study of dynamical aspects of our jet definition. A formal construction of optimal observables which takes into account the hadronization model (Eq. 4.20) provides a reference point for the constructions of observables based on jet algorithms. The conventional scheme for that is discussed in Sec. 4.28.

Section 5 discusses jet definitions. First in Sec. 5.1 the implicit conventional definition of an ideal jet algorithm is investigated. Then Sec. 5.6 introduces a definition of jets rooted in the formalism of the preceding sections. We then exhibit its connection with the concept of inversion of hadronization (Sec. 5.10). Then a quantitative version jet definition is described (Sec. 5.17). It is based on inequalities of a factorized form (Eq. 5.18) which estimate the loss of physical information content in the transition from events to jet configurations. We discuss how different jet algorithms can be compared on the basis of how well they conserve the information (Sec. 5.26). Then a universal dynamics-agnostic variant of jet definition is introduced (Sec. 5.27), and in Sec. 5.31 we explain how it can be modified to include dynamical information.

Section 6 is technical and devoted to the derivation of the factorized estimate. The main trick is the so-called recombination matrix (Sec. 6.1); finding the configuration of jets is equivalent to finding that matrix. The matrix can be regarded as a cumulative variant of the entire sequence of $2 \to 1$ recombinations in the conventional recombination jet finding scheme (cf. also Sec. 10.11) but now all particles are, so to say, recombined into jets democratically. (In this respect, OJD is equivalent to a prescription for determining the order of recombinations.)

In Section 7, the remaining ambiguities are fixed in such a way as to ensure a maximal computational convenience, momentum conservation, and Lorentz covariance. We consider both the spherical kinematics (c.m.s. annihilation of $e^+e^-$ pairs into hadrons) and hadron collisions kinematics (a boost-invariant formulation).

Section 8 clarifies the mechanism of the obtained jet definition and establishes its connection with shape observables of the conventional type (Sec. 8.11). Then we present simple analytical arguments which show that OJD is essentially a cone algorithm with dynamically determined positions and shapes of jet cones (Sec. 8.14).

An important tool we obtain as a subproduct is the so-called $Y$–$E_{soft}$ distribution (Sec. 8.19). It allows one to quantify the mechanism of hadronization in a model-independent fashion.

Section 9 considers the issues for a discussion of which the conventional schemes offer no framework whatever, namely, the problem of non-uniqueness of jet configurations which in the case of OJD takes the problem of multiple minima. The options naturally offered by the developed theory allow one to go beyond the restrictions of the conventional data processing scheme based on jet algorithms (4.38).

Section 10 compares OJD with the conventional cone and recombination schemes. We discuss the vicious circle in the conventional jet definitions (no principle to fix the initial cone configuration/order of recombinations; 10.2). Also derived is a curious variant of OJD (Eq. 10.7) which corresponds to the original cone algorithm of [8] rewritten in terms of IR safe shape observables. In Sec. 10.11 we discuss the connection of OJD with the conventional $2 \to 1$ recombination criteria.

A general conclusion is that the mechanism of OJD is rather similar to the conventional cone algorithms.

Section 11 summarizes our findings.

## Optimal observables, continuity and regularizations 2

For a meaningful discussion of jet algorithms it is essential to regard them as a special case of general data processing procedures. With that in mind, below are listed some basic facts of mathematical statistics which emerged as necessary for a systematic clarification of the issue of jet definition. Although the high-energy physics background affected the terminology and emphasis of the presentation below, it deals essentially with elementary notions of parameter estimation. However, the important prescription we arrive at in Sec. 2.25 seems to be missing from textbooks.

### Some generalities 2.1

One deals with a random variable $\mathbf{P}$ whose instances (specific values) are called events. Throughout most of this section, the nature of events $\mathbf{P}$ can be anything: they can be random points on the real axis or random measures on the unit sphere.

One always deals with a finite collection of experimentally observed events $\{\mathbf{P}_i\}_i$. In the context of applications of interest to us, events are obtained via rather complex measurement procedures, so that their probability distribution $\pi(\mathbf{P})$ reflects experimental imperfections.

Experimental imperfections are of two kinds to be called, respectively, *statistical errors* which are due to the finite number of events in the event sample, and *detector errors* i.e. distortions of individual events by measurement devices. Of course, the two cannot be strictly separated because detector errors may cause some events not to be seen at all but this is not important for our purposes.

Theory provides a model for $\pi(\mathbf{P})$ controlled by a small number of *fundamental parameters* such as the Standard Model's $\alpha_S$, $M_W$, etc.

Theoretical knowledge may also involve imperfections, e.g. the necessity to describe hadronic data in terms of quarks and gluons in perturbative quantum chromodynamics (pQCD).

Any data processing has, in the final respect, two purposes. One is to test the hypothesis of correctness of the underlying theoretical model, which we will not discuss. The other purpose is to extract the values of $\alpha_S$, $M_W$,… from given $\{\mathbf{P}_i\}_i$ and $\pi(\mathbf{P})$.

This can be represented as follows:

$$\left.\begin{matrix}\pi(\mathbf{P})\\ \{\mathbf{P}_i\}_i\end{matrix}\right\} \xrightarrow[\text{data processing algorithm}]{} \alpha_S,\ M_W,\ldots \qquad 2.2$$

It is convenient to call the collection of events $\{\mathbf{P}_i\}_i$ *raw physical information*. On the other hand, to obtain the parameters on the r.h.s., one has to interpret data in terms of a specific model, so such parameters are conveniently called *interpreted physical information*.

The scheme 2.2 represents the much studied basic problem of mathematical statistics [9], [10]. However, we would like to regard it in the light of specifics of the formalism of quantum field theory where a central role is played by quantum operators whose average values over ensembles of events are the quantum observables. In the language of mathematical statistics, this means that we are going to place emphasis on the generalized method of moments.

So we wish to consider the general scheme in which the transformation 2.2 is accomplished by choosing suitable functions $f(\mathbf{P})$ defined on events, and then finding the parameters by equating their theoretical average values,

$$\langle f\rangle_{\rm th} = \langle f\rangle = \int \mathrm{d}\mathbf{P}\,\pi(\mathbf{P})\, f(\mathbf{P})\,, \qquad 2.3$$

where $\pi$ is supposed to be known so that this can be computed for any values of fundamental parameters, with the corresponding experimental values:

$$\langle f\rangle_{\rm exp} = \frac{1}{N}\sum\nolimits_i f(\mathbf{P}_i)\,. \qquad 2.4$$

The scheme 2.2 becomes:

$$\left.\begin{array}{l}\pi(\mathbf{P}) \xrightarrow[\text{observable } f]{} \langle f\rangle_{\rm th} \\ \{\mathbf{P}_i\}_i \xrightarrow[\text{observable } f]{} \langle f\rangle_{\rm exp}\end{array}\right\} \xrightarrow[\text{fit}]{} \alpha_S,\, M_W,\ldots \qquad 2.5$$

In terms of mathematical statistics, the weight $f$ is a generalized moment. In the context of quantum field theory, to such functions there correspond quantum operators in terms of which the entire theory is formulated. We will be using the quantum-theoretic term *observable* for such functions, and call $\langle f\rangle$ its *observable value*.

The values of all possible observables $\langle f\rangle$ will be called *processed physical information*, which is a model-independent concept to be contrasted with the model-dependent notion of interpreted physical information (fundamental parameters).

With processed physical information, one simply deals with all possible functions on events. Their general properties such as continuity play an important role in the analysis of sensitivity of observables to experimental and theoretical imperfections. Such properties can be called *kinematical* because they depend only on the general structure of detector errors and of the underlying formalism (quantum field theory), and can be studied in a model-independent manner (Section 3).

All conventional data processing procedures (involving event selection, jet algorithms, histograms, etc.) are special cases of the scheme 2.5. In practice the fits of theoretical predictions to experimental data often involve many observables (e.g. each bin of a histogram represents one numeric-valued observable). Such collections can be regarded simply as observables that take non-numeric values (in the simplest interpretation, arrays, perhaps multidimensional; in a more sophisticated interpretation, the values may be functional objects).

For explicitness' sake, here is an obvious but key axiom:

> The best observables $f(\mathbf{P})$ are those which yield the best precision for fundamental parameters.
>
> 2.6

It turns out that there is a general prescription to construct such observables in a systematic fashion.

## Optimal observables 2.7

Suppose one needs to extract the value of the fundamental parameter $M$ on which depends the probability distribution $\pi(\mathbf{P})$.[b] We are going to study ways to choose an observable $f$ so as to determine $M$ to maximum precision. First we will obtain an ideal explicit formula for such an optimal observable (Eq. 2.17). The formula itself is essentially a translation of the method of maximum likelihood into the language of moments but our derivation is somewhat unconventional and it allows us to go further and study effects of small deviations from optimality (Eq. 2.22; it seems to be a new result), and then arrive at a prescription for a systematic practical construction of quasi-optimal observables (Sec. 2.25). The prescription seems to be both important and new[c].

In the context of precision measurements one can assume the magnitude of errors to be small. Under this assumption, one can relate variations in the values of $M$ with variations in the values of $\langle f\rangle$ as follows:

$$\delta M = \left(\frac{\partial\langle f\rangle}{\partial M}\right)^{-1}\delta\langle f\rangle\,, \qquad 2.8$$

where the derivative is applied only to the probability distribution ($M$ is unknown, so even though the solution, $f_{\rm opt}$, will depend on $M$, any such dependence is coincidental and therefore "frozen" in this calculation):

$$\frac{\partial\langle f\rangle}{\partial M} = \int \mathrm{d}\mathbf{P}\, f(\mathbf{P})\,\frac{\partial\pi(\mathbf{P})}{\partial M}\,. \qquad 2.9$$

The axiom 2.6 translates into the requirement of minimizing the expression 2.8 by an appropriate choice of $f$.

Then $\delta\langle f\rangle = N^{-1/2}\sqrt{\mathrm{Var}\langle f\rangle}$ , where:

$$\mathrm{Var}\langle f\rangle = \int \mathrm{d}\mathbf{P}\,\pi(\mathbf{P})\,\big(f(\mathbf{P})-\langle f\rangle\big)^2 \equiv \langle f^2\rangle - \langle f\rangle^2\,. \qquad 2.10$$

In terms of variances, Eq. 2.8 becomes:

$$\mathrm{Var}\, M = \left(\frac{\partial\langle f\rangle}{\partial M}\right)^{-2}\mathrm{Var}\, f\,. \qquad 2.11$$

We want to minimize this by a suitable choice of $f$.

A necessary condition for a minimum can be written in terms of functional derivatives:[d]

$$\frac{\delta}{\delta f(\mathbf{P})}\mathrm{Var}\, M = 0\,. \qquad 2.12$$

Substitute Eq. 2.11 into 2.12 and use the following relations:

---

[b] We assume that all mechanisms of distortion of observed events are included into the probability distribution $\pi(\mathbf{P})$. The problem of coping with insufficient knowledge of detector errors that distort individual events is discussed in Sec. 2.48.

[c] New to the extent that I've seen no trace in the literature of its being known to either theorists or experimentalists.

[d] An interesting mathematical exercise of casting the following reasoning (the functional derivatives, $\delta$-functionals, etc.) into a rigorous form is left to interested mathematical parties. For practical purposes it is sufficient to note that the range of validity of the prescriptions we obtain is practically the same as for the maximum likelihood method; see Sec. 2.32.

$$\frac{\delta}{\delta f(\mathbf{P})}\langle f\rangle=\pi(\mathbf{P}),\qquad \frac{\delta}{\delta f(\mathbf{P})}\langle f^2\rangle=2f(\mathbf{P})\,\pi(\mathbf{P}),$$
$$\frac{\delta}{\delta f(\mathbf{P})}\frac{\partial\langle f\rangle}{\partial M}=\frac{\partial\pi(\mathbf{P})}{\partial M}. \qquad 2.13$$

After some simple algebra one obtains:

$$f(\mathbf{P})=\langle f\rangle+\text{const}\,\frac{\partial\ln\pi(\mathbf{P})}{\partial M}, \qquad 2.14$$

where the constant is independent of **P**. The constant plays no role since $f$ is defined by this reasoning only up to a constant factor. Noticing that

$$\int d\mathbf{P}\,\pi(\mathbf{P})\frac{\partial\ln\pi(\mathbf{P})}{\partial M}=\frac{\partial}{\partial M}\int d\mathbf{P}\,\pi(\mathbf{P})\equiv\frac{\partial}{\partial M}1=0, \qquad 2.15$$

we arrive at the following general family of solutions:

$$f(\mathbf{P}) \;=\; C_1\frac{\partial\ln\pi(\mathbf{P})}{\partial M}+C_2, \qquad 2.16$$

where $C_i$ are independent of **P** but may depend on $M$.

For convenience of formal investigation we will usually deal with the following member of the family 2.16:

$$f_{\text{opt}}(\mathbf{P}) \;=\; \frac{\partial\ln\pi(\mathbf{P})}{\partial M}. \qquad 2.17$$

Then Eq. 2.15 is essentially the same as

$$\langle f_{\text{opt}}\rangle=0. \qquad 2.18$$

• As a practical prescription, one may drop multiplicative and additive **P**-independent constants from $f_{\text{opt}}(\mathbf{P})$ without violating optimality of the observable. However, Eq. 2.18 may then be violated, and the relations such as 2.22 would then have to be modified accordingly.

## The solution 2.17 is a local quadratic minimum 2.19

Consider 2.11 as a functional of $f$, $\text{Var}M[f]$. Assume $\varphi$ is a function of events such that $\langle\varphi^2\rangle<\infty$. We are going to evaluate the functional Taylor expansion of $\text{Var}M[f_{\text{opt}}+\varphi]$ with respect to $\varphi$ through quadratic terms:

$$\text{Var}\,M[f_{\text{opt}}+\varphi]=\text{Var}\,M[f_{\text{opt}}]$$
$$+\frac{1}{2}\int\left[\frac{\delta^2\,\text{Var}\,M[f]}{\delta f(\mathbf{P})\delta f(\mathbf{Q})}\right]_{f=f_{\text{opt}}}\varphi(\mathbf{P})\varphi(\mathbf{Q})\,d\mathbf{P}\,d\mathbf{Q}\;+\;\ldots \qquad 2.20$$

It is sufficient to use functional derivatives and relations such as 2.13 and

$$\frac{\delta}{\delta f(\mathbf{P})}f(\mathbf{Q})=\delta(\mathbf{P},\mathbf{Q}),\qquad \int\delta(\mathbf{P},\mathbf{Q})\,\varphi(\mathbf{P})\,d\mathbf{P}=\varphi(\mathbf{Q}). \qquad 2.21$$

We obtain the following result which appears to be new:

$$\text{Var}\,M[f_{\text{opt}}+\varphi]$$
$$=\frac{1}{\langle f_{\text{opt}}^2\rangle}+\frac{1}{\langle f_{\text{opt}}^2\rangle^3}\left\{\langle f_{\text{opt}}^2\rangle\times\langle\bar\varphi^2\rangle-\langle f_{\text{opt}}\times\bar\varphi\rangle^2\right\}+\ldots \qquad 2.22$$

where $\bar\varphi=\varphi-\langle\varphi\rangle$.

Non-negativity of the factor in curly braces follows from the standard Schwartz inequality. □

• The first term on the r.h.s. of 2.22, $\langle f_{\text{opt}}^2\rangle^{-1}$, is the absolute minimum for the variance of $M$ as established by the fundamental Rao-Cramer inequality [9], [10]. The latter is valid for all $\varphi$ and therefore is somewhat stronger than the result 2.22 which we have obtained only for sufficiently small $\varphi$. However, Eq. 2.22 gives a simple explicit estimate for the *deviation* from optimality and so makes possible the practical prescriptions of Sec. 2.25.

The quantity

$$I_{\text{opt}}=\langle f_{\text{opt}}^2\rangle \qquad 2.23$$

is closely related to Fischer's information [9], [10].

More generally, it will be convenient to talk about *informativeness* $I_f$ of an observable $f$ with respect to the parameter $M$, defined by

$$I_f=(\text{Var}\,M[f])^{-1}. \qquad 2.24$$

The smaller the error of the value of $M$ extracted using $f$, the larger the informativeness of $f$.

Then $I_{\text{opt}}$ is simply the informativeness of $f_{\text{opt}}$.

Note that Fisher's information is an attribute of data whereas the informativeness is a property of an observable.

It is also possible to talk about an optimal observable from a restricted class of observables. An example of such restriction is considered in Sec. 4.52.

## Quasi-optimal observables 2.25

The fact that the solution 2.17 is the point of a quadratic minimum means that any observable $f_{\text{quasi}}$ which is close to 2.17 would be practically as good as the optimal solution (we will call such observables *quasi-optimal*). A quantitative measure of closeness is given by comparing the $O(1)$ and $O(\varphi^2)$ terms on the r.h.s. of 2.22:

$$\frac{\langle f_{\text{opt}}^2\rangle\langle\bar\varphi^2\rangle-\langle f_{\text{opt}}\bar\varphi\rangle^2}{\langle f_{\text{opt}}^2\rangle^2}\ll 1, \qquad 2.26$$

where $\bar\varphi=f_{\text{quasi}}-\langle f_{\text{quasi}}\rangle-f_{\text{opt}}$.

The subtracted term in the numerator can be dropped, which only overestimates the l.h.s. and is safe. Assuming for simplicity of formulas that $\langle f_{\text{quasi}}\rangle=0$, the criterion 2.26 takes the following simple form:

$$\langle[f_{\text{quasi}}-f_{\text{opt}}]^2\rangle\ll\langle f_{\text{opt}}^2\rangle. \qquad 2.27$$

Here is the representation in terms of integrals:

$$\int d\mathbf{P}\,\pi(\mathbf{P})\left[f_{\text{quasi}}(\mathbf{P})-f_{\text{opt}}(\mathbf{P})\right]^2\ll\int d\mathbf{P}\,\pi(\mathbf{P})\,f_{\text{opt}}^2(\mathbf{P}). \qquad 2.28$$

The criterion 2.27 may be more useful in the practical construction of $f_{\text{quasi}}$, and since the latter would tend to oscillate around $f_{\text{opt}}$ causing $\langle f_{\text{opt}}\bar\varphi\rangle$ to be suppressed, the difference between 2.27 and 2.26 may be negligible.

As a rule of thumb, one would aim to minimize the bracketed expression on the l.h.s. of 2.28 for each (or "most") **P**:

$$\left[f_{\text{quasi}}(\mathbf{P})-f_{\text{opt}}(\mathbf{P})\right]^2\ll f_{\text{opt}}^2(\mathbf{P}). \qquad 2.29$$

One can talk about non-optimality of observables (i.e. their lower informativeness compared with the optimal observable) and also about _sources of non-optimality_. These have a simple interpretation in the case of quasi-optimal observables as the deviations of $f_{\rm quasi}(\mathbf{P})$ from $f_{\rm opt}(\mathbf{P})$ which give sizeable contributions to the integral in 2.27. The simplest example is when $f_{\rm opt}$ is a continuous smoothly varying function whereas $f_{\rm quasi}$ is a piecewise constant approximation. Then $f_{\rm quasi}$ would usually deviate most from $f_{\rm opt}$ near the discontinuities which, therefore, are naturally identified as sources of non-optimality.

It is practically sufficient to take Eq. 2.17 at some value $M = M_0$ close to the true one (which is unknown anyway). This is usually possible in the case of precision measurements. One could also perform an iterative procedure for $M$ starting from $M_0$, then replacing $M_0$ with the value newly found, etc. — a procedure closely related to the optimization in the maximum likelihood method.

So the _method of quasi-optimal observables_ is as follows:

(1) construct an observable $f_{\rm quasi}$ using 2.17 as a guide so that $f_{\rm quasi}$ were close to $f_{\rm opt}$ in the integral sense of Eq. 2.26;

(2) find $M$ by fitting $\langle f_{\rm quasi}\rangle_{\rm th}$ against $\langle f_{\rm quasi}\rangle_{\rm exp}$ ;

(3) estimate the error for $M$ via 2.11;

(4) $f_{\rm quasi}$ may depend on $M$ to find which one can optionally use an iterative procedure starting from some value $M_0$ close to the true one.

2.30

Furthermore, it is possible to use an approximate shape for the r.h.s. of 2.17 such as given by a few terms of a perturbative expansion. In terms of quantum-field-theoretic perturbation theory this means that it may be sufficient to construct $f_{\rm quasi}$ on the basis of the expressions for probability distribution (matrix elements squared) obtained in the lowest PT order in which the dependence on the parameter manifests itself: theoretical updates of radiative corrections need not be reflected in the quasi-optimal observables. It may also be convenient to use a piecewise linear $f_{\rm quasi}$ or even piecewise constant. The latter option actually corresponds to conventional procedures based on cuts (cf. Sec. 2.35); however, using piecewise linear approximations for $f_{\rm quasi}$ should yield noticeably better without incurring noticeable algorithmic complications.

If the dimensionality of the space of events is not large then it may be possible to construct a suitable $f_{\rm quasi}$ in a brute force fashion, i.e. build a multi-dimensional interpolation formula for $\pi(\mathbf{P})$ (via an adaptive routine similar to those used e.g. in [11]) for two or more values of $M$ near the value of interest, and perform the differentiation in $M$ numerically.

Also, one can use different expressions for $f_{\rm quasi}$: e.g. perform a few first iterations with a simple shape for faster calculations and then switch to a more sophisticated interpolation formula for best precision.

### Several parameters 2.31

With several parameters to be extracted from data there are the usual ambiguities due to reparametrizations but one can always define an observable per parameter according to 2.17. Then the informativeness 2.24 is a matrix (as is Fischer's information).

Since the covariance matrix of (quasi-) optimal observables is known (or can be computed from data), the mapping of the corresponding error ellipsoids for different confidence levels from the space of values of the observables into the space of parameters is straightforward.

### Connection with maximum likelihood 2.32

The prescription 2.17 is closely related to the standard method of maximum likelihood that prescribes to estimate $M$ by the value which maximizes the likelihood function:

$$\sum_i \ln \pi(\mathbf{P}_i) \, , \qquad 2.33$$

where summation runs over all events from the sample. The necessary condition for the maximum of 2.33 is

$$\frac{\partial}{\partial M}\Big[\sum_i \ln \pi(\mathbf{P}_i)\Big] = \sum_i \frac{\partial \ln \pi(\mathbf{P}_i)}{\partial M} \propto \big\langle f_{\rm opt}\big\rangle_{\rm exp} = 0 \, . \qquad 2.34$$

This agrees with 2.17 thanks to 2.18.

So the formula 2.17 can be regarded as a translation of the method of maximum likelihood (which is known to yield the theoretically best estimate for $M$ [9], [10]) into the language of the generalized method of moments.

Equivalents of the formula 2.17 can be found at intermediate stages of examples of derivations of estimators for parameters of standard (e.g. normal) probability distributions according to the maximum likelihood method.[e]

The method of quasi-optimal observables is expected to yield results on a par with the maximum likelihood method (because of their close relation; see Sec. 2.25) but it has the following advantages:

(i) applicability to situations with millions of events;
(ii) a greater flexibility in the case of complicated $\pi(\mathbf{P})$.

In such situations a direct minimization of the likelihood function 2.33 is unfeasible.

### Connection of Eq. 2.17 with event selection 2.35

As a simple consistency check, note that Eq. 2.17 agrees with the simplest procedures of event selection used to isolate the signal and suppress backgrounds.

For instance, suppose that most sensitivity of $\pi(\mathbf{P})$ to $M$ (i.e. the derivative $\partial_M \pi$ is largest) is localized in some region $\Pi$ of events (e.g. due to a superselection rule or if $M$ is the mass of a particle that predominantly decays into a certain number of jets). Then $f_{\rm opt}(\mathbf{P})$ vanishes outside $\Pi$:

$$f_{\rm opt}(\mathbf{P}) = 0 \quad \text{if} \quad \mathbf{P} \notin \Pi \, . \qquad 2.36$$

A popular procedure in such a situation is to introduce a selection criterion (a cut):

$$\mathbf{P} \text{ satisfies the selection criterion } \Longleftrightarrow \mathbf{P} \in \Pi \, , \qquad 2.37$$

and to compute the fraction of events from that region, i.e. the observable defined by

$$f_{\rm crude}(\mathbf{P}) = \theta(\ \mathbf{P} \text{ satisfies the selection criterion }) \, , \qquad 2.38$$

where the $\theta$-function is defined according to

$$\theta(\texttt{TRUE}) = 1; \quad \theta(\texttt{FALSE}) = 0 \, . \qquad 2.39$$

---

[e] Rather surprisingly, none of a dozen or so textbooks and monographs on mathematical statistics that I checked (including a comprehensive practical guide [9] and a comprehensive mathematical treatment [10]) explicitly formulated the prescription in terms of the method of moments.

In other words, with the observable 2.38 one simply ignores all non-trivial dependence of $f_{\rm opt}$ on **P** inside $\Pi$.

Furthermore, if in some region $\Pi'$ the magnitude of $\pi(\mathbf{P})$ is large and not offset by its sensitivity to $M$ (the situation of a "large background") then one introduces another selection criterion similar to 2.38, and so on. The net effect is that the observable takes the form

$$f_{\rm crude}(\mathbf{P}) = \prod_i \theta(\mathbf{P} \text{ satisfies } i\text{-th selection criterion}) \times \ldots \qquad 2.40$$

In general, $f_{\rm crude}$ may also contain a factor other than a $\theta$-function (shown with dots above). For instance, in the case of a histogram for some differential distribution of events, each bin corresponds to an observable of the form 2.40 (the last selection criterion is whether or not a value computed for the event belongs to the bin). Then the non-trivial factor (shown with dots) may take e.g. integer values such as the number of dijets from **P** that fall into the bin corresponding to an interval of invariant masses (with each bin representing one observable).

- An immediate practical prescription from the above concerns intermediate regions where either $\partial_M \pi$ is not small enough or $\pi(\mathbf{P})$ is not large enough. Then one should make $f$ interpolate between 0 and 1 over such intermediate regions.

It is clear from Eqs. 2.27–2.29 that such a procedure would increase informativeness of the observable. Simple prescriptions for that are considered in Sec. 2.52. The numerical effect here can be non-negligible (Sec. 2.62).

### Optimal observables and the $\chi^2$ method 2.41

The popular $\chi^2$ method makes a fit with a number of non-optimal observables (bins of a histogram). The histogramming implies a loss of information but the method is universal and implemented in standard software routines. On the other hand, the choice of $f_{\rm quasi}$ requires a problem-specific effort but then the loss of information can be made negligible by a suitable adjustment of $f_{\rm quasi}$.

The balance is, as usual, between the quality of custom solutions and the readiness of universal ones. However, once quasi-optimal observables are found, the quality of maximum likelihood method seems to become available at a lower computational cost.

The two methods are best regarded as complementary: One could first employ the $\chi^2$ method to verify the shape of the probability distribution and obtain the value of $M_0$ to be used as a starting point in the method of quasi-optimal observables in order to obtain the best final estimate for $M$.

A theoretical importance of the optimal observables is that the explicit (even if formal) expressions for optimal observables (cf. 4.20) shed light on the problem of optimal construction of complex data processing algorithms (such as jet finding algorithms). The concept of optimal observables offers specific guidelines for construction and comparison of such algorithms by simply regarding them as a tool for construction of quasi-optimal observables.

### Example. The Breit-Wigner shape 2.42

Let **P** be random real numbers distributed according to

$$\pi(\mathbf{P}) = \frac{1}{\pi\Gamma} \times \frac{1}{(M-\mathbf{P})^2 + \Gamma^2}\,. \qquad 2.43$$

There are two parameters here, and with more than one parameter in the problem there are the usual ambiguities due to reparametrizations. However, one still can define an observable per each parameter according to 2.17.

For 2.43, one obtains:

$$f_{M,\rm opt}(\mathbf{P}) = \frac{\partial}{\partial M} \ln \pi(\mathbf{P}) = -\frac{2(M-\mathbf{P})}{(M-\mathbf{P})^2+\Gamma^2}\,; \qquad 2.44$$

$$f_{\Gamma,\rm opt}(\mathbf{P}) = \frac{\partial}{\partial \Gamma} \ln \pi(\mathbf{P}) \to \frac{\pi\Gamma}{(M-\mathbf{P})^2+\Gamma^2}\,. \qquad 2.45$$

(Recall that there is an arbitrariness in the definition of optimal observables as described by 2.16. The arbitrariness can be used to simplify and conveniently normalize the optimal observables, as done in 2.45.)

The above two weights happen to be uncorrelated:

$$\langle f_{M,\rm opt}\, f_{\Gamma,\rm opt} \rangle = 0\,. \qquad 2.46$$

It is interesting to observe how $f_{M,\rm opt}$ emphasizes contributions of the slopes of the bump — exactly where the magnitude of $\pi(\mathbf{P})$ is most sensitive to variations of $M$ — whereas taking contributions from the two slopes with a different sign maximizes the sensitivity to the signal (i.e. information on $M$). At the same time it suppresses contributions from the middle part of the bump which generates mostly noise as far as $M$ is concerned.

- Unlike theoretical matrix elements which must include all known small corrections (cf. the programs for precision calculations of LEP1 processes [13]), the observables such as 2.44, 2.45 *need not* incorporate, say, loop corrections to $\Gamma$ although inclusion of some such information might be useful (e.g. by introducing simple shapes via linear splines, etc.; cf. comments after 2.30).

- CONNECTION WITH THE TECHNIQUES OF WAVELETS [12].

The form of 2.44 is reminiscent of a typical wavelet, which indicates that applying a wavelet filter to theoretical predictions and experimental formulas instead of the conventional binning prior to using the $\chi^2$ method would improve results. Since software implementations of the wavelet-based methods are available (e.g. on the Web), this could be a way to approach the quality of optimal observables via software routines as universal as those implementing the $\chi^2$ method.

## Continuity of observables 2.47

To directly use the prescription 2.17 may not be possible because of insufficient information about $\pi$. On the other hand, it is reasonable to ask what are the general properties of optimal observables which ensure the best control of uncertainties. With such a knowledge one could ensure that the pragmatically constructed observables at least possess those properties.

For instance, one could start with an ad hoc observable, identify sources of its non-optimality (Eq. 2.29 and remarks thereafter) and modify the observable to mitigate their effect.

From the above reasoning it follows that continuous observables are less sensitive to statistical and detector errors.

There are several reasons why one should prefer continuous observables:

(i) Optimal observables $f_{\rm opt}$ inherit continuity properties of $\pi(\mathbf{P})$. In the problems we consider the latter is always a continuous function of the particles' parameters.

(ii) The variance 2.10 is smaller for the more continuous and slower varying functions $f(\mathbf{P})$. It tends to be larger for $f(\mathbf{P})$ which have jumps or vary fast.

(iii) The error suppression effect of replacing a discontinuous observable by a continuous one (the so-called regularization) can be significant (Sec. 2.62).

(iv) Fluctuations induced by detector errors (distortions of the individual events $\mathbf{P}_i$) are best dampened in final results if the observables possess special continuity properties. Let us briefly discuss this.

### Taking into account detector errors 2.48

In reality the experimentally observed events $\mathbf{P}_i$ contain distortions due to detector errors. This is expressed by a convolution:

$$\pi(\mathbf{P}) = \int d\mathbf{P}' \pi_{\text{ideal}}(\mathbf{P}') D(\mathbf{P}', \mathbf{P}) , \qquad 2.49$$

where $D(\mathbf{P}', \mathbf{P})$ is the probability for the detector installation to see the event $\mathbf{P}$ if it is actually $\mathbf{P}'$. Then the optimal observables are built from 2.49 and must inherit the smearing induced by $D$.

The difficulty here is that it may be hard to take into account the exact form of $D$. Then the least one can say is that the smeared probability distribution 2.49 — and hence the optimal observables — are continuous.

This is not very informative in the simple case where $\mathbf{P}$ are, say, random points on the real axis. However, in high energy physics events $\mathbf{P}$ contain a fluctuating number of particles, each described by at least three numbers (energy, $\varphi$, $\theta$), so that one deals with $O(1000)$ degrees of freedom, i.e. the dimensionality of the space of events is practically infinite. In infinitely-dimensional spaces radically different notions of convergence/continuity[f] are possible (cf. the uniform convergence and integral convergences such as $L_2$ for functions on the real axis), and the significance of the different available options is often missed, so it may not be easy to make the correct choice. In jet-related problems, the relevant continuity is the so-called $C$-continuity (Sec. 3.18). However, for practical purposes it may be useful to keep in mind the following rule of thumb:

> If continuity turns out to be important, then *any* (non-pathological) kind of continuity is better than step-like discontinuities.
>
> 2.50

So, measurements based on conventional event selection procedures can often be improved via replacements of hard cuts by continuously varying observables. The simplest prescription for that is described below (Sec. 2.52). It is rather universal and insensitive to the specific nature of events and cuts one deals with in a particular application.

### On the concept of regularization 2.51

Such a prescription is a special instance of the general concept of _regularization_ (see [14] for a systematic treatment and history). A regularization is needed whenever there is a priori information about the exact solution (such as its continuity) which is not reflected in the approximations one's method yields. This can happen either when one uses crude heuristics (such as event selection procedures) or when one uses theoretical methods which are likely to yield singular solutions (such as those encountered in pQCD; cf. the discussion around Eq. 4.4).

[f] We use the terms convergence or continuity in place of the standard mathematical term topology as more suggestive and to avoid confusion with "topology of event".

Generally speaking, the regularization is a projection of the candidate solutions to the subspace where the exact solution is supposed to reside.

Regularizations may take very different forms depending on the specifics of the problem. One example is the Feier summation of Fourier series for continuous functions. Here the algorithmic simplicity of the method of Fourier expansion comes into conflict with the continuity of the solution, and it proves easier first to sacrifice continuity in order to take advantage of the power of Fourier method, and then recur to a special trick such as the Feier resummation to ensure a uniform convergence to the continuous solution.

Another example is the histogramming of events which, technically, is a transformation of a sum of $\delta$-functions corresponding to individual events into an ordinary function. In this case only a singular approximation (a finite set of events) for a continuous function (the probability distribution) is provided by Nature as a matter of principle.

### Empirical regularization of cuts 2.52

The simplest procedure to transform a typical observable corresponding to an event selection procedure, Eq. 2.40, into a continuous function consists in replacing the step functions $\theta$ with simple piecewise linear continuous functions. The simplest way is to regularize each $\theta$-function in 2.40 individually:

$$f^{\text{reg}}(\mathbf{P}) = \prod_i \theta_i^{\text{reg}}(\mathbf{P}) \times \ldots . \qquad 2.53$$

This can be accomplished as follows.

Each selection criterion in 2.40 can be reduced to the following generic form

$$\varphi(\mathbf{P}) > c_{\text{cut}} , \qquad 2.54$$

where the l.h.s. is a continuous function of the event. For instance, this could be a cut on the total energy of the observed event, then $\varphi(\mathbf{P})$ is the total energy.

Instead of a single parameter $c_{\text{cut}}$, one now chooses a _regularization interval_ specified by two values

$$c_{\text{lo}} < c_{\text{cut}} < c_{\text{hi}} , \quad r \equiv c_{\text{hi}} - c_{\text{lo}} > 0 . \qquad 2.55$$

Here $r$ is the so-called _regularization parameter_.

The simplest option is a symmetric choice:

$$r = 2(c_{\text{cut,hi}} - c_{\text{cut}}) = 2(c_{\text{cut}} - c_{\text{cut,lo}}) . \qquad 2.56$$

One defines:

$$\theta^{\text{reg}}(\mathbf{P}) = \begin{cases} 1 & \text{if } \varphi(\mathbf{P}) \geq c_{\text{hi}} , \\ 0 & \text{if } \varphi(\mathbf{P}) \leq c_{\text{lo}} , \\ \dfrac{\varphi(\mathbf{P}) - c_{\text{lo}}}{r} & \text{if } c_{\text{lo}} < \varphi(\mathbf{P}) < c_{\text{hi}} . \end{cases} \qquad 2.57$$

The linear form is chosen solely from considerations of simplicity. One could also use any other continuous (usually monotonic) shape which interpolates between the same values at the endpoints of the regularization interval. This is a useful option when $r$ is large.

For different selection criteria participating in 2.40, 2.53 the regularization interval and the shape of $\theta^{\text{reg}}$ can be chosen independently.

The most important parameter which controls the shape of $\theta^{\text{reg}}$ and therefore suppression of errors is $r$.

In the context of jet-related measurements one aims to achieve $C$-continuity of observables. Then $\theta^{\rm reg}$ and $f^{\rm reg}$ would also be $C$-continuous if such is $\varphi(\mathbf{P})$ in 2.54.

- WARNING It is possible, as a psychological crutch, to interpret the resulting weights as probabilities that the events carry the characteristics one uses for event selection. (Such an interpretation was mentioned in [4] in a footnote.) However, one should be explicitly warned against introducing a stochastic decision-making according to 2.57 (a procedure of this kind was suggested in [15]). Such additional stochasticity would only be an additional source of fluctuations and therefore increase variance (as can be easily verified in a formal manner), and thus not only defeat the purpose of regularization but exacerbate the problem.

### Choosing the regularization interval 2.58

A lower bound on useful values of $r$ is set by detector errors. Let $\sigma_{\rm meas}$ be the usual sigma for the errors induced in the values of $\varphi(\mathbf{P})$ by the distortions of $\mathbf{P}$ due to detector errors. The effect of error suppression is negligible unless

$$r \geq 2\sigma_{\rm meas}\,. \qquad 2.59$$

For larger $r$ the suppression factor increases as $O(r^{1/2})$ (see sec. 2.6 of [4]). The suppression effect for statistical errors is also greater for larger $r$, and so $r$ should be chosen as large as possible, in general. The best guidance here is Eq. 2.17. A high precision in the choice of regularization interval is not required. However, for large $r$ one may wish to choose more complex shapes than 2.57 (e.g. consisting of several linear pieces glued together).

The regularization interval may also be restricted by other considerations, especially for large $r$ (e.g. onset of a different physical mechanism which causes a large background). In such cases one may opt for more asymmetry than in 2.56.

Some idea about a potentially possible magnitude of suppression effects can be obtained from Sec.2.62.

### Algorithmic aspect 2.60

The simplest first step towards a systematic use of regularization is to introduce a special 4-byte real field for the weight, for each event. The field is initialized to 1. As the event passes selection stages, the weight is modified according to 2.57 and 2.53. If the weight becomes zero at some selection stage, the event is dropped as usual. In the end the selected events' weights are summed up instead of the simple counting of events. Similarly modified should be observables built from selected events:

$$\langle f \rangle_{\rm reg} = N^{-1} \sum w_i\, f(\mathbf{P}_i)\,. \qquad 2.61$$

In particular, what used to be an event fraction now becomes $N^{-1} \sum w_i$. The usual procedure corresponds to $w_i = 1$.

The algorithm described by 2.57 requires only a universal few-lines subroutine.

### Generic examples 2.62

Some idea about the effect of regularization on the sensitivity of observables to statistical errors is given by the following one-dimensional examples. However, the conclusions have a more general validity (see below).

Let $\mathbf{P}$ be a point from the segment $[0,1]$. Compare $f_1(\mathbf{P}) = \theta(x > 1/2)$ (a hard cut) and $f_2(\mathbf{P}) = x$ (a continuous analog). The suppression of relative errors is given by the ratio of the two factors $\sigma_i = \sqrt{\mathrm{Var}\, f_i}\, \langle f_i \rangle^{-1}$. One obtains:

$$\frac{\sigma_1}{\sigma_2} \approx 1.7 \quad \text{for } \pi(\mathbf{P}) = 1; \qquad \frac{\sigma_1}{\sigma_2} \approx 1.6 \quad \text{for } \pi(\mathbf{P}) = 2x. \qquad 2.63$$

The effect of error suppression is significant here for all probability distributions which can be approximated by linear polynomials — in fact significant enough to transform a $3\sigma$ discrepancy into a $5\sigma$ effect.

Although in more complex cases the suppression effect may be less than in 2.63, the above numbers are not as far as one might suppose from reality. Indeed, it is in general possible to change variables appropriately and integrate out inessential components to reduce a generic multidimensional case to the one-dimensional. A realistic example is discussed in Sec. 4.68.

### More on regularization of cuts 2.64

It appears necessary to state that there is absolutely no physical wisdom in preferring event selection (equivalent to dichotomic observables) over continuous weights. So, in view of the very general mechanism of error suppression in the case of continuous observables and the simplicity of regularization prescriptions, one has to explain not why one should regularize the cuts but why one does not do so.

Recall in this respect that the commonly used statistical methods such as histogramming originally emerged in the context of applications such as demography and agriculture, not high-precision particle physics experimentation where the proliferation of cuts in data processing elevates them to the level of a first-order algorithmic/mathematical phenomenon. For instance, the procedures for smoothing conventional histograms found in standard numerical packages are not the same as building histograms with regularized bins: the former entail a loss of numerical information, the latter enhance it by suppressing errors. (See e.g. sec. 12.9 of [4]. A closely related mathematical techniques is the wavelet analysis [12].)

Perhaps it ought to be considered an element of basic culture in data processing that an event should always be accompanied by a real weight. (There might be advantages in allowing the lower levels of the detector facility to yield events with weights not equal to 1 from the very beginning.) Computer memory is cheap enough that extra four bytes per event should not be a burden, and one can always revert to dichotomic weights — but one never quite knows what one looses precision-wise when one sequentially applies a dozen hard cuts to one's events loosing a few % of precision at each hard cut. Modest as the bang here may be, on a per buck basis it is certainly greater than with any hardware upgrade.

The widely spread way of thinking in terms of "event selection" as a primary tool of data processing is based on a mental attitude which could be explained by:

- The limitations of computer resources in the past — a factor which seems to be much alleviated thanks to Moore's law.
- The fact that standard textbooks teach probability in the spirit of the Kolmogorov axiomatics in terms of subsets and the corresponding probabilities. For such axiomatics, the issues of continuity in the cases when random events occur in a continuum, are extraneous.
- The penchant for thinking physics in terms of "regions of phase space" rather than continuously varying observables. This has some foundation in the cases when the events can be tagged somehow [e.g. in the case of (approximate) superselection rules] but not in QCD situations typically encountered in problems involving jet counting. Identifying the most interesting region of phase space is a useful heuristic but ought to be regarded as only a first step in the construction of the observable.

Interestingly, a similar way of thinking in terms of "regions" proved to be detrimental for the theory of Feynman diagrams (see the comments in the E-print posting of [16]). Apparently, the association of "regions" with "physics" is a piece of mythology deeply rooted in the intellectual culture of high-energy physics community. It may perhaps be partly connected to a subconscious rejection of the quantum mechanical notion that it is impossible as a matter of principle to tell which hole an interfering electron passed through.

This issue also seems to be psychologically related to the common insistence that Monte Carlo event generators produce events with unit weights. Even if one uses event selection, the most basic observables are probabilities, so neither individual events nor their (integer) number but only event fractions are fundamental simplest estimators of the corresponding probabilities. But then it is totally irrelevant whether the estimate is obtained by counting events or their fractional weights — the result will be fractional anyway.

Fractional weights accompanying events in the process of event selection would nicely mesh with other related experimental notions (such as the probabilities for a detected particle to be an electron) and with fractional weights accompanying MC-generated events. Theoretical estimates of the event fraction fluctuations for a given corner of phase space seem anyway to be best done by evaluating the variance of the corresponding weight function (because then adaptation techniques in integration routines may be used for greater computational efficiency).

Note that (pseudo) events with fractional weights occur naturally when one attempts to restore the partonic event (see Section 9). This is in fact similar to how experimentalists restore observed events from detector signals.

In short, the absolutization of event selection blinds one to some useful options in data processing.

## Observables in QCD. Kinematical aspects 3

In the preceding section we have introduced the notion of (quasi-) optimal observables for precision measurements of fundamental parameters. Such observables allow one to approach the theoretically possible precision for the parameters with a given event sample. We found that optimal observables are given by an explicit formula in terms of the probability density $\pi(\mathbf{P})$ (Eq. 2.17). In QCD, however, one may have a Monte Carlo event generator with a dependence on fundamental parameters built in, but no algorithm to evaluate $\pi(\mathbf{P})$ for a given event $\mathbf{P}$. In such a situation it is reasonable to construct observables incrementally by combining as many properties from the optimal ones as possible.

There are two types of such properties: kinematical and dynamical. The kinematical properties reflect requirements of two kinds: experimental (appropriate continuity to suppress sensitivity to statistical fluctuations and detector errors in data) and theoretical (conformance to structural properties of quantum field theory in general and QCD in particular, in order to enhance quality of theoretical predictions). Dynamical properties reflect the specific behavior of $\pi(\mathbf{P})$ such as predominant production of certain types of events. In general, the result 2.17 incorporates both kinematical and dynamical restrictions, with the former playing the role of a fine-tuning for the latter. However, the specifics of QCD dynamics (a fast variation of $\pi(\mathbf{P})$ between the points in the space of $\mathbf{P}$ which are close in the sense of $C$-continuity; see [4]) enhances the role of kinematical considerations (see the example in Sec. 4.68).

A systematic study of QCD observables from a kinematical viewpoint (continuity and sensitivity to errors, and compatibility with quantum field theory) was performed in [3]–[5]. In this section we review the findings of [3]–[5].

### Notations. Representations of events 3.1

The beam axis is $z$ and it corresponds to the 3rd component of 3-vectors. The polar angle $\theta$ is measured from the beam axis, and the azimuthal angle $\varphi$ is defined accordingly.

Let $E^{\mathrm{cal}}$ be the _calorimetric energy_ — the number measured by a calorimetric cell. It is usually interpreted as the time-like component of the 4-momentum of the particle which hit the cell. It is assumed sufficient to treat all particles as massless, so that their energies are not distinguished from absolute values of their 3-momenta.

In jet studies one deals with two physical situations in which slightly different kinematical aspects are emphasized. This is reflected in how jets are looked at:

When studying processes with c.m.s. jet production (mostly $e^+e^-$ annihilation), spherical symmetry is emphasized, and so one works within spherical kinematics, dealing with points of unit sphere represented either by the pair of angles $\theta$, $\varphi$ or by unit 3-vectors denoted as $\hat{\boldsymbol{p}}$, $\hat{\boldsymbol{q}}$, etc.

When studying hadron collisions, the colliding partons' rest frame is unknown so that invariance with respect to boosts along the beam axis has to be maintained. Then one works within cylindrical kinematics and introduces the so-called _transverse energy_

$$E^{\perp} \equiv E^{\mathrm{cal}} \sin\theta \cong \sqrt{p_1^2 + p_2^2}\ , \qquad 3.2$$

and _pseudorapidity_

$$\eta = \ln\cot(\theta/2), \quad -\infty < \eta < +\infty\ . \qquad 3.3$$

Then a massless 4-momentum $p = (E^{\mathrm{cal}}, p_1, p_2, p_3)$ is represented as

$$p = E^{\perp}(\cosh\eta,\ \cos\varphi,\ \sin\varphi,\ \sinh\eta). \qquad 3.4$$

Boosts along the beam axis correspond to shifts of $\eta$.

#### Particles and events 3.5

Let $\mathbf{P}$ be the event as seen by an ideal calorimetric detector installation. Then $\mathbf{P}$ is a collection of "particles" which can be just calorimetric cells lit up by the event. Particles in the event will be enumerated using the labels $a$, $b$. The $a$-th particle/cell is represented by its energy $E_a$ and direction $\hat{\boldsymbol{p}}_a$. Formally:

$$\mathbf{P} = \{E_a, \hat{\boldsymbol{p}}_a\}_{a=1\ldots N(\mathbf{P})}\ , \qquad 3.6$$

where $N(\mathbf{P})$ is the total number of particles in $\mathbf{P}$.

It is convenient to allow particles with zero energy in 3.6. This corresponds to the fact that a low-energy particle may not lit up the cell it hits.

In what follows we will be talking about partonic events, hadronic events, jet configurations, etc. They are all objects of the same type 3.6.

The meaning of the energies $E_a$ depends on the chosen kinematics:

$$E_a = \begin{cases} E_a^{\mathrm{cal}} & \text{spherical kinematics;} \\ E_a^{\perp} & \text{cylindrical kinematics.} \end{cases} \qquad 3.7$$

The directions $\hat{\boldsymbol{p}}$ can be represented in different ways (e.g. by $\varphi$ and $\theta$; by a unit 3-vector; etc.), but all the reasoning until Sec. 7 is independent of the representation. All we need is the usual angular distance between two directions, $|\hat{\boldsymbol p}-\hat{\boldsymbol q}|$, which is defined unambiguously.

- For definiteness, we will always be talking about spherical kinematics in what follows. Then $\mathrm{d}\hat{\boldsymbol p}$ is an infinitesimal element of the surface of the unit sphere. The final prescriptions for jet definition will be formulated independently of this assumption.

### Events as measures 3.8

We are actually interested in events as seen by a purely calorimetric detector installation, i.e. energy flows. Energy flow is insensitive to fragmentations of any particle of the event into any number of collinear fragments directed the same as the parent particle and carrying the same total energy. However, the representation 3.6 is defective in this respect in that it is not explicitly fragmentation-invariant.

The following representation of events-as-energy-flows was found to respect physical requirements to maximal degree (see [4] and the reasoning below):

$$\mathbf{P} \Longleftrightarrow \sum\nolimits_a E_a\,\delta(\hat{\boldsymbol p},\hat{\boldsymbol p}_a) \equiv \mathbf{P}(\hat{\boldsymbol p})\,. \tag{3.9}$$

Here the $\delta$-functions obey the usual rules of integration over the unit sphere:

$$\int_{\text{unit sphere}} \mathrm{d}\hat{\boldsymbol p}\;\delta(\hat{\boldsymbol p},\hat{\boldsymbol p}_a)\,d(\hat{\boldsymbol p}) = d(\hat{\boldsymbol p}_a) \tag{3.10}$$

for any continuous function on the unit sphere $d(\hat{\boldsymbol p})$.

In mathematical terms, the object 3.9 is a measure on the unit sphere. By definition, it acquires a numerical meaning after integrations with continuous functions:

$$\langle \mathbf{P}, d\rangle \equiv \int_{\text{unit sphere}} \mathrm{d}\hat{\boldsymbol p}\;\mathbf{P}(\hat{\boldsymbol p})\,d(\hat{\boldsymbol p}) = \sum\nolimits_a E_a\,d(\hat{\boldsymbol p}_a)\,. \tag{3.11}$$

In other words, Eq. 3.9 is essentially a convenient shorthand notation for the collection of values 3.11 for all such $d(\hat{\boldsymbol p})$:

$$\mathbf{P} \Longleftrightarrow \left\{\langle \mathbf{P}, d\rangle\right\}_{d(\hat{\boldsymbol p})\text{ are all continuous functions on unit sphere}} \tag{3.12}$$

- The expression 3.9 is explicitly fragmentation invariant, as are Eq. 3.11 and the r.h.s. of 3.12.

### Calorimetric detector cells 3.13

Elementary calorimetric cells are naturally represented by $d(\hat{\boldsymbol p})$ corresponding to their idealized angular acceptance functions: such $d(\hat{\boldsymbol p})$ takes the value 1 inside some small angular region, and continuously falls off to zero outside that region, so that if $E$, $\hat{\boldsymbol p}$ are the particle's energy and direction then the energy detected by the cell is $E\,d(\hat{\boldsymbol p})$ (the closer to the cell's boundary the particle hits the cell, the less the fraction of the energy registered by the cell). Then the energy which the cell $d$ sees when confronted with the event $\mathbf{P}$ is given by 3.11.

In view of this interpretation, it becomes physically transparent why the event-as-energy-flow is equivalent to the collection of values 3.12. In practice one deals with a finite collection of calorimetric modules $d_a$, and with the corresponding finite collection of numbers $d_a(\mathbf{P})$ for each event $\mathbf{P}$. These numbers constitute the experimentally measured approximation to the ideal information content of $\mathbf{P}$:

$$\mathbf{P}^{\exp} = \left\{\langle \mathbf{P}, d_a\rangle\right\}_a\,. \tag{3.14}$$

If the angular size of $d_a$ (= the size of the angular region in which $d_a \neq 0$) is sufficiently small, $d_a$ is represented by a direction $\hat{\boldsymbol p}_a$, and we come back to 3.6.

This proves that the representation of even-as-energy-flow by the collection 3.12 is equivalent to the conventional "particle" representation 3.6 — with one important improvement: unlike the numbers which constitute 3.6, each number in the collection 3.12 is fragmentation invariant.

- In what follows we will interpret events in the sense of 3.9 and 3.12 (the latter is just a shorthand notation for the former), treating the representation 3.6 as a bookkeeping device.

### The domain $\mathcal{P}$ 3.15

It is convenient to impose the following restriction on events:

$$\sum\nolimits_a E_a \le 1\,. \tag{3.16}$$

This is because the events' energies are bounded by a constant in any experiment, and the structure of energy flow is independent of the event's total energy, at the basic level of sophistication.

It would be sufficient to have the equality in the above restriction. The inequality is allowed only because of the formal convenience resulting from the use of the linear structure in the space of events represented as measures 3.9.

The following collection of events will be the arena of much of the subsequent mathematical action:

$$\mathcal{P} = \text{all events } \mathbf{P} \text{ which satisfy Eq. 3.16.} \tag{3.17}$$

## *C*-continuous observables 3.18

We are dealing with observables $f(\mathbf{P})$ defined on events $\mathbf{P}$ from the domain $\mathcal{P}$. We saw in Sec. 2.48 that smearings due to detector errors cause the probability distribution of observed events and, therefore, optimal observables 2.17 to possess special continuity properties which we are now going to study.

Note that the same notion of *C*-continuous observables will reemerge from analysis of predictive power of pQCD (see comments after 4.2). This is because the choice of calorimetric detectors for measurements is determined by the limitations of predictive power of pQCD in regard of hadronic events [8]. Therefore, *C*-continuity is a fundamental notion in the theory of jet observables.

Before we turn to precise formulations, note the following.

Any function $f(\mathbf{P})$ when considered on events with exactly $N$ particles, becomes an ordinary function of $N$ composite arguments:

$$f(\mathbf{P}) \to f_{(N)}\big(\{E_1,\hat{\boldsymbol p}_1\},\dots,\{E_N,\hat{\boldsymbol p}_N\}\big)\,. \tag{3.19}$$

Then $f(\mathbf{P})$ as a whole is a sequence of such component functions $f_N$, $N=1,\dots\infty$. Such a representation in terms of component functions is natural from the viewpoint of perturbative QCD where one deals with a small number of particles in each order of perturbation theory (cf. [17]).

However, similarly to how the naïve representation of events 3.6 is insufficient in that it is not explicitly fragmenta-tion invariant and thus potentially misleading in the construction of data processing algorithms, so the corresponding representation of observables 3.19 may also be insufficient.
In particular, it would be hard to formulate *C*-continuity in terms of 3.19.

### *C*-convergence of events 3.20

To define continuity of a function $f(\mathbf{P})$ one first has to establish the notion of convergence of its arguments, in our case the events **P**. The issue is non-trivial here because our events **P** run over the infinitely-dimensional domain $\mathcal{P}$, and in infinitely-dimensional spaces many radically non-equivalent notions of convergence are possible. So, when does a sequence of events $\mathbf{P}_n$ converge to an event **P**?

For instance, a naïve convergence defined on the basis of 3.6 would be to require convergence of all numerical "components" of 3.6. Namely, one would require that $N(\mathbf{P}_n) \to N(\mathbf{P})$, which would mean that $N(\mathbf{P}_n) = N(\mathbf{P})$ for all sufficiently large $n$. Then one would require that the energy and direction of each of the particle from $\mathbf{P}_n$ converged to the energy and direction of some particle in **P**. However, this is clearly inadequate because an event consisting of one narrow cluster of particles which gets narrower as $n \to \infty$ may converge in an intuitive physical sense to a one-particle event even if the distribution of energies between particles in $\mathbf{P}_n$ wildly fluctuates with changing $n$.

To obtain a correct answer one should realize that convergences such as the one being discussed are simply a mathematical way to describe the general structure of one's measurement devices, so that the corresponding continuity of observables would ensure their stability with respect to detector errors.

In our case the correct choice is the so-called <u>*C*-convergence</u>.[g, h] Its definition is directly connected to how calorimetric detector cells see events:

> The sequence of events $\mathbf{P}_n$ is said to *C*-converge to **P** if $\mathbf{P}_n$ in the limit of $n \to \infty$ become indistinguishable from **P** for any calorimetric detector cell $d$, i.e.
>
> $$\langle \mathbf{P}_n, d \rangle \to \langle \mathbf{P}, d \rangle \qquad 3.21$$
>
> in the usual numerical sense for each continuous function $d(\hat{\mathbf{p}})$ defined on the unit sphere.

One could use here special $d$ corresponding to realistic detector cells and described in Sec. 3.13 but the extension by linearity to arbitrary continuous functions is convenient and does neither restrict nor relax the definition.

The convergence 3.21 can be described in a more conventional fashion using an appropriately chosen measure of distance between events (Sec. 3.23).

### Formulation in terms of open sets 3.22

The above formulation is equivalent to the following one phrased in a canonical mathematical language. For simplicity we ignore statistical fluctuations of the errors; our purpose is only to show how the basic structure of detector errors uniquely determines the topology (convergence) in the space of events.

An elementary measurement device $\alpha$ (a calorimetric cell in our case) yields a non-empty interval of real numbers $(r', r'')$ for each instance of measurement. Consider the subset of all events which could have produced the same interval (which without loss of generality can be taken to be open) and denote it $O_{\alpha,r',r''}$.

A complex detector installation consists of a finite number of such devices, and each instance of measurements actually registers a subset of events which corresponds to the intersection of $O_{\alpha,r',r''}$ for all elementary devices which constitute the detector.

The sets $O_{\alpha,r',r''}$ constitute the so-called subbase which uniquely determines a topology in the space of events.

The convergence described by 3.21 is equivalent to the topology obtained in this way for elementary measurement devices described by 3.11 (cf. Sec. 3.13).

### Distance to quantify similarity of events 3.23

It may be helpful to point out a single numeric measure of distance between events **P** which would correspond to *C*-convergence. The distance is fully constructive (although a bit cumbersome) and corresponds to the intuitive notion of similarity of two events at various angular resolutions.

Define:

$$\psi(x) = \begin{cases} \exp\left[-x^{-2}\right] & \text{for } x < 1, \\ 0 & \text{for } x > 1; \end{cases} \qquad 3.24$$

$$d_{R,\hat{q}}(\hat{p}) = \psi\left(\theta_{\hat{p},\hat{q}} / R\right); \qquad 3.25$$

This describes an ideal calorimetric cell of radius $R$ centered at $\hat{q}$. (It would have been sufficient for each $R$ to restrict $\hat{q}$ to a finite grid of points so that each point of the unit sphere is no farther than $R/2$ from the nearest point of the grid.)

The following expression is interpreted as the distance between **P** and **Q** at the angular resolution $R$:

$$\mathrm{dist}_R(\mathbf{P},\mathbf{Q}) = \max_{\hat{q}} \left| \langle \mathbf{P}, d_{R,\hat{q}} \rangle - \langle \mathbf{Q}, d_{R,\hat{q}} \rangle \right| = \max_{\hat{q}} \left| \langle \mathbf{P} - \mathbf{Q}, d_{R,\hat{q}} \rangle \right|. \qquad 3.26$$

It is bounded by 1 if both events belong to $\mathcal{P}$.

To obtain a measure of distance for all angular resolutions, simply take a sum over increasingly better resolutions $R_n \to 0$:

$$\mathrm{Dist}(\mathbf{P},\mathbf{Q}) = \sum_{n=1,2\ldots} \zeta_n \, \mathrm{dist}_{R_n}(\mathbf{P},\mathbf{Q}). \qquad 3.27$$

The sequence $R_n$ is otherwise arbitrary, e.g. $R_n = 2^{-n}$.

The sum of positive coefficients $\zeta_n$ must be finite. We normalize them so that

$$\sum_{n=1,2\ldots} \zeta_n = 1. \qquad 3.28$$

This ensures the following normalization of Dist:

$$\mathrm{Dist}(\mathbf{P},\mathbf{Q}) \le 1 \quad \text{for any } \mathbf{P} \text{ and } \mathbf{Q} \text{ from } \mathcal{P}. \qquad 3.29$$

Verbally, each next term in the sum 3.27 describes the difference between **P** and **Q** at a higher angular resolution. The rate of decrease of $\zeta_n$ as $n \to \infty$ controls sensitivity of 3.27 to the differences between **P** and **Q** at higher angular resolutions. For instance, $\zeta_n = 2^{-n}$.

---

[g] *C* from "calorimetric"; we will also use the verb to *C*-converge, etc.

[h] In terms of pure mathematics, the *C*-convergence is an instance of the so-called $*$-weak topology in the space of linear functionals [18].

The decreasing sensitivity of the expression 3.27 to correlations between **P** and **Q** at smaller angular distances nicely reflects the decreasing physical importance of such correlations.

The usual definition of convergent sequences based on this measure of distance in $\mathcal{P}$,

$$\mathrm{Dist}\left(\mathbf{P}_n, \mathbf{P}\right) \to 0 , \qquad 3.30$$

is equivalent to the *C*-convergence, Eq. 3.21.

- The above definition of Dist resembles constructions of the wavelet analysis [12] with $\psi(x)$ corresponding to the mother wavelet. This is not the only place where the logical patterns of the wavelet analysis come to the surface in our theory (cf. the comments after 2.46).
- A mathematician would note that the closure of $\mathcal{P}$ is compact with respect to the *C*-convergence. This is a special case of the Banach-Alaoglu theorem [18]. This is important for the study of the structure of *C*-continuous observables (Sec. 3.35).
- Although one may be psychologically more comfortable with the definition of convergence in the space of events in terms of a single numeric measure of distance 3.30 rather than the seemingly more amorphous definition 3.21, the latter is deeper and is actually simpler. The possibility to express the convergence in terms of one distance 3.30 is accidental and its form exhibits too many inessential details. Eq. 3.21, on the other hand, goes to the heart of the matter by directly reflecting the structure of multimodule detectors and leading to the profound identification 3.43. The usefulness of the entire logical pattern rooted in the definition 3.21 is demonstrated by the derivation of jet definition in Section 6 — it is not clear what heuristics one would have been guided by should one decide to work in terms of the distance 3.30.

### *C*-continuity of observables 3.31

The formal definition is as follows:

An observable $f(\mathbf{P})$ defined on events from $\mathcal{P}$ is *C*-continuous if

$$f(\mathbf{P}_n) \to f(\mathbf{P}) \qquad 3.32$$

whenever $\mathbf{P}_n \to \mathbf{P}$ in the sense of *C*-convergence (3.21 or 3.30).

Qualitatively, *C*-continuity is the same as stability with respect to distortions of energy flow deemed physically less significant in jet-related measurements (such as due to minor rearrangements of detector cells, several particles hitting the same cell, detector errors, etc.). Such distortions may cause the numbers which constitute 3.6 (e.g. the observed number of particles) to jump erratically, whereas the values of *C*-continuous shape observables would exhibit continuous variations.

### *C*-continuity and fragmentation invariance 3.33

Since the definition of *C*-convergence is entirely in terms of the fragmentation-invariant representation of events 3.9, a function $f(\mathbf{P})$ that is *C*-continuous is automatically fragmentation invariant (if **Q** differs from **P** by exactly collinear fragmentations then $\mathrm{Dist}(\mathbf{Q}, \mathbf{P}) = 0$, so Eq. 3.32 implies that $f(\mathbf{Q}) = f(\mathbf{P})$).

Furthermore, each of the component functions $f_N$ (see 3.19) is continuous in all its arguments. However, the latter property is sufficient to ensure *C*-continuity of $f(\mathbf{P})$ (see sec. 6.9 of [4]). This is essentially because *C*-continuity imposes restrictions on allowed rate of variation of simultaneously *all* component functions $f_N$. From the viewpoint of pQCD, such a requirement connects all orders of perturbation theory, and therefore is inherently non-perturbative.

***C*-continuity** is a combination of fragmentation invariance and a special continuity in particles' parameters, formulated without reference to the structure of perturbative partonic states. 3.34

The usual shape observables such as thrust and the jet number discriminators (as well as the classes of observables described in [4]) are *C*-continuous whereas the thrust axis is not. Nor are *C*-continuous the number of jets and individual jets parameters — irrespective of the jet definition adopted.
On the other hand, the prescriptions of Section 9 eliminate (most) *C*-discontinuities from observables constructed on the basis of jet configurations found by the optimal jet definition introduced in this paper.

- Concerning the regularization prescriptions of Sec. 2.52, we note that if the l.h.s. of 2.54 is *C*-continuous (which is often the case in practical situations) then such is 2.57.

## Structure of the space of *C*-continuous functions 3.35

The simplest example of *C*-continuous functions is immediately deduced from the definitions 3.32 and 3.21. Suppose $f(\hat{\mathbf{p}})$ is continuous everywhere on the unit sphere. Then the function $f(\mathbf{P})$ defined on events[i] according to

$$f(\mathbf{P}) = \langle \mathbf{P}, f \rangle = \sum_a E_a\, f(\hat{\mathbf{p}}_a) . \qquad 3.36$$

(cf. 3.11) is *C*-continuous by definition. Such $f(\mathbf{P})$ will be called <u>*basic shape observables*</u>. They will be further discussed in Sec. 3.43.

Furthermore, arbitrary *C*-continuous functions can be approximated by algebraic combinations of the basic shape observables in a fashion similar to how arbitrary continuous functions on, say, unit cube can be approximated by ordinary polynomials.[j] This analogy is illustrated by the following table:

| vector $\mathbf{P} = (\mathbf{P}_1, \ldots)$ | event **P** |
|---|---|
| unit cube $0 \le \mathbf{P}_i \le 1$ | the domain $\mathcal{P}$ (3.17) |
| continuity | *C*-continuity |
| linear functions $\sum_i c_i \mathbf{P}_i$ | basic shape observables (Eq.3.36) |
| products of linear functions (monomials) | (multi-)energy correlators (Eq.3.40) |
| continuous functions $f(\mathbf{P})$ | *C*-continuous observables $f(\mathbf{P})$ (generalized shape observables) |

3.37

[i] Note a convenient abuse of notation: both the angular function and the corresponding shape observable are denoted by the same symbol $f$. Interpretation depends on the type of arguments.

[j] A well-known theorem due to Weierstrass. Its generalization needed for our purposes is known as the Stone-Weierstrass theorem [18]. A mathematician would easily supply the details which physicists, however, won't care about because they don't lead to useful algorithms.

The approximation meant here is in the usual uniform sense, i.e. for any $\varepsilon > 0$, an arbitrary *C*-continuous function $f(\mathbf{P})$ can be approximated by a sum of energy correlators $f'(\mathbf{P})$ so that:

$$\sup_{\mathbf{P}\in\mathcal{P}} \left| f(\mathbf{P}) - f'(\mathbf{P}) \right| < \varepsilon\,. \qquad 3.38$$

The two classes of observables shown in the right column (basic shape observables and energy correlators) play special roles from the viewpoint of the underlying physical formalism. We have already seen that the basic shape observables 3.36 are singled out by their relation to the structure of elementary detector modules (we will return to this in Sec. 3.42). Let us now discuss the energy correlators.

### Energy correlators 3.39

These have the form

$$f(\mathbf{P}) = \sum\nolimits_{a_1 \ldots a_n} E_{a_1} \ldots E_{a_n}\, f_n(\hat{\boldsymbol{p}}_{a_1}, \ldots, \hat{\boldsymbol{p}}_{a_n})\,, \qquad 3.40$$

where $f_n$ is a symmetric continuous function of $n$ arguments. Basic shape observables are special cases corresponding to $n = 1$.

The component functions 3.19 and the correlators 3.40 can be regarded as different bases in terms of which to express general *C*-continuous observables. One function $f_n$ in 3.40 corresponds to an infinite sequence of component functions 3.19. On the other hand, Eq. 3.40, unlike 3.19, is automatically fragmentation invariant.

Furthermore, the correlators 3.40 are singled out for two theoretical reasons which reflect the fundamental structures of, respectively, quantum field theory and QCD. This has far reaching consequences.

First, such correlators naturally fit into the general structure of quantum field theory where the apparatus of multiparticle correlators is intimately related to the fundamental formalism of Fock space and is central in quantum field theory and statistical mechanics [19] because it allows one to systematically describe systems with a fluctuating number of particles (as is the case e.g. with multiparticle events in high-energy physics experiments).

Second, the energy correlators 3.40 are directly expressed in terms of the fundamental energy-momentum tensor [5] (we do not need explicit expressions here). This allows one to directly address the well-known problem that predictions of pQCD are formulated in terms of quark and gluon fields whereas experimental data deal with the observed hadronic degrees of freedom. Indeed, the energy-momentum tensor is determined solely by the space-time symmetries of QCD. It is thus independent of a particular operator basis used to represent the theory (quark and gluon fields, or hadronic fields) and so absorbs all the unknown complexity of confinement and hadronization. Therefore, observables which are expressible in terms of the energy-momentum tensor can be computed either in terms of hadronic degrees of freedom or from perturbative quarks and gluons. For such observables, the criterion of infrared safety (cancellation of singular logarithms, etc.) reduces to verification of existence of the energy-momentum tensor in QCD as an operator object. We will return to this in Sec. 4.1, and here only note that the described way of reasoning clarifies the conjecture of [8] that observables for which pQCD predictions make sense are those for which infrared and collinear singularities cancel thus ensuring their insensitivity to non-perturbative physics. Such a cancellation is guaranteed for the energy correlators 3.40 (provided the angular function $f_n$ satisfies some additional regularity restrictions; see Sec. 4.1) whereas the general *C*-continuous functions are approximated by sums of energy correlators in such a way that the properties of fragmentation invariance etc. required for such cancellations are not compromised.

The direct connections of the energy correlators 3.40 with QFT and QCD reflect the physical nature of the phenomena concerned and ensure their superior amenability to theoretical studies (cf. the abundance and quality of theoretical calculations for the simplest shape observable thrust [20] and the method of a systematic study of power corrections outlined in [16]).

### Generalized shape observables 3.41

A few comments are in order concerning generalized shape observables. These are essentially arbitrary *C*-continuous functions. They are obtained from energy correlators using algebraic operations and appropriate limiting procedures which do not violate the property of *C*-continuity (theoretically, it is sufficient to ensure a uniform convergence on $\mathcal{P}$ in the sense of 3.38). Roughly speaking, such operations are applied to observables as a whole (i.e. after averaging over all events) and they should not allow arbitrary growth of the rate of variation of the component functions 3.19 for $N \to \infty$. An example of a correct limiting procedure is the minimization over the thrust direction involved in the definition of thrust; cf. 4.68. For an example of illegal sum see sec. 6.9 of [4].

It is clear a priori that if quantum field theory is a fundamental mechanics governing the phenomena observed in high energy physics, then it should be possible to express any truly observable phenomena (unlike artifacts such as instabilities) in a QFT-compatible language, i.e. via observables that can be approximated by energy correlators. This was the original rationale behind the theory of [3]–[5].

Unfortunately, even in one dimension simplest polynomial approximations in the spirit of the Weierstrass theorem (polynomial interpolation formulas) are seldom sufficient: spline approximations build by gluing local polynomials are generally more useful. This is even more so in infinitely many dimensions (as is the case with $\mathcal{P}$), whence the need for special tricks such as jet algorithms. An array of prescriptions allowing to simulate conventional jet-based observables such as dijet mass distributions in the language of *C*-continuous observables was described in [4]. Such prescriptions altogether circumvent representation of events in terms of jets.

However, the purpose of the examples presented in [4] was primarily to demonstrate mathematical mechanisms ensuring that information extracted via generalized shape observables contains features that can be directly related to the conventional procedures (such as $\delta$-spikes in the so-called spectral discriminators corresponding to multi-jet substates). For practical purposes, it may be more convenient to start from the conventional observables and try to eliminate *C*-discontinuities which spoil optimality of observables. Prescriptions for doing so are described later on in this paper.

In what follows we will be using the term *C*-continuous observables as less ambiguous than generalized shape observables.

### Basic shape observables and physical information 3.42

So, one can represent an event either in terms of particles, Eq. 3.6, or in terms of values of basic shape observables, cf. 3.12. In the absence of detector errors and other imperfections, the two representations are numerically equivalent.

However, the structure of detector errors is an essential part of physics, and in this respect the two representations differ: the numbers which constitute the r.h.s. of 3.12 *individually* possess the correct stability properties with respect to small distortions of the event — distortions of the kind specific to the type of measurements we deal with. The numbers which constitute Eq. 3.6, on the contrary, do not possess this property.

For clarity's sake, suppose the event has two sufficiently energetic particles $a$ and $b$ whose directions are close. Then replacing the pair $a$, $b$ with one particle $c$ whose 3-momentum is the sum of the 3-momenta of $a$ and $b$ is deemed to distort the calorimetric physical information carried by the event only a little (the less the difference between $\hat{\boldsymbol{p}}_a$ and $\hat{\boldsymbol{p}}_b$, the less the distortion). The individual numbers which constitute the r.h.s. of 3.6 do not have this property: they can exhibit non-negligible chaotic fluctuations even if the physical information content of the event varies negligibly.

A simple analogy may further help to understand the role of continuity: imagine a ruler marked randomly instead of the standard ordered numbering. Representing length by using a number obtained from such a ruler would be not dissimilar to representing the event via 3.6: it would be sufficient for book-keeping purposes, but it would require great care in construction of data processing algorithms such as computation of volumes, prices, etc.

Similarly, whereas the representation 3.6 is convenient for book-keeping purposes, one should avoid relying on its form in the design of data processing algorithms. Such algorithms should in general respect additional restrictions not reflected in 3.6, namely, the restriction of $C$-continuity. The difficulties encountered by the experts in jet definition (such as a lack of fragmentation invariance of some suggestions related to jet algorithms) are often artifacts due to a failure to reason about jets and energy flows in terms which correctly reflect the physical nature of the problem.

Note in this respect that all the seemingly abstract notions which we introduced (events as measures on the unit sphere, $C$-continuity, etc.) are essentially only notations, i.e. formulaic expressions of *what is*.

In fact, these notions are neither more abstract nor difficult than, say, the differential calculus. But they are usually taught as "advanced" topics in the abstract courses of functional analysis without link to applications, which earns them a bad reputation among physicists. Then when these notions are actually encountered there is a psychological tendency to reject them as too abstract to be useful in practical physics.

We are ready to take a philosophical look at Eq. 3.12:

> Ideal physical information content of the event **P** is identified with the collection of values of all basic shape observables, i.e. with the r.h.s. of Eq. 3.12.
> 3.43

The adjective "ideal" reminds us that in practice only a finite subset of the collection is used, as in 3.14. One should feel no more psychological discomfort with such a collection than with, say, a transcendental number such as $\pi$ which is completely specified by an infinite number of digits but in practice represented by their finite sequences.

By identifying the information content of the event **P** with the collection of expressions 3.36, not only have we not deviated from the experimental reality but we have actually returned closer to it compared with the r.h.s. of 3.6 (if only by allowing finite angular resolutions), at the same time giving it a systematic form which is convenient for the derivation of the jet definition (Section 5).

## Observables in QCD. Dynamical aspects 4

Now we turn to dynamical (i.e. QCD-specific) considerations in the construction of optimal observables according to Eq. 2.17 in the context of hadronic events produced in high energy physics experiments. The big problem is that such events contain $O(100)$ particles described by 3 degrees of freedom each. On the other hand, the underlying physics is controlled by a few Standard Model parameters, whereas all the complexity of hadronic events is supposed to be generated by the QCD Lagrangian that contains only one coupling $\alpha_S$ and quark and gluon fields most of which can be regarded as massless. This means that from the viewpoint of studies of both the Standard Model and QCD Lagrangian most of the observed degrees of freedom are physically not important. In the language of the theory of optimal observables (Sec. 2.7), one could say that the optimal observables for extraction of the Standard Model parameters are mostly sensitive to a few degrees of freedom which the conventional wisdom identifies with the representation of events in terms of jets.

The chain of reasoning presented below is intended to make more explicit, and thus help to clarify the argumentation of the theory of jets including the part about inversion of hadronization. Much of the argumentation is familiar but phrased in a more formal language to facilitate a systematic investigation.

Since we are interested in issues such as hadronization, the perturbation theory discussed below only concerns QCD. Electroweak effects are assumed to be taken into account in the theoretical amplitudes as necessary.

### The basic conjecture of the QCD theory of jets 4.1

The conjecture of Sterman and Weinberg [8] is that the property of infrared safety of observables ensure their calculability within the framework of pQCD. We would like to express it in a formal fashion and to connect the notion of IR safety with $C$-continuity (Sec. 3.31).

In the final respect, one needs (quasi-) optimal observables (Sec. 2.25) to extract the values of fundamental parameters such as the mass of the $W$ boson from hadronic data. Because of a large dimensionality of observed hadronic events **P**, one needs some specific structural information about the ideal probability density $\pi(\mathbf{P})$ of their production (ideal = not taking into account detector errors). Such information is obtained from pQCD which deals, however, not with hadronic but quark and gluon degrees of freedom.

The conclusions of [8], [5] (also see Sec. 3.39 above) can be summarized as follows.For any $C$-continuous observable $f(\mathbf{P})$ it is correct to compute theoretical predictions within the framework of pQCD. Formally:

$$\int \mathrm{d}\mathbf{P}\, \pi(\mathbf{P})\, f(\mathbf{P}) \;=\; \int \mathrm{d}\mathbf{p}\, \pi^{(N)}_{\mathrm{pQCD}}(\mathbf{p})\, f(\mathbf{p}) + O(\alpha_S^{N+1})\,. \qquad 4.2$$

Here $\pi(\mathbf{P})$ represents the exact probability density so that the l.h.s. is what experiments would see given ideal detectors and infinite statistics. The variable **p** on the r.h.s. represents perturbative quark and gluon final states. $\pi^{(N)}_{\mathrm{pQCD}}$ is the corresponding probability density computed within the shown precision in $\alpha_S$ from pQCD; it is a sum of contributions proportional to $\alpha_S^n$, $n = 0,\ldots,N$.

(i) The mathematical structure of events in the two expressions is essentially the same from the viewpoint of data processing (Eq. 3.6), the difference being in the number of particles ($O(100)$ for **P** and $O(1)$ for **p**).

(ii) The restriction of $C$-continuity is important for the validity of 4.2 in so far as fragmentation invariance and regularity properties (a continuity and related stronger regularity restrictions; cf. Sec. 4.3) have to be formulated non-perturbatively for the non-perturbative expression on the l.h.s.

(iii) QUARK-HADRON DUALITY. The proposition that it is possible to replace a sum over hadronic states by the corresponding partonic sum is known as the hypothesis of quark-hadron duality. The scenario of derivation of 4.2 described in Sec. 3.39 (see also Sec. 4.3) circumvents such direct replacement via an intermediate representation in which only an average over the initial state of a product of energy-momentum tensor densities is involved.

However, there is also the dynamical aspect, namely, that the perturbation theory would actually work in a numerically satisfactory fashion. This cannot be explained by reference to the energy-momentum tensor per se but is made possible by such a representation as it allows application of the usual renormalization group argument exactly as in the case of total cross sections.

Still, a reference to renormalization group is insufficient inasmuch as the convergence of the expansion on the r.h.s. of 4.2 depends on the behavior of the observable $f$. The easiest way to see this effect is by looking at the so-called power-suppressed corrections that are parametrized in terms of coefficients not predictable from perturbation theory.[k] A little experience with asymptotic expansions of integrals of perturbation theory[l] makes it obvious that such corrections are proportional to angular derivatives of the observables $f$ in 4.2: for $f$ which vary too fast at too many points of the phase space the perturbative expansion would not work. This confirms the notion that perturbation theory cannot predict small-scale angular correlations in observed events.

### A scenario of formal verification of Eq. 4.2 4.3

(Readers not interested in formal aspects may skip the technical details below and go directly to Sec.4.9.)

As long as the construction of perturbation theory is performed in an axiomatic fashion rather than derived from non-perturbatively formulated fundamental equations (as is done in conventional treatments [19]), any proof of 4.2 is bound to be only a more or less plausible scenario because the non-perturbative l.h.s. is, essentially, a theoretical fiction. So if it were possible to accurately verify 4.2, one would have to do so roughly as follows.

The first step would be to establish 4.2 for $f$ that are energy correlators, Eq.3.40. One would start with a non-perturbatively defined expression (the l.h.s. of 4.2), represent it in terms of correlators of the energy-momentum tensor densities as explained in [5], and then develop an expansion in $\alpha_S$, ending up with the r.h.s. of 4.2.

A technical subtlety is that owing to the singularities of pQCD the integrals on the r.h.s. of 4.2 are well defined only for $f$ that obey somewhat stronger regularity restrictions than a mere continuity.

The simplest illustration for this can be borrowed from pQCD where a typical object in theoretical answers is the so-called +-distribution, $(1-x)_+^{-1}$ (see e.g. [24]; here $x$ is the parton fraction but the analytical mechanism being demonstrated is completely general). This distribution is defined by its integration properties:

$$\int_0^1 \mathrm{d}x\, (1-x)_+^{-1} f(x) = \int_0^1 \mathrm{d}x\, \frac{f(x) - f(1)}{1-x}\,. \qquad 4.4$$

For the r.h.s. to be a well-defined integral, it is not sufficient that $f(x)$ is merely continuous, i.e. $f(x) \to f(1)$; one must also assume that $f(x)$ approaches $f(1)$ sufficiently fast, e.g.

$$f(x) - f(1) = O(|1-x|)\,. \qquad 4.5$$

This is satisfied e.g. if $f$ has continuous first derivatives.

The technical regularity restrictions on observables $f(\mathbf{P})$ required for the r.h.s. of 4.2 to be well-defined are multi-dimensional analogs of the restriction 4.5. For practical purposes it is sufficient to require e.g. that the angular functions $f_n$ in 3.40 have continuous first derivatives.[m] (Ref. [17] formulated the restrictions in a slightly more general form of the Hölder condition — but in the language of the component functions 3.19.)

That this regularity restriction does not become more stringent in higher orders of perturbation theory follows from the fact that the severity of neither soft nor collinear singularities in QCD increases in higher orders of perturbation theory (cf. [17], [25]; this property is related to renormalizability of QCD). But even if it did, it would not be an obstacle for the theory: one would only have to require that observables are smooth (i.e. belong to the class $C^\infty$).

The second step would be to extend Eq. 4.2 to more general observables than finite sums of energy correlators. This — as is usual in situations of this sort — would be accomplished by a limiting procedure with respect to $f$ which would commute with the limit $\alpha_S \to 0$. To this end, one has to rewrite the mentioned regularity conditions in a non-perturbative form. For instance, an analog of 4.5 could be

$$\left| f(\mathbf{P}) - f(\mathbf{P}')\right| \le K_f\, \mathrm{Dist}(\mathbf{P},\mathbf{P}') \quad \text{for any } \mathbf{P},\mathbf{P}' \in \mathcal{P}\,, \qquad 4.6$$

where Dist is defined in 3.27. One sees that if $f$ is an energy correlator 3.40 then Eq.4.6 implies that the angular function $f_n$ satisfies an analog of 4.5. Then one would define the norm

$$\| f \| = \max_{\mathbf{P}} \left| f(\mathbf{P}) \right| + K_f\,, \qquad 4.7$$

and define the space $C'(\mathcal{P})$ as the corresponding closure of the subspace spanned by energy correlators satisfying 4.6. This would be similar to the standard functional class $C^1$. Recall that functions from the class $C^1$ can be uniformly approximated by polynomials together with their first derivatives. Observables from $C'(\mathcal{P})$ can similarly be approximated by finite sums of energy correlators ex-cept that our formulation is in terms of 4.6 instead of derivatives in the argument **P** for purely technical reasons.[n]

[k] Cf. the studies of such corrections in the theory of QCD sum rules [21].

[l] Cf. a systematic scenario described in [16] based on the expansion method of the so-called asymptotic operation [22], [23] which is directly formulated in terms of $\delta$-functional counterterms, so that corrections suppressed by powers of the total energy involve derivatives of $\delta$-functions (power counting mechanisms ensure that higher power-suppressed corrections are accompanied by more derivatives on $\delta$-functions). After integrations with $f$, the derivatives are switched from $\delta$-functions to $f$.

[m] A similar technical assumption — existence of continuous derivatives of $f$ through second order — will be made in the derivation of the key bound for jet definition in Sec. 6.10.

Finally, one would need inequalities of the form

$$\left|\int \mathrm{d}\mathbf{P}\, A(\mathbf{P})\, f(\mathbf{P})\right| \le C_A \| f \| \qquad 4.8$$

for $A = \pi,\ \pi^{(N)}_{\text{pQCD}},\ \pi - \pi^{(N)}_{\text{pQCD}}$. In the first case ($A = \pi$) an even weaker inequality is expected to be true (with the norm defined without the second term on the r.h.s. of 4.7). In the second case the inequality is essentially equivalent to the proposition that the soft and collinear singularities in individual diagrams of pQCD are never more severe than logarithmic.[o] In the third case one would also have to verify that $C_A = O(\alpha_S^{N+1})$ (such a proposition is unlikely not to be true).

All in all, there does not seem to exist any analytical mechanism which might invalidate any of the listed propositions because of the intrinsic analytical naturalness of the described scheme. Although some technical details (e.g. the description of regularity conditions) might need to be made more precise, the basic requirement of *C*-continuity fits into the general scheme of things so tightly and naturally, from the viewpoints of both physics and mathematics, that it seems unlikely that it could require a modification.

### The mechanism behind Eq. 4.2 — 4.9

Let us now try to understand the structural reasons behind Eq. 4.2.

The reasoning below will be more transparent if one bears in mind that discussing *C*-continuous functions defined on $\mathcal{P}$ is rather similar to discussing ordinary continuous real functions $f(x)$ defined on the simplex $\mathcal{S}$, the part of the euclidean space $\mathbf{R}^n$ described by $x_i \ge 0$, $\sum_i x_i \le 1$ (the latter restriction is analogous to 3.16). The distance 3.27 is similar in general properties to the usual euclidean distance in $\mathbf{R}^n$ although the explicit expression is rather different. However, it is exactly this difference that masks the dissimilarity of the infinitely dimensional space of measures on the unit sphere from the ordinary euclidean space $\mathbf{R}^n$ and thus makes possible the analogy between the events $\mathbf{P} \in \mathcal{P}$ and the vectors $x \in \mathcal{S}$.

The *C*-continuous observable *f* in Eq. 4.2 is in principle arbitrary and so can probe any small region in $\mathcal{P}$ (smallness can be measured using the distance 3.27). Let $\Pi$ be a region of $\mathcal{P}$ such that:

(i) $\Pi$ is small, i.e. any two events from $\Pi$ differ by slightly acollinear fragmentations into/recombinations of, any number of particles. Formally, one can say that the distance Dist between any two events from $\Pi$ is small (i.e. $\ll 1$).

(ii) Events from $\Pi$ are produced with a relatively significant probability formally given by $\int \mathrm{d}\mathbf{P}\pi(\mathbf{P})\theta(\mathbf{P} \text{ is from } \Pi)$.

The condition (i) means that for any fixed event **Q** from $\Pi$, one would have:

$$f(\mathbf{P}) \approx f(\mathbf{Q}) \quad \text{for any } \mathbf{P} \in \Pi \text{ and for any } C\text{-continuous } f. \qquad 4.10$$ [p]

$\Pi$ contains events with an arbitrarily large number of particles but the number of particles is always positive and so limited from below by a minimum value. Choose any **Q** from $\Pi$ so that its number of particles is equal to the minimum value. (This need not fix **Q** uniquely.) Then the condition (ii) implies that $\pi^{(N)}_{\text{pQCD}}(\mathbf{Q})$ is significant, i.e. that **Q** cannot contain more than a few particles if perturbation theory works because emission of each additional particle is then suppressed by an additional factor $\alpha_S$.

That events from $\Pi$ are close to **Q** in the sense of the *C*-convergence (as measured e.g. by the distance 3.27) means that such events consist of a few more or less narrow energetic sprays of particles (each spray roughly corresponding to a particle of **Q**) and perhaps some soft background, i.e. randomly directed particles which together carry a small fraction of the event's energy.

### Formal model of hadronization — 4.11

One adopts the following theoretical model for the probability distribution of events **P** which is built into any Monte Carlo event generator:

$$\pi(\mathbf{P}) \approx \int \mathrm{d}\mathbf{p}\ \pi^{(n)}_{\text{pQCD}}(\mathbf{p}) \times H^{(n)}(\mathbf{p}, \mathbf{P}), \qquad 4.12$$

where $H^{(n)}(\mathbf{p}, \mathbf{P})$ is the probability for the parton event **p** to develop into the observed event **P**.

- The approximate equality in 4.12 is meant to indicate that it is not a theorem that the probability $\pi(\mathbf{P})$ can be exactly represented in such a convolution form. (With $O(100)$ free parameters in $H^{(n)}$ the error can be made very small, of course.)

Note the following normalization restriction:

$$\int \mathrm{d}\mathbf{P}\ H^{(n)}(\mathbf{p}, \mathbf{P}) \equiv 1. \qquad 4.13$$

The hadronization kernel $H^{(n)}$ must depend on *n* if the r.h.s. of 4.12 as a whole is to represent the exact non-perturbative answer. In practice *n* is fixed and small.[q]

- Higher perturbative terms are to be added to $\pi^{(n)}_{\text{pQCD}}$, whereas $H^{(n)}$ — which is supposed to express the effect of the entire sum of missing terms — acts on $\pi^{(n)}_{\text{pQCD}}$ multiplicatively. It would be interesting to clarify this point in a systematic manner.

Represent the l.h.s. of 4.2 in terms of 4.12:

$$\int \mathrm{d}\mathbf{P}\ \pi(\mathbf{P})\, f(\mathbf{P}) = \int \mathrm{d}\mathbf{p}\ \pi^{(n)}_{\text{pQCD}}(\mathbf{p}) \int \mathrm{d}\mathbf{P}\ H^{(n)}(\mathbf{p}, \mathbf{P})\, f(\mathbf{P}). \qquad 4.14$$

This agrees with the r.h.s. of 4.2 if

$$\int \mathrm{d}\mathbf{P}\ H^{(n)}(\mathbf{p}, \mathbf{P})\, f(\mathbf{P}) = f(\mathbf{p}) + O(\alpha_S^{n+1}). \qquad 4.15$$

The approximate equality here is supposed to be valid for any *C*-continuous function *f*. For this reason, Eq. 4.15 can be conveniently represented as follows:

---

[n] For instance, note the curious fact that elements of the tangent space to $\mathcal{P}$ at any point **P** are distributions on the unit sphere. In other words, the tangent space (a natural habitat of the differentials d**P**) is different from the complete linear space to which it is tangent. One would probably have to develop a differential calculus for functions on $\mathcal{P}$ and reformulate 4.6 in terms of the space $C^1(\mathcal{P})$ if QCD were non-renormalizable because then one might need to require smoothness (the property $C^\infty$) of all the functions $f(\mathbf{P})$ involved.

[o] This has the same power-counting reasons behind it as the renormalizability of pQCD.

[p] At this point we don't discuss how the approximation error depends on *f*, etc. See Sec. 5.17.

[q] Strictly speaking, the convolution 4.12 ought to be performed at the level of quantum amplitudes rather than probabilities but we ignore such details here.

$$H^{(n)}(\mathbf{p},\mathbf{P}) = \delta(\mathbf{p},\mathbf{P}) + O(\alpha_S^{n+1}) \,, \qquad 4.16$$

where the $\delta$-function on the r.h.s. is defined in the usual way with respect to integration over $\mathbf{p}$ or $\mathbf{P}$.

• Eqs. 4.15, 4.16 are simply a convenient formulaic expression of the verbal statement that observed events generated from the partonic event $\mathbf{p}$ mostly consist of narrow jets that resemble the parent partons.

Taking into account the normalization 4.13 and the fact that $f$ is arbitrary (apart from the general restriction of $C$-continuity), one deduces from 4.15 that for $\mathbf{P}$ typically generated from $\mathbf{p}$ according to $H^{(n)}$, one has

$$f(\mathbf{P}) = f(\mathbf{p}) + O(\alpha_S^{n+1}) \,. \qquad 4.17$$

This is another form of the proposition that $\mathbf{P}$ is similar to $\mathbf{p}$. The meaning of similarity is established by the restriction of $C$-continuity of $f$ that are allowed in 4.17.

• If one defines a configuration of jets $\mathbf{Q}$ to roughly correspond to the partonic event $\mathbf{p}$ then Eq. 4.17 implies that $\mathbf{Q}$ should satisfy the relation $f(\mathbf{Q}) \approx f(\mathbf{P})$. We will see it again in Sec. 5.6

### Sensitivity to hadronization and $C$-continuity 4.18

We saw in Sec. 2.48 that optimal observables are $C$-continuous as a result of the smearing caused by detector errors described by 2.49. $C$-continuity made observables less sensitive to such errors. In the present dynamical context, we note that hadronization is described by a similar convolution 4.12, and for $C$-continuous observables fluctuations induced by the stochastic hadronization are suppressed too.

• Actually, Eq. 4.2 means that $C$-continuity makes observables insensitive (within the precision of perturbative approximation) to the hadronization effects which transform the perturbative $\pi^{(N)}_{\mathrm{pQCD}}$ into the hadronic $\pi(\mathbf{P})$. Remember that the mentioned precision of perturbative approximation depends on the magnitude of derivatives of the observable.

## Formal construction of optimal observables 4.19

Suppose we wish to measure a fundamental parameter $M$ such as the mass of the $W$ boson. All the dependence on such a parameter is localized within $\pi_{\mathrm{PT}}$. Then we can combine 2.17 and 4.12 and write down a formal expression for the corresponding optimal observable:

$$f_{\mathrm{opt}}(\mathbf{P}) = \partial_M \ln \pi_{\mathrm{th}}(\mathbf{P}) = \frac{\int d\mathbf{p} \left[\partial_M \pi^{(n)}_{\mathrm{pQCD}}(\mathbf{p})\right] \times H^{(n)}(\mathbf{p},\mathbf{P})}{\int d\mathbf{p}' \left[\pi^{(n)}_{\mathrm{pQCD}}(\mathbf{p}')\right] \times H^{(n)}(\mathbf{p}',\mathbf{P})} \,. \qquad 4.20$$

• The philosophical importance of this expression is that it corresponds to the fundamental Rao-Cramer limit on the attainable precision for the values of $M$ extracted from a given data set (recall Sec. 2.7 and the comments after 2.22). Therefore Eq. 4.20 is an ideal starting point for deliberations about any data processing algorithms (including jet algorithms) geared towards specific precision measurement applications.

The key difficulty is that neither the probability distribution 4.12 nor the formula 4.20 can be evaluated for a given $\mathbf{P}$ due to the huge number of degrees of freedom in $\mathbf{P}$ (in reality, theoretical versions of $\pi(\mathbf{P})$ as a whole are materialized only in the form of Monte Carlo event generators). This means that a careful choice of parametrization of events is needed before the construction of good approximations to $f_{\mathrm{opt}}$ becomes possible.

• With a suitable parametrization, Eq. 4.20 could be used in a brute force fashion: one would map the events into a multidimensional domain of the chosen parameters (say, $\mathbf{q}$), build a multi-dimensional interpolation formula for $\pi(\mathbf{P}(\mathbf{q}))$ (via an adaptive routine similar to those used e.g. in [11]) for two or more values of $M$ near the value of interest, and perform the differentiation in $M$ numerically. The resulting multidimensional interpolation formula would represent the optimal observable mapped to $\mathbf{q}$ and could be used for the processing of experimental events to complement the standard $\chi^2$ method based on histogramming (recall the comments in Sec. 2.41).

The difficulty is to find a parametrization that would not involve a significant loss of information about $M$.

Usually employed are parametrizations obtained by describing the events $\mathbf{P}$ in terms of a few jets, which is made possible by the specific structure of 4.20, namely:

(i) Eq. 4.16 means that observed events $\mathbf{P}$ are close (in the sense of $C$-continuity as measured e.g. by the distance 3.27) to their parent parton events $\mathbf{p}$.

(ii) The dimensionality of $\mathbf{p}$ is small.

Finding such a $\mathbf{p}$ for each observed event $\mathbf{P}$ amounts to an approximate inversion of hadronization. This will be further discussed in Sec. 4.28. Here we would like to take a slightly different view on the problem.

If one could restore $\mathbf{p}$ from $\mathbf{P}$ uniquely, then the optimal observable would be identified with its perturbative version:

$$f^{(n)}_{\mathrm{opt}}(\mathbf{p}) = \partial_M \ln\left(\pi^{(n)}_{\mathrm{pQCD}}(\mathbf{p})\right). \qquad 4.21$$

However, the perturbative probability density $\pi_{\mathrm{pQCD}}(\mathbf{p})$ contains singular expressions (generalized functions such as the one represented by Eq. 4.4) that are not positive-definite, beyond the leading order[r]. This means that the perturbative expression $\pi_{\mathrm{pQCD}}(\mathbf{p})$ cannot be immediately interpreted as a probability density. As a result, the derivation of optimal observables described in Sec. 2.7 is inapplicable. In other words, the expression 4.21 is formal beyond the tree approximation of pQCD.

Nevertheless, it is not impossible to use Eq. 4.21 for the construction of quasi-optimal observables provided one could find a natural way to extend it (or some its simplified version) to all events $\mathbf{P}$ by $C$-continuity. (Remember that the formal nature of $\mathbf{p}$ and $\mathbf{P}$ is the same.) Such an extension can sometimes be accomplished in such a way that the problem of restoring $\mathbf{p}$ from $\mathbf{P}$ does not occur. Here is an example.

### Constructing observables via extension by $C$-continuity. Precision measurements of $\alpha_S$ 4.22

Consider measurements of the strong coupling $\alpha_S$ in the process $e^+e^- \to$ hadrons. We are going to show how the concept of optimal observables could have been employed to obtain shape observables that best suit this purpose.

[r] If $\varphi(x) > 0$ and $\varphi(x) = \varphi_0 + \varphi_1 x + \varphi_2 x^2 + \ldots$ then $\varphi_0 > 0$ (if it is nonzero) but the sign of $\varphi_1$ etc. may be arbitrary.

At a very crude level of reasoning, the probability density can be represented as a direct sum

$$\pi_{\text{pQCD}}(\mathbf{p}) = \{\pi_2(\mathbf{p})\}_2 \oplus \{\alpha_S \pi_3(\mathbf{p})\}_3 \oplus \{\alpha_S^2 \pi_4(\mathbf{p})\}_4 \oplus \dots, \quad 4.23$$

where each term corresponds to the $n$-particle sector of the space of **p**. Each $\pi_n$ also contains $O(\alpha_S)$ corrections.

Use the prescription 4.21 with $M \to \alpha_S$, multiply the r.h.s. by $\alpha_S$ (because $f_{\text{opt}}$ is defined up to a constant), and drop higher-order terms in each sector, which is not prohibited by the prescriptions of Sec. 2.25. Then one obtains:

$$f_{\text{pragm}}(\mathbf{P}) \sim \{0\}_2 \oplus \{1\}_3 \oplus \{2\}_4 \oplus \dots \quad 4.24$$

In other words, a minimal requirement is that the observables should vanish on 2-particle events. This is exactly the requirement which was used in [3], [4] to derive the so-called jet-number discriminators $\mathbf{J}_m[\mathbf{P}]$ by assuming the simplest analytical form for $f_{\text{quasi}}$ (a 3-particle correlator). The simplest $C$-continuous expression corresponding to the above requirement then is

$$f_{\text{quasi}}(\mathbf{P}) = \mathbf{J}_3[\mathbf{P}]. \quad 4.25$$

(See [3], [4] for exact expressions.)

Remember, however, that there is an arbitrariness in the construction of $\mathbf{J}_m[\mathbf{P}]$ in [3], [4]: the factors $\Delta_{ij}$ involved in the construction are only required to behave as $O(\theta_{ij}^c)$ with a positive $c$ as $\theta_{ij} \to 0$ ($\theta_{ij}$ is the angle between $i$-th and $j$-th particles of the event). The simplest analytical behavior corresponds to $c=1$ whereas the simplest covariant expressions correspond to $c=2$.

- It might be possible to fix this arbitrariness, as follows. The perturbative expression for $\pi_3$ is singular and not strictly non-negative exactly in situations corresponding to $\theta_{ij} \to 0$. To rectify this one could perform a resummation of perturbation series thus introducing a non-trivial dependence on $\alpha_S$. Then the differentiation in 2.17 would replace the 1, 2, … in 4.24 with something more interesting (i.e. dependent on $\alpha_S$) in the region $\theta_{ij} \to 0$. Then by examining how such a dependence affects the result of differentiation in $\alpha_S$ in the definition of $f_{\text{opt}}$, one might be able to modify the observable 4.25 accordingly. This interesting theoretical problem seems to require a kind of pQCD expertise similar to that behind the $k_T$-algorithm [32].

The increasing integer weight in each sector in 4.24 corresponds to the simple fact that higher powers of $\alpha_S$ are increasingly more sensitive to its variations. So the expression 4.24 suggests to replace 4.25 with a sum similar to the following one:[s]

$$f_{\text{quasi}}(\mathbf{P}) = \tilde{\mathbf{J}}_3[\mathbf{P}] + 2\tilde{\mathbf{J}}_4[\mathbf{P}] + 3\tilde{\mathbf{J}}_5[\mathbf{P}] + \dots \quad 4.26$$

Of course, the series cannot contain more terms than the number of theoretically known corrections to $\pi_{\text{pQCD}}$.

Actually, any conventional shape observable that vanishes (only) on 2-particle configurations meets the above requirement. For instance, one such shape observable is the combination $1-T$, where $T$ is the so-called thrust (eq.(46) in [1] and refs. therein). In our notations (we assume that the event's total energy is normalized to 1), the explicit expression is

$$1 - T(\mathbf{P}) = 1 - \max\left[\sum\nolimits_a E_a \,|\cos\theta_a|\right] = \min\left[\sum\nolimits_a E_a \left(1 - |\cos\theta_a|\right)\right], \quad 4.27$$

where $\theta_a$ is the angle between the $a$-th particle's direction and an axis (the thrust axis). The optimizations are performed with respect to the directions of the thrust axis that determines orientation of the angular function as a whole but does not affect the magnitude of derivatives, which ensures $C$-continuity of $1-T(\mathbf{P})$.

In terms of the classification of Sec. 3.35, the definition 4.27 belongs to the class of generalized shape observables because it involves an optimization procedure on top of a basic shape observable.

The above reasoning can be regarded as an argument for quasi-optimality of the observables such as thrust and jet-number discriminators for precision measurements of $\alpha_S$. We will have to say more on this in Sec. 4.68.

## Constructing observables via jet algorithms. The conventional approach 4.28

Let us explore how one could construct quasi-optimal observables that would approximate 4.20 using the fact that the majority of hadronic events **P** resemble their partonic parents **p**, as formally expressed by 4.16 (4.15). Although it is impossible to exactly restore the parton parent **p** for each observed event **P** (see after 4.32, Sec. 4.47 and Sec. 5.10), the idea is a useful heuristic to start from.

The conventional approach to construction of observables involves three elements: a jet algorithm, an event selection procedure which we call the jet-number cut, and a function on jet configurations.

We will focus only on the general structure and properties of the conventional data processing scheme based on jet algorithms, and the specific form of the jet algorithm will play no role in the following discussion.

### General structure of jet algorithms 4.29

Assume there is a so-called *jet algorithm* that somehow accomplishes an approximate inversion of hadronization. Formally, such algorithm is a mapping of arbitrary events **P** into similar (pseudo) events **Q**:

$$\mathbf{P} \xrightarrow[\text{jet algorithm}]{} \mathbf{Q} = \mathbf{Q}[\mathbf{P}]. \quad 4.30$$

**Q** usually has many fewer (pseudo) particles than **P**.

Recall that partonic events **p** have the same formal nature as the hadronic events **P**. This implies, first, that **Q** is an object of the same nature as **P** and **p**; second, that the mapping 4.30 is defined on both hadronic and partonic events.

We will call **Q** *jet configurations* and their pseudoparticles, *jets*. For clarity's sake, we distinguish jets the mathematical objects (the pseudoparticles of **Q**) from jets the collections of particles (hadrons or partons) in which case we will use the terms *spray* or *cluster*, usually in informal reasoning.

Jets in **Q** will be labeled by the index $j$, and the $j$-th jet is characterized similarly to particles of the event **P** (cf. 3.6), i.e. by its energy and direction denoted as $\mathcal{E}_j$ and $\hat{\boldsymbol{q}}_j$:

[s] Recall that jet-number discriminators are normalized so that their maximal value reached on configurations with no less than $m$ widely separated particles, is equal to 1.

$$\mathbf{Q} = \{\mathcal{E}_j, \hat{\boldsymbol{q}}_j\}_{j=1\ldots N(\mathbf{Q})} . \qquad 4.31$$

In practice both particles and jets are endowed with additional attributes, e.g. Lorentz 4-momenta, and the jet algorithm evaluates them along the way, but at this point we ignore such complications.

That the mapping 4.30 is supposed to be an approximate inversion of the hadronization described by the kernels $H^{(n)}$ in Eq. 4.12, means that **Q** should be close to **p**. This can be represented as

$$\mathbf{Q} \approx \mathbf{p} . \qquad 4.32$$

The exact meaning of the approximate equality is yet to be specified, and it may be impossible to identify a single partonic configuration which hadronized into **P**. Still, there is a class of events for which the relation 4.32 is unambiguous (at least in the asymptotic limit of high energies), and this provides a minimal requirement which any jet algorithm must satisfy and which serves as a sort of boundary condition for jet definition which we present for explicitness' sake:

For events which consist of a few energetic well isolated narrow sprays of particles, each spray is associated with a jet whose energy and direction coincide with those of the spray. 4.33

The ambiguity of jet definition concerns how jet algorithms handle fuzzy events that do not fall into the above category.

Another important condition usually imposed on jet algorithms is that the mapping 4.30 should be fragmentation invariant. In the context of our theory this is essentially superfluous since the interpretation of events and functions on them modulo *C*-continuity (which incorporates fragmentation invariance; see 3.34) is built into our formalism at a linguistic level: If all the arguments are expressed in the language of 3.12 rather than the particle representation 3.6 then the resulting jet definition will be automatically fragmentation invariant.

Note that any reasonable jet algorithm sets, explicitly or implicitly, a lower limit on the angular distances between jets in **Q**. The limit may depend on jets' energies.

A related observation is that the mapping 4.30 cannot (unless it is trivial, i.e. $\mathbf{Q}[\mathbf{P}] = \mathbf{P}$) be continuous in any non-pathological sense for some **P**. The points of discontinuity usually correspond to the events whose different small deformations result in jet configurations with different numbers of jets.

## The jet-number cut 4.34

Another element of the conventional data processing scheme is the so-called *jet-number cut*, which is a selection procedure (similar to any other event selection procedure; see Sec. 2.35) based on the number of jets the chosen jet algorithm finds.

It is convenient to introduce a notation for the collection of events with a given number of jets (the *K-jet sector*):

$$\mathcal{P}_K = \{\mathbf{P} \in \mathcal{P} : \ \mathbf{Q}[\mathbf{P}] \text{ has } K \text{ jets}\} . \qquad 4.35$$

Then the space of events $\mathcal{P}$ is sliced into a sum of $\mathcal{P}_K$ for different $K$. The exact shapes of $\mathcal{P}_K$ depend on the chosen jet algorithm.

The jet-number cut is equivalent to inclusion into observables of a dichotomic factor of the form

$$\theta(\mathbf{Q}[\mathbf{P}] \text{ has } K \text{ jets}) \equiv \theta(\mathbf{P} \in \mathcal{P}_K) . \qquad 4.36$$

(The $\theta$-function is defined in 2.39.)

The value of $K$ is chosen to enhance sensitivity to $M$ and to suppress backgrounds. It is usually determined using the approximate relation jets ≈ partons (Eq. 4.32). Then $K$ is the number of partons in the final states in the lowest order of the QCD perturbation theory in which the dependence on the parameters one is interested in is manifest.

## Observables 4.37

The last element of the conventional approach is a function defined on jet configurations **Q** which passed the jet-number cut (Sec. 4.34); denote it $\varphi_{\text{ad hoc}}(\mathbf{Q})$.

In practice $\varphi(\mathbf{Q})$ is chosen in ad hoc fashion although once the jet algorithm is chosen then it is possible in principle to construct optimal observables for the probability distribution mapped to **Q** (Sec. 4.52).

The observable on events **P** is then defined as follows:

$$\mathbf{P} \xrightarrow[\text{j.a.}]{} \mathbf{Q} \xrightarrow[\text{j. number cut}]{} \mathbf{Q} \xrightarrow[\varphi]{} f(\mathbf{P}) , \qquad 4.38$$

where $f(\mathbf{P}) = \theta(\mathbf{Q}[\mathbf{P}] \text{ has } K \text{ jets})\, \varphi(\mathbf{Q}[\mathbf{P}])$.

The data processing scheme 2.5 becomes

$$\left.\begin{array}{l} \pi(\mathbf{P}) \xrightarrow[\text{j.a. + cut}]{} \pi^*(\mathbf{Q}) \xrightarrow[\varphi]{} \langle f \rangle_{\text{th}} \\ \{\mathbf{P}_i\}_i \xrightarrow[\text{j.a. + cut}]{} \{\mathbf{Q}_i\}_i \xrightarrow[\varphi]{} \langle f \rangle_{\text{exp}} \end{array}\right\} \xrightarrow[\text{fit}]{} \alpha_S, M_W, \ldots \qquad 4.39$$

It is quite obvious that the optimal observable 4.20 cannot be represented in the form 4.38 with any non-trivial jet algorithm in realistic situations. This means that with such observables it is impossible to achieve the theoretical Rao-Cramer limit on the precision of determination of fundamental parameters. We will come back to this in Sec. 4.43.

## Examples 4.40

Two typical examples are as follows.

The first example is the so-called 3-jet fraction in the process $e^+e^- \to$ hadrons which used to be one of the observables employed for measurements of $\alpha_S$ at LEP1. Here one simply has:

$$f_{3\text{ jets}}(\mathbf{P}) = \theta(\mathbf{Q} \text{ has 3 jets}) . \qquad 4.41$$

The second example is a simplified (but sufficient for the purposes of illustration) version of what might be used at LEP2 to measure the mass of $W$ in the process $e^+e^- \to W^+W^- \to$ hadrons above the $W^+W^-$ threshold where each $W$ decays into two jets. Here one would select events with 4 jets and choose $\varphi(\mathbf{Q})$ to yield an array of numbers, each being the number of jet pairs from **Q** whose invariant mass falls into the corresponding interval of the mass axis (bin):

$$f_{\text{dijets}}(\mathbf{P}) = \theta(\mathbf{Q} \text{ has 4 jets}) \times [\text{no. of dijets from } \mathbf{Q} \text{ in the } m\text{-th bin}]_{m=1\ldots N_{\text{bins}}} . \qquad 4.42$$

## Understanding the observables 4.38 4.43

Substitute $f(\mathbf{P})$ defined by 4.38 into the l.h.s. of 4.2 and use 4.12. Simple formal changes of the order of integrations yield:

$$\int \mathrm{d}\mathbf{P}\, \pi(\mathbf{P})\, f(\mathbf{P}) \equiv \int_{\mathbf{q} \text{ has } K \text{ jets}} \mathrm{d}\mathbf{q}\, \pi^*(\mathbf{q})\, \varphi(\mathbf{q}) , \qquad 4.44$$

where

$$\pi^*(\mathbf{q}) = \int d\mathbf{p}\, \pi^{(n)}_{\text{pQCD}}(\mathbf{p})\, h^{(n)}(\mathbf{p},\mathbf{q}) , \qquad 4.45$$

with the kernel $h^{(n)}$ given by

$$h^{(n)}(\mathbf{p},\mathbf{q}) = \int d\mathbf{P}\, H^{(n)}(\mathbf{p},\mathbf{P})\, \delta(\mathbf{q},\mathbf{Q}[\mathbf{P}]) . \qquad 4.46$$

The $\delta$-function on the r.h.s. is similar to the one in 4.16.

If one drops the jet-number cut from the definition 4.38 then the only change to be made is to drop the restriction on $\mathbf{q}$ in the integral on the r.h.s. of 4.44.

Note that Eq. 4.45 differs from 4.12 by the replacements $H^{(n)} \to h^{(n)}$, $\mathbf{P} \to \mathbf{q}$.

If the mapping $\mathbf{P} \to \mathbf{Q}$ corresponds to a typical jet algorithm then the domain of $\mathbf{q}$ is, generally speaking, the same as for $\mathbf{P}$, i.e. $\mathcal{P}$ (Sec. 3.15). However, most of the probability density $\pi^*(\mathbf{q})$ is now concentrated on pseudoevents $\mathbf{q}$ with fewer particles than was the case with $\pi(\mathbf{P})$. Second, $\pi^*(\mathbf{q})$ is zero on jet configurations with some pairs of jets sufficiently close (the corresponding events $\mathbf{P}$ are then mapped to jet configurations with a single jet instead of such a pair; recall the comments after 4.33).

The kernel $h^{(n)}(\mathbf{p},\mathbf{q})$ is interpreted as the probability for the partonic event $\mathbf{p}$ to generate any hadronic event that would yield the jet configuration $\mathbf{q}$ after application of the jet algorithm.

### On inversion of hadronization 4.47

Eq. 4.45 means that the kernel $h^{(n)}(\mathbf{p},\mathbf{q})$ effects a smearing of the perturbative expression. If the complete $\pi(\mathbf{P})$ given by 4.12 is strictly non-negative then such must also be $\pi^*(\mathbf{q})$.

The latter fact has the following consequence:

> Since the pQCD probability density $\pi^{(n)}_{\text{pQCD}}(\mathbf{p})$ is not strictly non-negative near some $\mathbf{p}$, the non-negativity of its smeared analog $\pi^*(\mathbf{q})$ implies that an exact inversion of hadronization is impossible with any jet algorithm in the form of the mapping 4.30.
> 4.48

This impossibility can be quantified; see Sec. 5.10.

Furthermore, the hadronization kernel $H^{(n)}$ depends on $n$, the order of pQCD corrections included into the perturbative probabilities $\pi^{(n)}_{\text{pQCD}}(\mathbf{p})$ in 4.12. It is not clear which $n$ the inversion of hadronization should be geared to.

For instance, consider radiation of a gluon by a quark. If $n$ corresponds to the leading order (LO) approximation then the mechanism of gluon radiation is described by the hadronization kernel $H^{(n)}$. If $n$ corresponds to the next-to-leading order (NLO) then $\pi^{(n)}_{\text{pQCD}}(\mathbf{p})$ is a sum of LO and NLO terms, and then $H^{(n)}$ should contain contributions which dress the LO and NLO terms. This in fact is a different aspect of the same problem 4.48: Jets have to be defined at the level of perturbative quarks and gluons before a connection with observed data can be established.

Still another aspect of the same problem is in terms of non-uniqueness of inversion of hadronization. In general, different configurations of partons may result in the same hadronic event. This is seen e.g. from the collective nature of hadronization (a single colored parton cannot develop into a jet of colorless hadrons). It is even more true if partonic cross sections are evaluated in NLO approximation where a quark can radiate an almost collinear gluon, etc. Then for some events one must rely on a convention about whether such an event is a hadronized LO quark, or a hadronized NLO configuration of the same quark and a gluon. We will come back to this point in Sec. 4.70.

Lastly, from a computational viewpoint, inversion of a convolution like 4.12 is in general an ill-posed problem. This means that even if a solution formally exists, numerical instabilities may be encountered in practice. In the present case, such instabilities occur near the discontinuity of the mapping $\mathbf{P} \to \mathbf{Q}$, as already discussed.

### Understanding $h^{(n)}(\mathbf{p},\mathbf{q})$ 4.49

The importance of the kernel $h^{(n)}(\mathbf{p},\mathbf{q})$ is due to the fact that it characterizes the combined effect of the chosen jet algorithm and the hadronization mechanism represented by $H^{(n)}$.

$h^{(n)}(\mathbf{p},\mathbf{q})$ may be non-zero even if the numbers of particles in $\mathbf{p}$ and $\mathbf{q}$ do not coincide (two close partons from $\mathbf{p}$ may hadronize into overlapping sprays of hadrons which the jet algorithm maps into a single jet). This motivates introduction of the following quantities. Define

$$h^{(n)}(\mathbf{p},K) = \int_{\mathbf{q}\text{ has }K\text{ jets}} d\mathbf{q}\, h^{(n)}(\mathbf{p},\mathbf{q}) . \qquad 4.50$$

This is interpreted as the probability for the partonic event $\mathbf{p}$ to hadronize into hadronic events recognized by the jet algorithm as having $K$ jets. Then the fraction of $L$-parton events which hadronized into $K$-jet events is formally given by

$$h(L,K) = \frac{\int_{\mathbf{p}\text{ has }L\text{ partons}} d\mathbf{p}\, \pi^{(n)}_{\text{pQCD}}(\mathbf{p})\, h^{(n)}(\mathbf{p},K)}{\int_{\mathbf{p}\text{ has }L\text{ partons}} d\mathbf{p}\, \pi^{(n)}_{\text{pQCD}}(\mathbf{p}) \int d\mathbf{q}\, h^{(n)}(\mathbf{p},\mathbf{q})} . \qquad 4.51$$

(The integral over $\mathbf{q}$ in the denominator yields 1 as is seen from the definition 4.46 and the normalization 4.13.)

The quantity $h(K,K)$ is the fraction of events $\mathbf{P}$ generated from partonic events with $K$ partons and recognized by the algorithm as having $K$ jets. The quantities $h^{(n)}(\mathbf{p},\mathbf{q})$, $h^{(n)}(\mathbf{p},K)$ and $h(L,K)$ give a more differential information. They can in principle be studied numerically using Monte Carlo event generators. In particular, it is interesting to compare the spread of $\mathbf{q}$ around $\mathbf{p}$ for a few typical $\mathbf{p}$ and for different jet algorithms.

It might be useful (certainly interesting) to have a reasonably detailed empirical information of the kernels $h^{(n)}(\mathbf{p},\mathbf{q})$, $h^{(n)}(\mathbf{p},K)$, etc.

### Optimal observables in the class 4.38 4.52

From a general mathematical viewpoint the smearing 4.45 can be regarded as an example of a regularization of a singular approximation (Sec. 2.51; i.e. the pQCD approximation $\pi^{(n)}_{\text{pQCD}}(\mathbf{p})$ of the exact probability density $\pi(\mathbf{P})$), transforming it into a physically meaningful form. This implies that whereas the perturbative expression 4.21 is formal, it is entirely meaningful to construct an optimal observable defined on $\mathbf{q}$ from $\pi^*(\mathbf{q})$ according to the standard recipe 2.17:

$$\varphi_{\text{opt}}(\mathbf{q}) = \partial_M \ln \pi^*(\mathbf{q}) . \qquad 4.53$$

In terms of events $\mathbf{P}$, the observable 4.53 is

$$\hat f_{\rm opt}(\mathbf{P}) = \varphi_{\rm opt}(\mathbf{Q}[\mathbf{P}]) = \frac{\int d\mathbf{p}\left[\partial_M \pi^{(n)}_{\rm pQCD}(\mathbf{p})\right]\times h^{(n)}(\mathbf{p},\mathbf{Q}[\mathbf{P}])}{\int d\mathbf{p}'\left[\pi^{(n)}_{\rm pQCD}(\mathbf{p}')\right]\times h^{(n)}(\mathbf{p}',\mathbf{Q}[\mathbf{P}])}. \qquad 4.54$$

The kernel $h^{(n)}$ is given by 4.46.

If the dimensionality of $\mathbf{q}$ for which 4.54 is non-negligible is not too high then a brute force construction of a numerical interpolation formula to represent 4.54 might be feasible.

The formula 4.54 is valid for any jet algorithm (cone, $k_{\rm T}$, etc.), and it describes a way to achieve the theoretically best precision for the parameter $M$ with a given jet algorithm within the conventional scheme 4.38. Of course, the functions defined by 4.54 differ for different jet algorithms.

It is easy to take into account the jet-number cut. Then all one has to do is restrict considerations to the $K$-jet sector:

$$\pi^*(\mathbf{q}) \to \pi^*_K(\mathbf{q}) = Z_K^{-1}\,\pi^*(\mathbf{q})\,\theta(\,\mathbf{q}\text{ has }K\text{ jets}\,), \qquad 4.55$$

where $Z_K$ is an appropriate normalization factor (it may depend on $M$). Then Eq. 4.53 is modified as follows:

$$\varphi_{{\rm opt},K}(\mathbf{q}) = \partial_M \ln \pi^*_K(\mathbf{q}) \times \theta(\,\mathbf{q}\text{ has }K\text{ jets}\,) = \left[\varphi_{\rm opt}(\mathbf{q}) - \partial_M \ln Z_K\right]\times\theta(\,\mathbf{q}\text{ has }K\text{ jets}\,). \qquad 4.56$$

Note that in practical constructions the subtracted term in square brackets may be dropped (the comment after 2.18).

The corresponding observable defined on events $\mathbf{P}$ is

$$\hat f_{{\rm opt},K}(\mathbf{P}) = \varphi_{{\rm opt},K}(\mathbf{Q}[\mathbf{P}]) = \left[\hat f_{\rm opt}(\mathbf{P}) - \partial_M \ln Z_K\right]\times\theta(\,\mathbf{Q}[\mathbf{P}]\text{ has }K\text{ jets}\,). \qquad 4.57$$

This construction remains valid for any $K$, i.e. one can construct an optimal observable in any jet-number sector.
Of course, usually one sector (which corresponds to the "canonical" value of $K$; see the remarks after 4.36) would yield a more informative observable than others.

### Inclusion of adjacent jet-number sectors 4.58

There is nothing to prevent inclusion into consideration in 4.55–4.57 of additional jet-number sectors. Quite obviously, this would increase informativeness of the resulting aggregate observable. If the $K$-jet sector was the most informative one then it is natural first to include one or both adjacent sectors which correspond to $(K \pm 1)$ jets.

## Comparison of different classes of observables 4.59

In the following discussion we assume that a jet algorithm is fixed (unless indicated otherwise).

We can compare different kinds of observables for measurements of a fundamental parameter $M$:

(1) A conventional observable of the form

$$\hat f_{{\rm ad\ hoc},K}(\mathbf{P}) = \theta(\,\mathbf{Q}[\mathbf{P}]\text{ has }K\text{ jets}\,)\times\varphi_{{\rm ad\ hoc},K}(\mathbf{Q}[\mathbf{P}]), \qquad 4.60$$

which involves a jet-number cut and an ad hoc function $\varphi_{{\rm ad\ hoc},K}(\mathbf{q})$ usually defined on jet configurations with $K$ jets only, as described by the $\theta$-function in 4.60.

(2) The observable $\hat f_{{\rm opt},K}$ given by 4.57, which yields the best precision among observables of the form 4.60, i.e. defined via intermediacy of the chosen jet algorithm in the $K$-jet sector.

(3) The observable $\hat f_{{\rm opt},K,K\pm1}$ defined by inclusion of the adjacent jet-number sectors (Sec. 4.58; one could include only one of the two adjacent sectors.)

(4) The ideal observable $\hat f_{\rm opt}$ 4.54 which yields the best precision among observables defined via intermediacy of a jet algorithm in all jet-number sectors.

(5) The ideal observable $f_{\rm opt}$ (4.20) defined without jet algorithms. It yields the absolutely best (Rao-Cramer) precision for the parameter $M$.

The observables are listed in increasing informativeness: Quite obviously, each additional restriction on the form of observables is an extra obstacle for achieving the Rao-Cramer limit of precision.

Furthermore, it is clear that one can, at least in principle, construct quasi-optimal observables (Sec. 2.25) for any of the observables $\hat f_{{\rm opt},K}$ and $\hat f_{{\rm opt},K,K\pm1}$.

The following figure illustrates the relation between the various observables which we discuss:

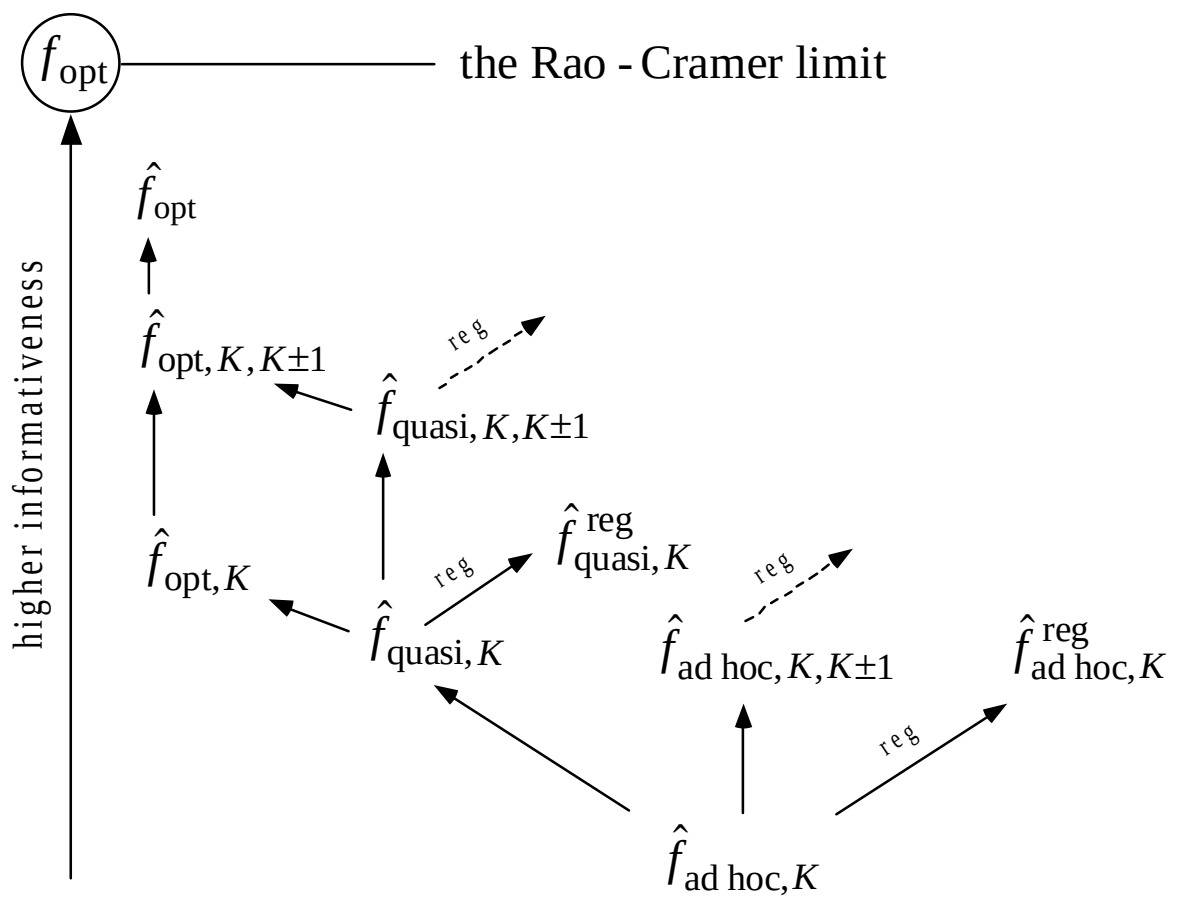

4.61

Hats denote observables defined via intermediacy of a jet algorithm (remember that the reasoning in this section is valid for any fixed jet algorithm). Arrows indicate an increase in informativeness (neither absolute nor relative magnitudes of the increase can be predicted a priori). The absence of arrows between two observables means their informativeness cannot be compared a priori (except for the case of $f_{\rm opt}$ which has the absolutely highest informativeness).

The "reg" arrows correspond to the option of regularization of cuts which will be discussed separately (Sec. 4.68 and Section 9).

### Ways to increase informativeness of ad hoc observables 4.62

Consider a conventional ad hoc observable $\hat f_{{\rm ad\ hoc},K}$ (Eq. 4.60). There are at least the following ways to improve it (cf. Fig. 4.61):

(i) Replacing the ad hoc observable with a quasi-optimal one (the prescriptions of Sec. 2.25).

(ii) Inclusion of the adjacent $(K \pm 1)$-jet sectors (Sec. 4.63).

(iii) Regularization of discontinuities (Sections 4.68 and 9).

(iv) Adjustment of the underlying jet algorithm (to be discussed in Sec. 5.3).

These options may be combined. In subsequent subsections we will discuss them in more detail together with some related issues.

### Inclusion of additional jet-number sectors 4.63

If, say, the $K$-jet sector was the most informative one (usually because the lowest PT order where the dependence on the desired parameter first manifests itself corresponds to final states with $K$ partons) then it is natural to first include one or more of the adjacent, $(K\pm 1)$-jet sectors.

The best precision is obtained if one uses a quasi-optimal observable in each sector, otherwise an increase of precision is not guaranteed.

The simplest way to include information from additional jet-number sectors is to map events from each additional sector into one point (the scheme 4.44–4.46, 4.53 is valid irrespective of the physical or mathematical nature of the mapping $\mathbf{P}\to\mathbf{Q}$). Then it is sufficient to determine the value of the corresponding optimal observable at that point (this means that all events from this sector receive the same weight). The magnitude of the resulting increase of informativeness could be regarded as a signal of whether or not a more detailed treatment might be warranted. Such a procedure might be a useful way to control the loss of information due to the restriction of the jet-number cut.

Inclusion of additional jet-number sectors seems to become useful whenever the quantity $1-h(K,K)$ (Eq. 4.51 and the comments thereafter) is appreciably non-zero. The difficulty here would be if adding even one jet increases the dimensionality of phase space too much to allow a meaningful construction of observables following 4.53. Then one may be satisfied with defining reasonable ad hoc observables in the adjacent jet-number sectors. In such a situation one may find inspiration e.g. in the constructions of [4] such as spectral discriminators. Computation of spectral discriminators may be prohibitively expensive for raw hadronic events but some similar observables defined on jet configurations with, say, no more than 10 jets should not be difficult to compute.

For instance, in the context of example 4.42, one could include into consideration the 5-jet sector and define a similar observable by allowing both di- and tri-jets (an additional jet may have been radiated from one of the partons originally forming a dijet). And/or one could include the 3-jet sector and define an observable in it based on the fact that some pairs of partons may generate overlapping jets which may be seen by the jet algorithm as a single jet (e.g. the invariant mass distribution of single jets).

### Sources of non-optimality of the observable 4.54 4.64

With a fixed jet algorithm, a conventional ad hoc observable 4.60 can always be improved, in principle, by a transition to the optimal observable 4.57, and by inclusion of all jet-number sectors (the combined effect of both tricks is represented by 4.54). So the truly fundamental limitations of the conventional scheme are those associated with the sources of non-optimality (i.e. loss of precision of the extracted values for $M$) of the observable 4.54 compared with the ideal expression 4.20.

There are two sources of such non-optimality in the jets-mediated optimal observable 4.54 compared with 4.20:

1) The variation of the expression 4.20 over the collections of events $\mathbf{P}$ which correspond to the same jet configuration $\mathbf{q}$. Each such collection is described by the equation $\mathbf{q}=\mathbf{Q}[\mathbf{P}]$.

2) The discontinuities of 4.38 at the boundary of the regions $\mathcal{P}_K$ (defined in 4.35).

Concerning the first source, the guidance here is provided by the general criteria 2.26, 2.27 with $f_{\text{quasi}}=\hat{f}_{\text{opt}}$. One can make the following simple observation:

> The faster the variation of $f_{\text{opt}}$ near some $\mathbf{P}$, the more fine-grained should be the mapping $\mathbf{P}\to\mathbf{Q}$ there.
> 4.65

### Non-optimality due to discontinuities 4.66

The second source of non-optimality is due to discontinuities at the boundaries of $\mathcal{P}_K$. Fig. 4.67 gives an illustration of what happens near such a boundary.

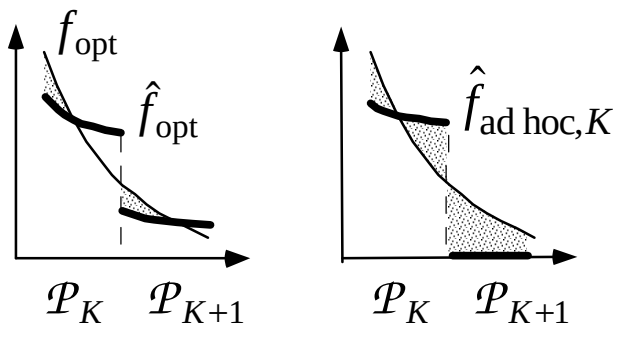

4.67

The left figure shows $\hat{f}_{\text{opt}}$ against $f_{\text{opt}}$. It is assumed that the latter is small outside the $K$-jet sector. The shaded areas corresponds to the non-optimality of $\hat{f}_{\text{opt}}$ (recall the criterion 2.27 and the rule of thumb 2.29). $\hat{f}_{\text{opt},K}$ differs from $\hat{f}_{\text{opt}}$ by being equal to zero outside $\mathcal{P}_K$ (apart from an inessential additive constant). The right figure shows an ad hoc variable against $f_{\text{opt}}$. If the variable where the $K$-jet fraction, it would be constant in $\mathcal{P}_K$.

The problem is exacerbated if the boundary of $\mathcal{P}_K$ passes through the region of a fast variation of $f_{\text{opt}}$. Note e.g. that the probability density (from which $f_{\text{opt}}$ is constructed) in QCD varies by an order of magnitude between the regions corresponding to $K$ and $K+1$ jets because radiation of an additional jet is accompanied by the factor $\alpha_S\sim 0.1$.

It is clear from Fig. 4.67 that forcing the observable to continuously interpolate between its different branches (represented by fat lines) would eat away at the non-optimality (the shaded area) and thus increase precision of determination of the parameter $M$.

The relevant notion of continuity (among the many possible ones in an infinitely-dimensional space of events $\mathcal{P}$) is the $C$-continuity discussed at length in Sec. 3.18. (However, remember 2.50.)

Next we consider an example which shows that elimination of the discontinuities that are typical of the conventional observables may result in a noticeable improvement of precision of measurements.

### The role of continuity (shape observables *vs*. 3-jet fraction in measurements of $\alpha_S$) 4.68

The effect of non-optimality due to the jet-number cut is seen in the precision measurements of $\alpha_S$ at LEP. Here one usually employs observables $f$ such that

$$\langle f \rangle = O(\alpha_S) \; . \qquad 4.69$$

One example is the shape observable thrust defined by 4.27. Another example of an observable which satisfies 4.69 is the 3-jet fraction 4.41. Since the latter is a discontinuous observable of the conventional kind (4.60) whereas the former is continuous (even $C$-continuous; Sec. 3.18) and smoothly interpolates between different jet-number sectors, it is interesting to compare them in regard of the quality of results they yield.

We have seen (Sec. 4.22) that shape observables such as the thrust are nearly optimal for measurements of $\alpha_S$ (which is quite obvious already from 4.69). On the other hand, the boundary between the 3-jet and 2-jet regions is located where the probability density and $f_{\text{opt}}$ vary fast: as was pointed out in [3], 1% of 2-jet events incorrectly interpreted due to detector errors and statistical fluctuations in hadronization as having 3 jets induces a 10% error in the 3-jet fraction because the corresponding probabilities differ by a factor of $O(\alpha_S)$.

Note that $1-T(\mathbf{P})$ smoothly interpolates between the points in phase space where it takes its minimal and maximal values (0 and 1). This should be contrasted with the discontinuities of the 3-jet fraction 4.41.

Another kinematic property of the shape observables such as the thrust is that they are rather simple energy correlators and thus fit into the structure of quantum field theory. This property ensures their superb amenability to theoretical investigations such as the sophisticated higher order calculations for the thrust reviewed e.g. in [20].

By now it has been accepted that the $\alpha_S$ measurements (at least of the LEP type) are done best via shape observables rather than the 3-jet fraction.[t]

### The boundaries of $\mathcal{P}_K$ and non-uniqueness of inversion of hadronization 4.70

We conclude that the discontinuities at the boundaries of different jet-number sectors in the space of observed events may be a major source of non-optimality of conventional observables. The events near the discontinuities have two or more jets that are hard to resolve reliably.

This has a simple physical interpretation in terms of non-uniqueness of inversion of hadronization: There is no way to tell whether an event with overlapping jets was generated by $K$ hard partons dressed by a hadronizing QCD radiation, or by $K+1$ hard partons with two of them close enough to make the resulting jets overlap.

So, in general, there may be more than one candidate partonic events that can be regarded as parents for a given hadronic event. The best one can do is provide weights for each such candidate; the weight reflects the expected probability for the hadronic event to have been generated from a particular partonic candidate.

In Section 9 we will discuss ways to assign to the same event several different jet configurations with suitably chosen weights, together with prescriptions for regularization of the jet-number discontinuities.

[t] A large table presented in the lecture [26] did not contain a line for the 3-jet fraction which used to be a standard feature of such tables. In response to a query, the speaker mentioned unsatisfactory experimental errors.

## Definitions of jets 5

The use of jet algorithms in data processing following the scheme 4.39 is motivated by the specifics of QCD dynamics. The arguments of Sections 2–4 provide a framework to discuss jet definitions. A jet algorithm is a tool for construction of observables for specific precision measurement applications (Sec. 4.28), and the resulting observables can be compared using the notion of informativeness of observables (2.24). The usefulness of jet algorithms is due to the dynamics of pQCD (Sec. 4.1) which allows one to regard hadronic events as similar to their hard partonic parents (Eq. 4.16). This justifies the point of view that jet algorithms effect an (approximate) inversion of hadronization (Eq. 4.32).

First in Sec. 5.1 we discuss the conventional criterion used to compare jet algorithms. Then in Sec. 5.6 we introduce a jet definition based on the informational abstractions of Section 3 (the identification 3.43). The explicit purpose of such a jet definition is to serve as a tool for a systematic construction of quasi-optimal observables defined on hadronic states. In Sec. 5.10 we examine how the dynamical considerations complement the picture.

### The conventional approach to jet algorithms 5.1

A common way to judge suitability of a jet algorithm to a particular application (a precision measurement of a fundamental parameter $M$; Sec. 2.1) is as follows. One chooses $K$ as described after Eq. 4.36 and evaluates the fraction of events generated from partonic events with $K$ partons and recognized by the jet algorithm as having $K$ jets. This fraction is formally given by $h(K,K)$ defined by Eq. 4.51. (Note that $0 < h(K,K) < 1$.) The larger this fraction, the better the jet algorithm is deemed to be.

> This criterion amounts to an implicit definition of an ideal jet algorithm as the one which maximizes $h(K,K)$. The various jet algorithms are then regarded as candidate approximations constructed empirically. 5.2

This definition can be related to the notion of optimal observables as follows. Consider the example 4.42. Then at the level of partons, the optimal observable is entirely localized on 4-parton events. At the level of hadrons, the optimal observable $f_{\text{opt}}$ is mostly localized in the 4-jet sector $\mathcal{P}_4$. If it were a constant there then it is entirely specified by $\mathcal{P}_4$, and the conventional criterion simply attempts to find the shape of $\mathcal{P}_4$. Note that there can be many parameters one may wish to measure and so many different $f_{\text{opt}}$. Since their shapes are all different, focusing only on the shape of $\mathcal{P}_4$ is a convenient compromise.

The advantage of the conventional criterion 5.2 is its simplicity and naturalness.

The disadvantages are as follows:

(i) Beyond the leading PT order, the signal is non-zero in other jet-number sectors.

(ii) $f_{\text{opt}}$ is not piecewise constant.

(iii) The criterion is based on a convention which although plausible is not based on a precise argument (see however the

reasoning in Sec. 5.3).

(iv) There is no clear way to improve upon the conventional scheme 4.38 should one find the leading order PT arguments insufficient.

The concept of optimal observables allows one to be a little more precise:

### Improving upon the conventional jet definition 5.3

Within the limitations of the conventional scheme 4.38, the best precision for $M$ is achieved with the observable $\hat{f}_{\mathrm{opt},K}$ (Eq. 4.57) which is entirely determined once the jet algorithm is fixed. So it is legitimate to ask which jet algorithm maximizes the informativeness of $\hat{f}_{\mathrm{opt},K}$. The definition is meaningful because the informativeness of $\hat{f}_{\mathrm{opt},K}$ is given by the following integral:

$$\int d\mathbf{q}\ \pi_K^*(\mathbf{q}) \times \theta(\,\mathbf{q}\text{ has }K\text{ jets}\,) \times \varphi^2_{\mathrm{opt},K}(\mathbf{q})\,. \qquad 5.4$$

The best jet algorithm would then maximize this.

It is interesting to find a way to connect this with the conventional criterion 5.2. Suppose one aims at a universal jet definition, then it is natural to replace the last factor by a constant. Then recall Eqs. 4.55 and 4.45. In the latter, restrict the integration to $K$-parton events. The resulting integral coincides, up to a normalization, with the numerator of $h(K,K)$; cf. Eq. 4.51.

Unfortunately, it is not clear how to derive from this a specific jet algorithm.

Furthermore, the conventional framework 4.38 per se imposes a restriction on attainable precision for fundamental parameters. If one seeks to alleviate it by, say, an inclusion of the adjacent jet-number sectors then the best jet algorithm should minimize the informativeness of $\hat{f}_{\mathrm{opt},K,K\pm1}$ rather than $\hat{f}_{\mathrm{opt},K}$.

Furthermore:

> It is not clear whether or not imperfections of the conventional jet algorithms are more important than the intrinsic limitations of the conventional scheme 4.38 as a whole. 5.5

The answer probably depends on the problem. A priori one cannot exclude that for some applications, an improvement of the scheme 4.38 as a whole via relaxation of the jet-number cut in the spirit of regularizations of Sec. 2.52 could be more important than improvements of the jet algorithm only.
A relaxation of the jet-number cut implies an inclusion into consideration of events with at least the adjacent numbers of jets, $K \pm 1$.

The example of Sec. 4.68 lends credibility to this point of view: the transition from 3-jet fractions to shape observables in the measurements of $\alpha_S$ can be regarded as a trick to take into account events from all jet-number sectors. This example is special in that jet algorithms can be avoided altogether in the improved observables. In general such luck may not occur, so jet algorithms are bound to remain a part of the answer.

But if one includes into consideration events with "wrong" numbers of jets and/or finds a way to regularize the discontinuities at the boundaries between different jet-number sectors in order to make the resulting observable continuous at those boundaries, then the details of how the space of events is sliced into jet-number sectors may become less important.

We conclude that a desirable property of a good jet algorithm is to provide options for a systematic improvement upon the conventional scheme 4.38. The jet algorithm we derive below offers such options.

## The optimal jet definition. Qualitative aspects 5.6

The jet definition we are going to introduce deserves to be called optimal for two reasons:

THE KINEMATICAL REASON: It involves an optimization that has a well-defined meaning in terms of information content of events and the corresponding jet configurations (Sec. 5.7).

THE DYNAMICAL REASON: It possesses a property naturally interpreted as an optimal inversion of hadronization (Sec. 5.10).

- The two properties are logically independent (at least I don't see a formal connection), and both lead to exactly the same definition 5.9. The only common element is the formal language in which both are phrased (the language of generalized [$C$-continuous] shape observables, Sec. 3).

The equivalence of the two approaches came about as a complete surprise. The order of presentation is determined by historical reasons.

### Informational definition 5.7

From the most general point of view, the jet algorithm, operationally, is a data processing tool whose purpose is to facilitate extraction of physical information. The resulting simplifications come at a price — a loss of information in the transition from events to jet configurations. The most basic and general requirement for any data processing tool — jet algorithms not excluded — is that the distortions it induces in the physical information should be minimized.

So it is natural to require that the best algorithm should minimize such an information loss:

> The jet configuration $\mathbf{Q}[\mathbf{P}]$ must inherit maximum information from the original event $\mathbf{P}$. 5.8

This in fact is similar to the conventional criterion 5.2 but now we would like to be more systematic in regard of interpretation of the information loss. To this end we will rely on the kinematical analysis of Sec. 3.

Note that the criterion 5.8 is applicable both to experimentally observed hadronic events and to theoretical multiparton events in situations where radiative QCD corrections need to be taken into account.

The analysis performed in Section 3 led us to the identification 3.43. This immediately allows us to translate the criterion 5.8 into the following form:

$$f(\mathbf{P}) \approx f(\mathbf{Q}[\mathbf{P}]) \quad \text{for any basic shape observable } f. \qquad 5.9$$

The less the discrepancy between the left and right hand sides, the more information from $\mathbf{P}$ is inherited by the jet configuration $\mathbf{Q}[\mathbf{P}]$.

The definition requires comments.

(i) The exact equality can always be achieved in 5.9 for $\mathbf{Q}[\mathbf{P}] = \mathbf{P}$, so for the replacement $\mathbf{P} \to \mathbf{Q}$ to make sense, one requires, heuristically speaking, that **Q** should have fewer jets than **P** has particles. Thanks to the *C*-continuity of the participating observables $f$, this can be achieved via two mechanisms: a replacement of sufficiently narrow sprays of particles by single jets (pseudoparticles from **Q**), and dropping particles which carry sufficiently small fractions of event's total energy (the so-called soft particles).

(ii) The replacement of a narrow spray of particles by one pseudoparticle implies that the detailed structure of the events at small correlation angles is less important than its structure grosso modo (what is called "topology of jets"). This makes natural the eventual occurrence in the jet definition of a parameter interpreted as the maximal jet radius ($R$). On the other hand, different $f$ in 5.9 are differently sensitive to replacements of sprays of particles with a single jet. The induced error will be greater for the observables whose angular functions (see 3.36) vary faster. The angular resolution parameter $R$ will then control the subclass of $f$ for which the error is minimized (Sec. 6.25).

(iii) The criterion 5.9 is formulated for individual events, and the error may also depend on **P**, so that the approximate equality must hold in some integral sense (Sec. 5.21). This is where dynamical considerations may enter into the picture (Sec. 5.31).

(iv) With 5.9, the jet algorithm can be interpreted as a trick for approximate evaluation of (or for construction of approximations for) complicated *C*-continuous observables such as the optimal observables 4.20. The trick is unusual in that here one simplifies the *arguments*, whereas normally one would simplify the *expression* of the function to be computed.

(v) It is clear that the optimal jet configuration for a given event need not be defined uniquely (more than one jet configurations may ensure 5.9 with a comparable error). Physically, this corresponds to the fact that different hard parton events may hadronize into the same hadronic event. This is an important option completely missing from conventional discussions. We will come back to this in Sec. 9.1.

(vi) A definition such as Eq. 5.9 would be genuinely useful only if one could control the approximation error via an estimate which would be both simple and precise. The general form of such estimates is discussed in Sec. 5.17.

## Inversion of hadronization 5.10

A remarkable fact is that the same criterion 5.9 also ensures what can be described as an optimal inversion of hadronization.

How well a given jet algorithm inverts hadronization is measured by how well the kernel 4.46 is approximated by the $\delta$-function:

$$h^{(n)}(\mathbf{p},\mathbf{q}) \approx \delta(\mathbf{p},\mathbf{q}). \tag{5.11}$$

The only way to interpret this is via integrals with *C*-continuous functions (cf. 3.43).

So, integrate both sides with an arbitrary *C*-continuous function $f(\mathbf{q})$. For the r.h.s. we obtain $f(\mathbf{p})$. For the l.h.s., use the definition 4.46 and obtain

$$\int d\mathbf{P}\, h^{(n)}(\mathbf{p},\mathbf{q})\, f(\mathbf{q}) = \int d\mathbf{P}\, H^{(n)}(\mathbf{p},\mathbf{P})\, f(\mathbf{Q}[\mathbf{P}]) \,. \tag{5.12}$$

Then consider the resulting difference:

$$\begin{aligned} & f(\mathbf{p}) - \int d\mathbf{P}\, H^{(n)}(\mathbf{p},\mathbf{P})\, f(\mathbf{Q}[\mathbf{P}]) \\ &\equiv f(\mathbf{p}) - \int d\mathbf{P}\, H^{(n)}(\mathbf{p},\mathbf{P})\, f(\mathbf{P}) \\ &+ \int d\mathbf{P}\, H^{(n)}(\mathbf{p},\mathbf{P}) \big[ f(\mathbf{P}) - f(\mathbf{Q}[\mathbf{P}]) \big] , \end{aligned} \tag{5.13}$$

where $f(\mathbf{P})$ was subtracted from and added to $f(\mathbf{Q}[\mathbf{P}])$.

The first line on the r.h.s.,

$$f(\mathbf{p}) - \int d\mathbf{P}\, H^{(n)}(\mathbf{p},\mathbf{P})\, f(\mathbf{P}) \,, \tag{5.14}$$

is independent of the jet algorithm. Its smallness is described by 4.15. The subtracted term is the average value of $f$ on hadronic events generated by the partonic event **p**. We can draw the first conclusion:

> The contribution 5.14 sets a jet definition-independent limit on how well hadronization can be inverted. 5.15

A consequence is that measuring quality of jet algorithms by percentage of restored parton events is meaningless beyond a certain limit. To go beyond that limit, it is necessary to go beyond the restrictions of the scheme 4.38.

The dependence on the jet algorithm only appears in the second line on the r.h.s. of 5.13. Therefore, to minimize 5.13 (and so the error in 5.11) it is sufficient to minimize the following expression:

$$\int d\mathbf{P}\, H^{(n)}(\mathbf{p},\mathbf{P}) \big[ f(\mathbf{P}) - f(\mathbf{Q}[\mathbf{P}]) \big]. \tag{5.16}$$

Such a minimization has to be accomplished for any **p** but since the jet algorithm cannot depend on the unknown **p**, the only meaningful general option is to minimize the expression in square brackets for each **P**, and we come back to the criterion 5.9.

An interesting property is that for a fixed **P**, the obtained criterion is independent of the hadronization kernel $H^{(n)}$, i.e. of any dynamical information. This conclusion holds independently of $n$, the order of pQCD corrections included into the parton-level probabilities.

- Dynamical information, however, may affect one's decisions about the allowed error for *different* **P**. We will turn to this in Sec. 5.31.

## A quantitative definition 5.17

Our analysis of the qualitative definition 5.9 is based on inequalities of the following factorized form to be obtained in Section 6:

$$\big| f(\mathbf{P}) - f(\mathbf{Q}) \big| < C_f\, \Omega[\mathbf{P},\mathbf{Q}] \,, \tag{5.18}$$

where the constant $C_f$ is independent of **P** and $\mathbf{Q} = \mathbf{Q}[\mathbf{P}]$, whereas the expression $\Omega[\mathbf{P},\mathbf{Q}]$ is independent of $f$.

Existence of a factorized estimate 5.18 could not have been postulated a priori. Another surprise is that $\Omega$ turns out to be an infrared-safe shape observable of a conventional kind and, moreover, closely related to the thrust (see Sec. 8.11).

An estimate of the form 5.18 would be sufficient for defining jet configurations in such a way as to control the errors induced in the observables via 5.24. The simplest option is to specify a small positive $\omega_{\text{cut}}$ (which in general may be chosen differently for different events **P**) and then define **Q** by demanding that it ensures that

$$\Omega[\mathbf{P},\mathbf{Q}] \le \omega_{\text{cut}}(\mathbf{P}) \,. \qquad 5.19$$

Since the purpose of replacing **P** by **Q** is to simplify calculations, one would seek to satisfy the restriction 5.19 with a minimal number of jets in **Q**. On the other hand, in order to minimize the actual error induced in the transition to jets, one would seek to minimize $\Omega[\mathbf{P},\mathbf{Q}]$. To summarize:

An optimal configuration of jets $\mathbf{Q}_{\text{opt}}$ for a given event **P** minimizes $\Omega[\mathbf{P},\mathbf{Q}]$ while satisfying the restriction 5.19 with a minimal number of jets.
5.20

See Sec. 9.1 for a discussion of important implications of the fact that the optimal jet configuration on which the minimum of $\Omega[\mathbf{P},\mathbf{Q}]$ is reached is, in general, not unique.

### Errors induced in observed numbers 5.21

In the final respect, the observed physical value is $\langle f \rangle$ and not $f(\mathbf{P})$. So, we must study how the errors induced in $f(\mathbf{P})$ for each **P** propagate to the level of $\langle f \rangle$.

To this end, recall Eqs. 2.3–2.4. The replacement of events **P** by the corresponding jet configurations **Q** results in the following expressions:

$$\langle f \rangle_{\text{th, jets}} = \int d\mathbf{P}\, \pi(\mathbf{P})\, f(\mathbf{Q}) \,, \qquad 5.22$$

$$\langle f \rangle_{\text{exp, jets}} = \frac{1}{N} \sum_i f(\mathbf{Q}_i) \,, \qquad 5.23$$

where $\mathbf{Q} = \mathbf{Q}_{\text{opt}}$ is a function of **P** as defined by 5.20, so that $\mathbf{Q}_i$ is its value on $\mathbf{P}_i$. Using the bound 5.18, one obtains

$$\left| \langle f \rangle - \langle f \rangle_{\text{th, jets}} \right| \le C_f\, \omega, \qquad 5.24$$

where

$$\omega = \int d\mathbf{P}\pi(\mathbf{P}) \times \Omega[\mathbf{P},\mathbf{Q}] \le \int d\mathbf{P}\pi(\mathbf{P}) \times \omega_{\text{cut}}(\mathbf{P}). \qquad 5.25$$

This expression controls the errors inherited by all interpreted physical information ($M_W$, etc.) extracted via jets according to 5.22–5.23.

- The quantity $\omega$ together with its fluctuations can be estimated like any other observable as the mean value and variance of $\Omega[\mathbf{P},\mathbf{Q}]$ which is computed for each event **P** in the process of minimization according to 5.20.

### (Non) optimality of jet definitions 5.26

The above reasoning shows that a jet algorithm can be regarded as a tool for approximate evaluation of at least basic shape observables. However, recall that general *C*-continuous observables — including the optimal observables 4.20 — can be approximated by algebraic combinations of basic shape observables (Sec. 3.35). This means that the error estimate 5.18 will be inherited by a class of general *C*-continuous observables which have appropriate regularity properties. (From the derivation in Section 6 this should be a *C*-continuous analog of continuous second order derivatives; cf. the discussion in Sec. 4.3.) The optimal observables 4.20 cannot be reasonably expected to have discontinuities in any derivatives, so they fall into this class.

With the bound 5.18 valid for the optimal observables, the jet algorithm based on it can be regarded as a trick for approximate computation of — or, equivalently, constructing approximations for — such observables. This allows one to compare different jet definitions on the basis of the magnitude of the errors they induce in the relation 5.9 (more precisely, one looks at 5.18).

We will use the term *optimal* and its derivatives in connection with various jet definitions in the following sense:

A jet finding prescription *A* is less optimal than another prescription *B* if with a given number of jets (which is a measure of computational economy) the jet configurations produced by *A* inherit less information from the original event than is the case with *B*. In other words, the use of the scheme *A* makes it computationally harder compared with *B* to approximate optimal observables 4.20 and thus to achieve the best possible precision for fundamental parameters such as $\alpha_S, M_W$, etc.

(It is possible to make this more precise via inequalities for different $\Omega$ by analogy with the standard techniques for comparison of norms in vector spaces. We skip this exercise because the conventional algorithms cannot be easily represented in the spirit of 5.18.)

We will use this notion in Section 10 to compare the jet definition we will derive with the conventional algorithms.

- An obvious conclusion from the above reasoning is that the estimate 5.18 should be as precise as possible, i.e. its construction should not involve tricks which would overestimate the error. This would ensure optimality of the resulting jet definition. We will pay heed to this in Section 6.
- To avoid confusion, note that the optimality of jet algorithms is a different (although metaphysically related) thing from the optimality of observables (Sec. 2.25), in particular from the optimality of observables within the restrictions of the scheme 4.38 with a fixed jet algorithm.

### The universal jet definition 5.27

The simplest universal option is to choose $\omega_{\text{cut}}(\mathbf{P})$ to be independent of the event **P**:

$$\omega_{\text{cut}}(\mathbf{P}) = \omega_{\text{cut}} = \text{const} \,. \qquad 5.28$$

Then because the probability distribution is normalized to 1,

$$\int d\mathbf{P}\pi(\mathbf{P}) \equiv 1 \,, \qquad 5.29$$

Eq. 5.24 would be ensured with some $\omega \le \omega_{\text{cut}}$. So:

The parameter $\omega_{\text{cut}}$ of the universal jet definition directly controls the errors induced in the physical information by the replacement of events with the corresponding configurations of jets.
5.30

### Inclusion of dynamical information 5.31

As is clear from Eq. 5.25, one can include dynamics into consideration by simply making $\omega_{\rm cut}$ depend on **P**.

All the dynamics is expressed by the probability density $\pi(\mathbf{P})$. Suppose it has enhancements for certain types of events (as indeed it does in QCD owing to collinear singularities). Then it would be sufficient to choose $\omega_{\rm cut}(\mathbf{P})$ to anticorrelate with $\pi(\mathbf{P})$. For instance,

$$\omega_{\rm cut}(\mathbf{P}) = \overline{\omega}_{\rm cut} \times \zeta(\mathbf{P}) \,, \qquad 5.32$$

where the factor $\zeta(\mathbf{P}) \le 1$ (which should, strictly speaking, be a shape observable) contains all the dependence on events. Then Eq. 5.25 becomes

$$\omega \;\le\; \overline{\omega}_{\rm cut} \times \int \mathrm{d}\mathbf{P}\, \pi(\mathbf{P})\, \zeta(\mathbf{P}) \,. \qquad 5.33$$

Choosing $\zeta$ to anticorrelate with $\pi$ would reduce the integral thus suppressing the overall error.

From the point of view of minimization of induced errors, the following observation makes such a modification less attractive. Indeed, the computational savings due to larger $\omega_{\rm cut}$ for the events which are produced less often may turn out to be simply not worth the trouble: one could simply take a smaller event-independent $\omega_{\rm cut}$ from the very beginning.

However, a non-constant $\omega_{\rm cut}(\mathbf{P})$ would affect the shape of the $k$-jet subregions in the space of events similarly to the difference between how, say, cone and $k_{\rm T}$ algorithms see jets. So one may wish to keep this option open for ultimate flexibility.

Note that if one modifies the conventional scheme 4.38 as a whole — and the most important modifications seem to correspond to relaxation of the jet-number cut (some such options are discussed in Section 9) — then the details of how $K$-jet sectors are defined in the space of events may become less important.

Determining a specific form for $\zeta(\mathbf{Q})$ is left to experts in the dynamics of QCD. In practice high precision is not needed here, and one could choose $\zeta(\mathbf{P})$ to depend on **Q** found e.g. using the universal jet definition with $\omega_{\rm cut} = \overline{\omega}_{\rm cut}$ or some other value. Then $\zeta(\mathbf{Q})$ could be chosen so that $\zeta^{-1}(\mathbf{Q})$ roughly imitates the structure of dominant terms in $\pi(\mathbf{P})$. Note that the quantities such as the invariant masses of the jets, and transverse momenta of particles in each jet — along with new interesting characteristics such as the fuzziness of each jet; cf. 8.19 — are easily computed from the output of the optimal jet definition which we will derive.

Lastly, an effect essentially equivalent to a modification of $\omega_{\rm cut}$ according to 5.32 can be achieved via keeping $\omega_{\rm cut}$ **P**-independent but replacing $\Omega[\mathbf{P},\mathbf{Q}]$ in 5.19 by another function such that

$$\overline{\Omega}[\mathbf{P},\mathbf{Q}] \ge \Omega[\mathbf{P},\mathbf{Q}] \,. \qquad 5.34$$

Then the control of information loss in the transition from events to jets would still be ensured but one could choose $\overline{\Omega}[\mathbf{P},\mathbf{Q}]$ to meet some additional requirements. The difficulty here is to keep $\overline{\Omega}[\mathbf{P},\mathbf{Q}]$ simple and suitable for numerical implementation.

A detailed investigation of these options is beyond the scope of the present paper.

## Derivation of the factorized estimate 6

In this section we are going to obtain a factorized estimate of the form 5.18 which would satisfy the criterion of optimality of Sec. 5.26.

Surprisingly, all one needs to obtain such an estimate is essentially an angular Taylor expansion through second order.

### Recombination matrix $z_{aj}$ 6.1

Recalling 3.36, the quantity to be estimated becomes

$$\left| f(\mathbf{P}) - f(\mathbf{Q}) \right| = \left| \sum_a E_a\, f(\hat{\boldsymbol{p}}_a) - \sum_j \mathcal{E}_j\, f(\hat{\boldsymbol{q}}_j) \right| . \qquad 6.2$$

To construct a bound for the r.h.s., one can only compare the values of $f$ at some $\hat{\boldsymbol{p}}_a$ with its values at some $\hat{\boldsymbol{q}}_j$. But which $\hat{\boldsymbol{p}}_a$ to compare with which $\hat{\boldsymbol{q}}_j$ may not be decided a priori.

Introduce the *recombination matrix* $z_{aj}$ which is heuristically interpreted as the fraction of $a$-th particle's energy that goes into the $j$-th jet (this interpretation will be justified below; cf. 6.16). Impose the following restrictions on $z_{aj}$:[u]

$$z_{aj} \ge 0 \quad \text{for any } a, j; \qquad 6.3$$

$$\bar{z}_a \overset{\text{def}}{=} 1 - \sum_j z_{aj} \ge 0 \quad \text{for any } a. \qquad 6.4$$

One can see from the derivation that removing the restrictions on $z_{aj}$ does not expand the eventual range of options. All one has to do is replace $z_{aj},\, \bar{z}_j \to |z_{aj}|, |\bar{z}_j|$ in 6.8 and 6.20, so that configurations not satisfying 6.3, 6.4 are automatically disfavored compared with the corresponding boundary points.

Non-zero values of the quantity $\bar{z}_a$ correspond to some energy being left out of the formation of jets (the so-called *soft energy*). We will see that this corresponds to exclusion of some soft stray particles (the soft component of the event's energy flow) from the formation of jets.

Allowing fractional values for $z_{aj}$:

a) fully agrees with the physical picture of production of colorless hadrons as a result of collective interaction of the underlying hard colored partons;

b) is extremely convenient algorithmically because the space of all possible jet configurations for a given event is then path-connected, so that any jet configuration can be reached from any other via a continuous path, allowing efficient shortest-path search algorithms [7].

We will say that the $a$-th particle belongs to the $j$-th jet if $z_{aj} = 1$. If $\bar{z}_a = 1$, the particle is said to belong to soft energy.

With the recombination matrix, rewrite 6.2 as follows (the first line is an identical transformation of 6.2, which explains the restriction 6.4):

$$\begin{aligned}
&\left| \sum_a \bar{z}_a E_a\, f(\hat{\boldsymbol{p}}_a) + \sum_a \Big(\sum_j z_{aj}\Big) E_a\, f(\hat{\boldsymbol{p}}_a) - \sum_j \mathcal{E}_j\, f(\hat{\boldsymbol{q}}_j) \right| \\
&\le \left| \sum_a \bar{z}_a E_a\, f(\hat{\boldsymbol{p}}_a) \right| + \left| \sum_j \Big( \sum_a z_{aj} E_a\, f(\hat{\boldsymbol{p}}_a) - \mathcal{E}_j\, f(\hat{\boldsymbol{q}}_j) \Big) \right| \\
&\le \sum_a \bar{z}_a E_a \left| f(\hat{\boldsymbol{p}}_a) \right| + \sum_j \left| \sum_a z_{aj} E_a\, f(\hat{\boldsymbol{p}}_a) - \mathcal{E}_j\, f(\hat{\boldsymbol{q}}_j) \right| .
\end{aligned} \qquad 6.5$$

One sees why we split particles into fragments corresponding to jets rather than vice versa: we target situations with fewer jets than particles, so it is desirable to arrange cancellations between as many terms as possible (the inner sum), and to minimize the number of positive terms in the outer sum.

[u] Formulas in solid boxes are part of the final result; they represent all the information needed for algorithmic implementations.

Estimating the effects of soft energy 6.6

The first sum on the r.h.s. of 6.5 can be estimated as follows:

$$\sum_a \bar{z}_a E_a \left| f(\hat{\boldsymbol{p}}_a) \right| \le C_{f,1} \, \mathrm{E}_{\mathrm{soft}}[\mathbf{P},\mathbf{Q}], \quad 6.7$$

where $C_{f,1}$ is the maximal value of $|f|$ over all directions and

$$\mathrm{E}_{\mathrm{soft}}[\mathbf{P},\mathbf{Q}] = \sum_a \bar{z}_a E_a \, . \quad 6.8$$

This quantity will play a central role in the optimal jet definition. It is interpreted as the event's energy fraction left out of the formation of jets (the soft energy, as we agreed to call it). It can be visualized as a background from which jets stick out.

Understanding the form of $\mathrm{E}_{\mathrm{soft}}$ 6.9

Mathematically possible are many other ways to obtain a factorized estimate of the form 6.7. The variant with 6.8 is singled out by the following properties:

(i) Analytical simplicity which leads to fast algorithms.

(ii) Linearity in energies of all particles which ensures infrared safety of the resulting jet definition.

(iii) The property that can be called *maximal inclusiveness*.

For instance, also possible is a bound in terms of $\max_a(\bar{z}_a E_a)$ but that would require comparison of particles' energies, which is physically meaningless if their directions are close.

A somewhat more meaningful option would be to perform a smearing of the soft energy over some angular radius thus transforming the soft energy flow into a continuous function, and then using the maximal value of that function as an alternative to 6.8 (the constant $C_{f,1}$ would change accordingly). This would be similar to the so-called 'f'-cut [2], i.e. a lower cut on the energy of the jets retained in the final jet configuration.[v] However, there are three reasons why such alternatives seem to be undesirable:

1) On the measurement side, they introduce non-optimality into the bound implying a further loss of information in the transition from the event to jets.

2) Computationally, they introduce a complexity into our jet definition unwarranted by physical considerations.[w]

3) On the QCD side, they are less inclusive than the expression 6.8, i.e. they introduce into consideration subregions of phase space. It is a well-known fact that exclusiveness of observables anti-correlates with the predictive power of QCD. For instance, a totally inclusive treatment of soft energy was built already into the jet definition of [8].

Taylor expansion in angles 6.10

To Taylor-expand $f(\hat{\boldsymbol{p}}_a)$ near $\hat{\boldsymbol{q}}_j$ is just a little tricky, and we proceed as follows. Consider the plane $\Pi$ which is normal to the unit 3-vector corresponding to $\hat{\boldsymbol{q}}_j$. Then map the directions to $\Pi$: $\hat{\boldsymbol{p}} \to \hat{\boldsymbol{p}}^{\Pi}$, so that the angular distances between directions are preserved near $\hat{\boldsymbol{q}}_j$:

$$\left| \hat{\boldsymbol{p}}^{\Pi} - \hat{\boldsymbol{q}}_j^{\Pi} \right| = O\left( \left| \hat{\boldsymbol{p}} - \hat{\boldsymbol{q}}_j \right| \right), \quad \hat{\boldsymbol{p}} \to \hat{\boldsymbol{q}}_j, \quad 6.11$$

where the l.h.s. is a euclidean distance in $\Pi$. (An example of such a mapping is given in Sec. 7.2.)

Then $f(\hat{\boldsymbol{p}})$ becomes a function on $\Pi$ which we denote as $f_{\Pi}(\hat{\boldsymbol{p}}^{\Pi}) \equiv f(\hat{\boldsymbol{p}})$. We will use the Taylor expansion in the form of the following inequality:

$$\left| f_{\Pi}(\hat{\boldsymbol{p}}_a^{\Pi}) - f_{\Pi}(\hat{\boldsymbol{q}}_j^{\Pi}) - [\hat{\boldsymbol{p}}_a^{\Pi} - \hat{\boldsymbol{q}}_j^{\Pi}] \cdot f'_{\Pi}(\hat{\boldsymbol{q}}_j^{\Pi}) \right| \le \Delta_{aj} C_{f,3}, \quad 6.12$$

where $\hat{\boldsymbol{p}}_a^{\Pi} - \hat{\boldsymbol{q}}_j^{\Pi}$ is a vector in $\Pi$ and $f'_{\Pi}$ is the gradient of $f_{\Pi}$. The constant $C_{f,3}$ hides maximal values of some combinations of $f$ and its derivatives through second order. The maximum is taken over all directions $\hat{\boldsymbol{q}}_j^{\Pi}$ because we will deal with a sum over unspecified $\hat{\boldsymbol{q}}_j^{\Pi}$.

The only properties which we require the factor $\Delta_{aj}$ to have are that it is a monotonic function of the angular distance $|\hat{\boldsymbol{p}} - \hat{\boldsymbol{q}}_j|$, and it is such that

$$\Delta_{aj} \approx |\hat{\boldsymbol{p}} - \hat{\boldsymbol{q}}_j|^2, \quad \hat{\boldsymbol{p}} \to \hat{\boldsymbol{q}}_j. \quad 6.13$$

It may otherwise be arbitrary. A modification of $\Delta_{aj}$ within these restrictions is compensated for by an appropriate change of $C_{f,3}$. This observation effectively decouples the form of the r.h.s. of 6.12 from the concrete choice of the mapping $\Pi$.

The result 6.12 can be used to estimate the second term on the r.h.s. of 6.5 (add and subtract terms as needed to apply 6.12). Take into account the fact that the values of $f$ at different points are in general independent, so the corresponding expressions have to be bounded independently. Obtain the following upper bound for the second sum on the r.h.s. of 6.5:

$$C_{f,4}\left( \sum_j \left| \sum_a z_{aj} E_a - \mathcal{E}_j \right| \right)$$
$$+ C_{f,5}\left( \sum_j \left| \sum_a z_{aj} E_a [\hat{\boldsymbol{p}}_a^{\Pi} - \hat{\boldsymbol{q}}_j^{\Pi}] \right| \right) + C_{f,3}\left( \sum_{j,a} z_{aj} E_a \Delta_{aj} \right). \quad 6.14$$

Minimizing 6.14 6.15

The task is to minimize 6.14 using the freedom to choose $\mathcal{E}_j$, $\hat{\boldsymbol{q}}_j^{\Pi}$, $z_{aj}$. The arbitrariness associated with $\Pi$, and $\Delta_{aj}$ will require additional consideration to be eliminated (Section 7).

The first term is suppressed if

$$\mathcal{E}_j = \sum_a z_{aj} E_a \quad \text{for each } j. \quad 6.16$$

This fixes $\mathcal{E}_j$ in terms of $z_{aj}$ and is immediately interpreted as energy conservation in the formation of jets.

The second term is suppressed if

$$\mathcal{E}_j \hat{\boldsymbol{q}}_j^{\Pi} = \sum_a z_{aj} E_a \hat{\boldsymbol{p}}_a^{\Pi} \quad \text{for each } j, \quad 6.17$$

where we used 6.16. This determines $\hat{\boldsymbol{q}}_j$ (via $\hat{\boldsymbol{q}}_j^{\Pi}$) in terms of $z_{aj}$. Note the arbitrariness due to the arbitrary $\Pi$ which will be fixed in Section 7. Anyhow:

[v] The discussion in the first posting of this paper interpreted conventional procedures incorrectly. The present version owes much in this respect to ref. [2].

[w] In contrast, the conventional algorithms seem to favor the 'f'-cuts because, apparently, there is no simple recipe to identify the particles to be relegated to soft energy prior to recombinations.

Eqs. 6.16 and 6.17 fix the parameters of jets in terms of the recombination matrix $z_{aj}$ which, therefore, is the fundamental unknown in this scheme. 6.18

• The described trick actually differs from conventional schemes only by explicit presence of $z_{aj}$ which fully describes the distribution of particles between jets in any jet finding scheme with energy-momentum conservation.

With 6.16 and 6.17, only the last sum survives in 6.14. So, redefining the inessential $f$-dependent constants and recalling Eq. 6.7, we arrive at the following estimate:

$$\left|\{f,\mathbf{P}\}-\{f,\mathbf{Q}\}\right| \le C_{f,1}\mathrm{Y}[\mathbf{P},\mathbf{Q}]+C_{f,2}\,\mathrm{E}_{\mathrm{soft}}[\mathbf{P},\mathbf{Q}]\,, \qquad 6.19$$

where

$$\mathrm{Y}[\mathbf{P},\mathbf{Q}] = \sum_{j,a} z_{aj}E_a\Delta_{aj}\,. \qquad 6.20$$

Note that Y is linear in all particles' energies as is $\mathrm{E}_{\mathrm{soft}}$; cf. the comments after 6.8. Recall also that $\Delta_{aj}=O(\theta_{aj}^2)$ for small $\theta_{aj}$, which is the angle between the $a$-th particle and the $j$-th jet.

• Although the bound 6.19 falls short of the desired factorized form 5.18, its derivation did not involve any arbitrariness that would deserve any further discussion.

The following two points do deserve a detailed discussion:

(i) The arbitrariness in the choice of the angular factors $\Delta_{aj}$ in 6.20. This will be fixed in Section 7 from simple kinematical considerations, resulting in considerable algorithmic simplifications.

(ii) Transition to the factorized form 5.18.

## Obtaining a factorized estimate[x] 6.21

### General options 6.22

Mathematically speaking, the basic bound 6.19 can be reduced to the required factorized form 5.18 in a variety of ways. Consider Y and $\mathrm{E}_{\mathrm{soft}}$ as components of a two-dimensional vector $\upsilon=(\upsilon_1,\upsilon_2)\equiv(\mathrm{Y},\mathrm{E}_{\mathrm{soft}})$. Then for a wide class of non-negative functions $W(\upsilon)$ one can obtain inequality of the form

$$C_{f,1}\upsilon_1+C_{f,2}\upsilon_2\le C_{f,W}\times W(\upsilon)\,, \quad \text{for all } \upsilon\,. \qquad 6.23$$

For instance, one can take

$$W(\upsilon)=(\alpha\upsilon_1^p+\beta\upsilon_2^p)^{1/p} \qquad 6.24$$

with $\alpha,\beta,p>0$. In any event it is reasonable to restrict $W$ to satisfy the condition $W(k\upsilon)=kW(\upsilon)$ for all positive $k$ (or even to be a norm in the mathematical sense, i.e. also satisfy $W(\upsilon_1+\upsilon_2)\le W(\upsilon_1)+W(\upsilon_2)$ ).

[x] The first posting of this paper described a somewhat simplistic way to take into account $\mathrm{E}_{\mathrm{soft}}$ (then called $\mathrm{E}_{\mathrm{miss}}$) in which one would minimize Y while keeping $\mathrm{E}_{\mathrm{soft}}$ fixed to a constant. It was justified by a somewhat vague reference to "the physical meaning of jet counting" — and, although not incorrect, was the only step of the derivation not clarified by a precise argument. The systematic approach outlined below attains an ultimate analytical simplicity for the criterion, exhibits a deep connection with the conventional cone algorithms, and results in a much faster algorithmic implementation thanks to elimination of the algorithmically cumbersome restriction $\mathrm{E}_{\mathrm{soft}}$ = const.

From 6.23 one obtains Eq. 5.18 with $\Omega_W=W(\mathrm{Y},\mathrm{E}_{\mathrm{soft}})$. (Note that although thus defined Ω is not a linear function of all particles' energies, all the dependence on the event is via two such functions, Y and $\mathrm{E}_{\mathrm{soft}}$. So infrared safety is not an issue here.)

### The linear choice 6.25

However, there is one choice of $W(\upsilon)$ which is singled out by its nice properties, namely,

$$W(\upsilon)=R^{-2}\upsilon_1+\upsilon_2\,. \qquad 6.26$$

The coefficients of the linear combination must be positive, and the overall normalization is inessential. The specific form of the coefficient, $R^{-2}$, is chosen for convenience of interpretation. Its introduction makes explicit the arbitrariness in the choice of measurement unit for angles (the role of $R$ is discussed in Sec. 8.14).

With $W$ given by 6.26, one obtains 5.18 with $C_f=\max(R^2C_{f,1},C_{f,2})$ and with Ω replaced by

$$\Omega_R=R^{-2}\mathrm{Y}+\mathrm{E}_{\mathrm{soft}}\,. \qquad 6.27$$

This choice is singled out by the following properties:

(i) analytical simplicity resulting in transparency of the corresponding jet definition and simplicity of implementation;

(ii) the inequality 6.23 becomes an identical equality for

$$C_{f,2}=R^2C_{f,1}\,. \qquad 6.28$$

This last fact means that for observables satisfying 6.28, the transition from the basic estimate 6.19 to the factorized one, Eq.5.18, via 6.26 does not entail any further loss of information about the event — for any event. Only the linear form 6.27 has this property.

We will consider linear form 6.27 as a standard reference point for comparison of alternatives. This issue will be further discussed in the context of the so-called $\mathrm{Y}-\mathrm{E}_{\mathrm{soft}}$ distribution in Sec. 8.19.

## Existence of the optimal jet configuration 6.29

We have obtained the factorized estimate 5.25 with Ω given by 6.27. This allows us to define optimal jet configurations according to the prescription 5.20.

Such a configuration $\mathbf{Q}_{\mathrm{opt}}$ always exists. Indeed, the quantity $\Omega_R$ is a non-negative continuous function of $z_{aj}$, and the domain of $z_{aj}$ is compact for each fixed $N(\mathbf{Q})$ (cf. 6.4, 6.3). So the l.h.s. always has a global minimum in this domain. Furthermore, the minimum value is a monotonically decreasing function of $N(\mathbf{Q})$ because each extra jet in **Q** adds new degrees of freedom for minimization, driving down the minimal value which reaches zero for all $N(\mathbf{Q})\ge N(\mathbf{P})$. So $\mathbf{Q}_{\mathrm{opt}}$ exists for any **P** and $\omega_{\mathrm{cut}}>0$.

The global minimum need not be unique even modulo renumberings of jets.

## Fine-tuning the angular factors in Y[**P**,**Q**] 7

The form of the angular factors $\Delta_{aj}$ in 6.20 is fixed within the arbitrariness of the scheme by simple additional considerations: (a) conformance to relativistic kinematics; (b) momentum conservation. The true elegance — and final justification — of the resulting construction is in the considerable computational simplifications resulting from a representation of the jet finding criterion in terms of 4-vectors and Lorentz scalar products (see after 7.10).

First, with each pair $E_a, \hat{\boldsymbol{p}}_a$ one associates a massless 4-vector $p_a$, $p_a^2 = 0$ (specific expressions depend on the representation of $\hat{\boldsymbol{p}}_a$; see below). Then define:

$$q_j = \sum_a z_{aj} p_a \,. \tag{7.1}$$

This object occurs in a natural way in our construction, and we will call it *the jet's physical 4-momentum*.

### Spherical geometry ($e^+e^- \to$ hadrons in c.m.s.) 7.2

Here one emphasizes spherical symmetry. The directions $\hat{\boldsymbol{p}}$ are interpreted as points of the unit sphere, i.e. unit 3-vectors: $\hat{\boldsymbol{p}}^2 = 1$. Then the 4-momentum $p_a$ associated with the pair $E_a, \hat{\boldsymbol{p}}_a$ has the energy component $p_a^0 = E_a$ and the 3-momentum component $\boldsymbol{p}_a = E_a \hat{\boldsymbol{p}}_a$.

#### 3-momentum conservation 7.3

We must choose a mapping of the unit sphere to the plane $\Pi$ which is normal to $\hat{\boldsymbol{q}}_j$. A simple choice is the stereographic projection from the point $-\hat{\boldsymbol{q}}_j$:

$$\hat{\boldsymbol{p}}_a^{\Pi} = \hat{\boldsymbol{p}}_a + t_{aj}\left(\hat{\boldsymbol{p}}_a + \hat{\boldsymbol{q}}_j\right) = \hat{\boldsymbol{p}}_a + O(\theta_{aj}^2), \tag{7.4}$$

where $t_{aj} = (1-c)/(1+c)$ with $c = \hat{\boldsymbol{p}}_a \cdot \hat{\boldsymbol{q}}_j$. Then Eq. 6.17 is rewritten as

$$\hat{\boldsymbol{q}}_j \mathcal{E}_j = \boldsymbol{q}_j + \sum_a z_{aj} E_a t_{aj}\left(\hat{\boldsymbol{p}}_a + \hat{\boldsymbol{q}}_j\right) = \boldsymbol{p}_j + O(E_a\theta_{aj}^2)\,. \tag{7.5}$$

where $\boldsymbol{q}_j$ is the space-like component of 7.1.

The arbitrariness in the choice of the mapping manifests itself through the terms $O(E_a\theta_{aj}^2)$ in 7.5. The simplest choice is to drop those terms altogether but then one would have to impose a correct normalization on $\hat{\boldsymbol{q}}_j$:

$$\hat{\boldsymbol{q}}_j = \boldsymbol{p}_j / |\boldsymbol{p}_j|\,. \tag{7.6}$$

The direct normalization here cannot take one outside the $O(E_a\theta_{aj}^2)$ arbitrariness in 7.5 (because ensuring a correct normalization of $\hat{\boldsymbol{q}}_j$ was part of the job of the $O(E_a\theta_{aj}^2)$ terms). This can also be verified directly.

#### Fixing $\Delta_{aj}$ 7.7

$\Delta_{aj}$ can be chosen in such a way as to eliminate one cumbersome summation over all particles in the event (which has to be performed with 6.20 after evaluation of $\hat{\boldsymbol{q}}_j$) and reduce all the complexity in the computation of the criterion to evaluation of the 4-vectors $q_j$.[y] The choice is this:

$$\tfrac{1}{2}\Delta_{aj} = 1 - \hat{\boldsymbol{p}}_a \cdot \hat{\boldsymbol{q}}_j \equiv 1 - \cos\theta_{aj} \equiv \tfrac{1}{2}[\hat{\boldsymbol{p}}_a - \hat{\boldsymbol{q}}_j]^2 \equiv E_a^{-1} p_a \tilde{q}_j\,, \tag{7.8}$$

where

$$\tilde{q}_j = (1, \hat{\boldsymbol{q}}_j)\,, \qquad \tilde{q}_j^2 = 0\,. \tag{7.9}$$

This is a light-like Lorentz vector with unit energy uniquely associated with the jet's spatial direction $\hat{\boldsymbol{q}}_j$. Then one uses 7.1 to perform the summation over $a$ and obtains:

$$\mathrm{Y}[\mathbf{P},\mathbf{Q}] = \sum_j \mathrm{Y}_j[\mathbf{P},\mathbf{Q}] \equiv 2\sum_j q_j \cdot \tilde{q}_j\,, \tag{7.10}$$

where the r.h.s. contains only Lorentz scalar products.
Note that in this kinematics $q_j \cdot \tilde{q}_j = \mathcal{E}_j - |\boldsymbol{q}_j|$ but the covariant form 7.10 is more general as we are going to see shortly.

### Cylindrical geometry (hadron collisions) 7.11

According to the standard Snowmass conventions (cf. [2]), here one direction (the beam axis) is singled out, and one emphasizes invariance with respect to Lorentz boosts along the beam axis. Therefore one should use the representation 3.4 for particles' 4-momenta. In particular, one has to interpret energies according to 3.7 in all the formulas related to jet definition. Then a reasoning similar to the spherically symmetric case leads one to the following results:

The $j$-th jet's transverse direction $\hat{\boldsymbol{q}}_j^{\perp}$ is determined similarly to 7.6 from conservation of transverse momentum:

$$\hat{\boldsymbol{q}}_j^{\perp} = \boldsymbol{q}_j^{\perp} / |\boldsymbol{q}_j^{\perp}|, \tag{7.12}$$

with $\boldsymbol{q}_j^{\perp}$ taken from 7.1. (At this point we choose to differ from the Snowmass definition which postulates conservation of energy-weighted azimuthal angle in jet formation. For narrow jets the two definitions are equivalent. On the other hand, our definition leads to a simpler code; cf. the remark after 7.10.)

For the jet's pseudorapidity[z] one has the Snowmass definition which is invariant with respect to boosts along the beam axis:

$$\mathcal{E}_j \eta_j = \sum_a z_{aj} E_a \eta_a\,. \tag{7.13}$$

For $\Delta_{aj}$ there is the following simple choice (this structure is borrowed from [27] where it appeared in the context of conventional jet algorithms):

$$\tfrac{1}{2}\Delta_{aj} = \cosh(\eta_a - \eta_j) - \cos(\varphi_a - \varphi_j)\,. \tag{7.14}$$

Then — surprise! — one recovers 7.10 with

$$\tilde{q}_j = (\cosh\eta_j, \sinh\eta_j, \hat{\boldsymbol{q}}_j^{\perp})\,, \qquad \tilde{q}_j^2 = 0\,, \tag{7.15}$$

where $\tilde{q}_j$ is also a light-like Lorentz vector uniquely associated with the jet's spatial direction specified in this case by the rapidity $\eta_j$ and the transverse direction $\hat{\boldsymbol{q}}_j^{\perp}$ — and also with unit energy — but now it is the unit *transverse* energy!

---

[y] The described choice allows a simple incremental update of $q_j$ after a modification of a particle's splitting between jets, which results in a major speedup (by two orders of magnitude) of the minimum search algorithm; for more see [7].

[z] Note that one can compute the jet's physical rapidity directly from $q_j$.

**Summary. The optimal jet definition (OJD) 7.16**

Finding an optimal configuration of jets **Q** (Eq. 4.31) for a given event **P** (Eq. 3.6) is equivalent to finding the recombination matrix $z_{aj}$ (Sec. 6.1) that determines jets' parameters via 7.1 and 7.6 (for spherical [c.m.s.] kinematics) or via 7.12 and 7.13 (for cylindrical [hadron collisions] kinematics).

The matrix elements $z_{aj}$ are found according to the prescription 5.20 with $\Omega_R[\mathbf{P},\mathbf{Q}]$ specified by 6.27 where $\mathrm{E_{soft}}[\mathbf{P},\mathbf{Q}]$ and $\Upsilon[\mathbf{P},\mathbf{Q}]$ are defined, respectively, by 6.8 and 7.10.

The light-like Lorentz vectors $\tilde{q}_j$ are given by 7.9 (spherical kinematics) or 7.15 (cylindrical kinematics).

This is the simplest universal dynamics-agnostic jet definition. Dynamical considerations can be accommodated as described in Sec. 5.31.

## Understanding the mechanism of OJD 8

To understand how the optimal jet definition (OJD; Sec. 7.16) "finds" jets, it is sufficient to understand what jet configurations yield minima for the criterion $\Omega_R$ depending on the structure of the original event **P**.

### "Fuzziness" of the event 8.1

For each integer $m \geq 1$, compute the quantity

$$\mathbf{J}_m^R(\mathbf{P}) = \min_{N(\mathbf{Q}')=m} \Omega_R[\mathbf{P},\mathbf{Q}'] \geq 0 . \qquad 8.2$$

For each fixed **P** and $R$, this sequence monotonically decreases with increasing $m$.

As will become clearer from what follows, the observable $\mathbf{J}_m^R(\mathbf{P})$ is best described as the event's <u>*cumulative fuzziness relative to m axes at the angular resolution R*</u>. It receives contributions of two kinds as seen from 6.27:

- a contribution from each of the $m$ jets, $\Upsilon_j = 2q_j \cdot \tilde{q}_j$ ; this can be conveniently called <u>*the fuzziness of the j-th jet*</u>;
- a contribution from soft stray particles which is simply the soft energy $\mathrm{E_{soft}}$.

One can describe the mechanism as follows:

OJD minimizes the cumulative fuzziness of the event by balancing contributions from each of the jets and from the soft energy. 8.3

The functions $\mathbf{J}_m^R(\mathbf{P})$ are shape observables similar to thrust (8.11). Observables similar to 8.2 were first introduced on the basis of conventional algorithms [28] but in our case they are specified by explicit analytical expressions. Even simpler analytical expressions (not involving optimization of any kind) were introduced in [3], [4] (the so-called jet-number discriminators) but they avoid identification of individual jets altogether.

In order to understand the mechanism of minimization, one notes that the analytical structure of OJD is very simple and regular, so it is sufficient to consider a few simple examples.

### A simplified jet definition (the Y-criterion) 8.4

First it is convenient to ignore $\mathrm{E_{soft}}$ in 6.19. This is valid for events without soft particles outside a few energetic jets and is equivalent to restricting the jet configurations **Q** used to minimize the error 6.2 by requiring that all particles are included into the formation of jets, with none relegated to the soft energy. Formally, this is described as follows:

$$\bar{z}_a \equiv 0 \iff \mathrm{E_{soft}}[\mathbf{P},\mathbf{Q}] \equiv 0 . \qquad 8.5$$

Such a restriction makes the error estimate less precise entailing a non-optimal loss of information in the transition from **P** to **Q**, but it is otherwise admissible.

The corresponding simplified definition is as follows:

A sub-optimal configuration of jets $\mathbf{Q}_{\mathrm{sub}}$ for a given event **P** minimizes $\Upsilon[\mathbf{P},\mathbf{Q}]$ and meets the following criterion with a minimal number of jets:

$$\Upsilon[\mathbf{P},\mathbf{Q}_{\mathrm{sub}}] \leq y_{\mathrm{cut}} . \qquad 8.6$$

It will be convenient to refer to this as <u>*the Y-criterion*</u>.

Note that this type of the criterion corresponds to $R \to \infty$ in 6.27. (For very large $R$, any contributions to soft energy would be disfavored. See also the discussion in Sec. 8.14.)

### Minimizing Y 8.7

Let us verify that the Y-criterion satisfies the boundary condition 4.33.

The quantity $\Upsilon[\mathbf{P},\mathbf{Q}]$, Eq. 6.20, is sensitive to presence of sprays of particles in the event **P** due to the angular factors $\Delta_{aj}$. Consider the simplest event **P** with two particles carrying equal energy. Then the criterion will see either one or two jets depending on whether or not

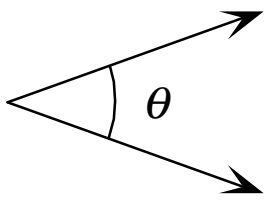

8.8

$$\tfrac{1}{4}\theta^2 \lesssim y_{\mathrm{cut}} \qquad 8.9$$

(remember that we are always dealing with fractions of the total energy of the event). For configurations with energy distributed between particles in a less symmetric fashion, a wider jet will be allowed for the same $y_{\mathrm{cut}}$.

Next suppose one has two pairs of particles, with a narrow angular separation between particles of each pair, and the angular separation between the pairs denoted as $\Theta$. Assume $\Theta^2 \gg y_{\mathrm{cut}}$. Then if one minimizes $\Upsilon[\mathbf{P},\mathbf{Q}]$ on the configurations **Q** with two jets, there is a global minimum corresponding to the configuration with each pair combined into a jet, and the minimum is unique up to a renumbering of the jets.

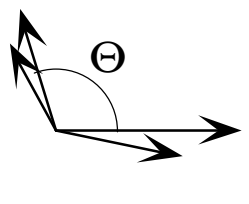

8.10

In other words, the angular factors in the expression for Y ensure a maximal suppression of a contribution from a spray of particles if the particles of the spray are made to constitute a jet (i.e. the corresponding $z_{aj} = 1$ with $z_{aj'} = 0$ for all $j' \neq j$ ), so that the jet's axis is automatically inside the spray.

This conclusion extends to more than two jets: if $N(\mathbf{Q})$ (the number of jets in **Q**) matches the number of sprays of particles in the event, then the global minimum of Y is reached on the configuration with jets and sprays in one-to-one correspondence so that each jet comprises exactly all the particles from the corresponding spray. If sprays are not narrow enough then the allocation of particles between jets is effected in a more dynamic fashion.

### Y-criterion and thrust 8.11

Recall the definition of the shape observable thrust, 4.27. Suppose all particles of the event are localized within a sufficiently narrow solid angle. Then the maximum is achieved for some axis inside the angle so that for all particles $\theta_a < \frac{\pi}{2}$. Recall 3.16 and obtain:

$$1 - T = \min \sum_a E_a \left(1 - \cos\theta_a\right) \approx \tfrac{1}{2} \min \sum_a E_a \theta_a^2 . \qquad 8.12$$

Comparing this with 6.20 and 7.10, we see that finding the thrust axis in this case is equivalent to finding the single jet direction according to the Y-criterion. Then $1 - T$ is equivalent to $\mathbf{J}_2^\infty(\mathbf{P})$ with the two jet directions restricted to be exactly opposite (each forming one half of the thrust axis). We conclude:

> The Y-criterion generalizes $1 - T$, where $T$ is the thrust, to the case of any number of thrust semi-axes which in the case of the Y-criterion become jet directions.
>
> 8.13

The same can be said about OJD because it is a modification of the Y-criterion.

### From the Y-criterion to OJD. Connection with cone algorithms 8.14

OJD differs from the Y-criterion by inclusion of $\mathrm{E}_{\mathrm{soft}}$ into the function to be minimized.

Let us discuss how OJD determines the optimal jet configuration compared with the Y-criterion 8.6. Thanks to the analytical simplicity and regularity of 6.27 (just two degrees of freedom, Y and $\mathrm{E}_{\mathrm{soft}}$, both with a simple structure and a clear meaning) it is sufficient to consider a few simple examples.

Fix a configuration $\mathbf{P}_{(0)}$ that consists of only one hard parton with the 3-momentum represented by the left figure 8.15. Then both the simplified criterion 8.6 and the optimal one 6.27 would find one jet exactly equal to the parton ($\mathbf{Q} = \mathbf{P}_{(0)}$), with $\mathrm{Y} = \mathrm{E}_{\mathrm{soft}} = 0$.

Now deform $\mathbf{P}_{(0)}$ by: (1) splitting the parton into almost collinear fragments; (2) radiating a soft fragment; (3) both. The three configurations are as follows:

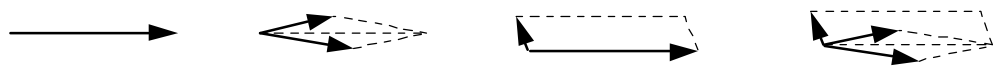

(0) (1) (2) (3) 8.15

What OJD and the Y-criterion would see depends on $\omega_{\mathrm{cut}}$ and $R$ and the magnitude of deformations (the acollinearity angle and the energies of fragments).

In the case of (1), OJD will be yielding exactly the same configuration as the Y-criterion (at least for not too large acollinearity), i.e. all $\bar{z}_a = 0$. This is because it is more advantageous to have the particles' energy contribute to Y where it will be suppressed by the acollinearity angle squared (cf. 6.13, 7.8, 7.14), rather than relegate any fraction of it to $\mathrm{E}_{\mathrm{soft}}$ (i.e. have the corresponding $\bar{z}_a > 0$) where no angular suppression is present. It is also clear that $R$ directly controls the threshold angle beyond which the configuration with both particles included into the jet yields a larger value of $\Omega_R$ than the configuration with the less energetic particle relegated to the soft energy. The exact relation between the threshold angle and $R$ depends on how energy is distributed between particles (see below).

In the case of (2) the Y-criterion must include the soft fragment into the jet. However, OJD would relegate the fragment to the soft energy (the corresponding $\bar{z}_a = 1$) to avoid enhancement by the angular factors (unless $R$ is very large). As a result the jet will consist of the hard parton only.

A similar conclusion is reached for the case (3) where the jet direction as found by OJD would include only the two hard fragments.

Furthermore, inclusion of an infinitesimally soft particle $(\varepsilon, \hat{\mathbf{p}})$ into an event changes $\Omega_R$ (apart from the overall renormalization by $1+\varepsilon$) by $\sim \varepsilon \theta_{\varepsilon j}^2 R^{-2}$ if the particle is included into the $j$-th jet (with $\theta_{\varepsilon j}$ the angle between the jet and the particle), and by $\varepsilon$ if the particle is relegated to the soft energy. So if the particle's angular distance from the nearest jet is $\lesssim R$ then OJD includes it into that jet. Otherwise the particle is relegated to the soft energy.

For non-infinitesimally soft particles the threshold angle is $< R$. For instance, if an isolated hard parton is split into two equal-energy fragments separated by $2\theta$ then OJD would include them both into one jet or relegate one to soft energy depending on whether or not $\theta \lesssim R/\sqrt{2}$. Note that either one of the two fragments can be relegated to soft energy, which simply means that the global minimum is not unique. However, the probability of occurrence of events for which the criterion has a degenerate global minimum is theoretically zero. These issues will be discussed in more detail in Sec. 9.1.

Given the generality of the described mechanism, we arrive at the following conclusion:

> OJD forms jets on the basis of local structure of energy flow within the correlation angle $R$.
>
> Quantitatively, $R$ is the maximal angular jet radius as probed by infinitesimally soft particles.
>
> 8.16

Furthermore, the above examples allow us to relate the parameter $R$ to the jet radius of the conventional cone algorithms $R_{\mathrm{cone}}$:

$$R_{\mathrm{cone}} \approx R/\sqrt{2}, \qquad R_{\mathrm{cone}} = 0.7 \longleftrightarrow R = 1. \qquad 8.17$$

The value 0.7 is preferred in the practice of cone algorithms on empirical grounds (e.g. [29]).

To conclude:

> Sensitivity of OJD to the presence of soft stray particles is controlled by the two parameters $R$ and $\omega_{\mathrm{cut}}$:
>
> $R$ controls which particles are expelled into the soft energy because they are too far from jets' axes (the decision also depends on the particle's energy), and
>
> $\omega_{\mathrm{cut}}$ effectively imposes an inclusive upper bound on the soft energy.
>
> 8.18

Remember that the primary role of $\omega_{\mathrm{cut}}$ is to control the loss of information in the transition from events to the corresponding jet configuration (Sec. 5.21).

### The Y–$E_{soft}$ distribution 8.19

From the discussion in 8.14 it follows that it would be interesting to consider contributions to OJD of the two components, Y and $E_{soft}$, separately. For different events 8.15 this is shown on the left figure below:

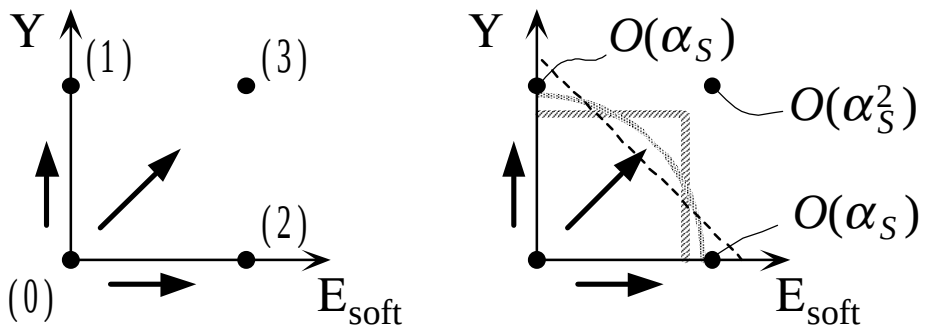

8.20

Acollinear fragmentations shift the point along the Y-axis, and soft stray radiation, along the $E_{soft}$ axis.

Then consider the figure 8.20 in the context of QCD. One sees that a fragmentation (1) or emission of a stray soft parton (2) are, from perturbative viewpoint, effects of relative order $\alpha_S$ whereas their combination (3) is an effect of relative order $\alpha_S^2$. Similarly, if one attributes these effects to non-perturbative "power-suppressed" corrections (i.e. suppression by an extra power of the unnormalized total [transverse] energy of the event) then one arrives at a similar conclusion with $\alpha_S$ replaced by $E^{-1}$.

On the theoretical side, one can make the following observations. For definiteness consider the process $e^+e^- \to$ jets and the distribution constructed for $N(\mathbf{Q})=2$. Then the lowest order quark-antiquark events are concentrated at $\mathrm{Y} = \mathrm{E}_{soft} = 0$ (the distribution is $\delta(\mathrm{Y})\delta(\mathrm{E}_{soft})$ ). Emission of one gluon ( $\bar{q}qg$ events in the perturbative order $\alpha_S$) creates two $\delta$-functional terms localized along the two axes:

$$p_0\,\delta(\mathrm{Y})\delta(\mathrm{E}_{soft}) + \alpha_S\big[p_{1,\mathrm{Y}}(\mathrm{E}_{soft})\times\delta(\mathrm{Y}) + p_{1,soft}(\mathrm{Y})\times\delta(\mathrm{E}_{soft})\big], \qquad 8.21$$

where $p_{1,\mathrm{Y}}(\mathrm{E}_{soft})$ and $p_{1,soft}(\mathrm{Y})$ are continuous functions. This is because the third parton is either included as a whole into one of the two jets or relegated to soft energy (configurations with the third parton exactly at the boundary of the two corresponding phase space regions occur with probability zero).

Emission of further gluons gives rise (apart from modifications of the coefficients of 8.21) to configurations which populate the internal region $\mathrm{Y}>0,\ \mathrm{E}_{soft}>0$, corresponding to a continuous distribution:

$$\alpha_S^2\, p_2(\mathrm{Y},\mathrm{E}_{soft})\,. \qquad 8.22$$

Such a picture, with the $\delta$-functions appropriately smeared and deformed into the internal region $\mathrm{Y}>0,\ \mathrm{E}_{soft}>0$, is expected to be seen in the data (assuming correctness of pQCD). It may be possible to theoretically describe the smearing of $\delta$-functions by taking into account power-suppressed corrections as well as resummation of large collinear logarithms. Given that the Y–$E_{soft}$ distributions can be constructed for any $N(\mathbf{Q})$ and for any process involving jet production, whereas the mechanisms behind, say, power corrections seem to be rather universal, studying such distributions may prove to be a valuable test of our understanding of the dynamics of QCD.

To summarize:

> The distribution of events in the Y–$E_{soft}$ plane [aa] provides a direct model-independent way to quantify the two different effects in the mechanism of hadronization, namely, collinear fragmentation and soft radiation. So the Y–$E_{soft}$ distribution is a **window on non-perturbative QCD effects**.
> 8.23

- Note that an even more detailed information is provided by the values of fuzziness $\mathrm{Y}_j$ of individual jets. One can e.g. study the fluctuations of $\mathrm{Y}_j$ within the same event, correlations, etc.
- Also, the values of $\mathrm{Y}_j$ together with $E_{soft}$ can be used as additional parameters on top of jets' 4-momenta. This is a natural extension of the jet-related degrees of freedom in terms of which to parameterize the events, e.g. in the construction of event selection procedures of the conventional type or quasi-optimal observables (Sec. 2.25) for specific precision measurement applications.

A word of caution: the values Y and $E_{soft}$ may not be always stable with respect to data errors (unlike the minimum value of $\Omega_R$). This is similar to how positions of global minima may be unstable under deformations of the function's shape. It results in a smearing of the event distribution along the lines $\Omega_R$ = const, and may impose limitations on the precision of such tests of QCD. However, the precision requirements here are not as high as in the Standard Model studies.

### On alternative forms of the criterion 8.24

At this point it is convenient to discuss the ambiguity involved in how Y and $E_{soft}$ are combined to obtain a factorized estimate of the form 5.18. After that we will also discuss a similar ambiguity with combining contributions from different jets into one expression Y (Sec. 8.30).

#### Combining Y and $E_{soft}$ 8.25

As was already pointed out (see before Eq. 6.23), the form of the criterion which is linear in Y and $E_{soft}$, Eq. 6.27, is, mathematically, not unique. On the other hand, the qualitative conclusions about how the criterion organizes particles into jets and soft energy (as discussed above in connection with 8.15), remain valid for any $\Omega$ based on any valid choice of $W(\upsilon)$ in 6.26. In particular, the arguments around Eqs. 8.21–8.22 remain valid. This makes it worthwhile to examine whatever further arguments one may find in favor of, or against the simple linear form 6.27.

First of all note that the physically most important degree of freedom in $W(\upsilon)$ is adequately represented by the free parameter $R$. To discuss the remaining ambiguities it is convenient to limit the discussion to the degree of freedom represented by the parameter $B \geq 1$ in the following alternative expression for $\Omega_R$:

$$\Omega_{R,B} = \left([R^{-2}\mathrm{Y}]^B + \mathrm{E}_{soft}^B\right)^{1/B}. \qquad 8.26$$

This expression is infrared safe and leads to only marginally slower code (the formulas for derivatives used in the algorithm [7] become more complex though, but this affects only a small part of the entire code which is executed not often provided the

[aa] One fixes the number of jets, and for each event finds the corresponding optimal jet configuration by minimizing $\Omega_R$. Y and $E_{soft}$ are obtained as a by-product.

covariant form of Y is used). In the limit $B \to \infty$ the function becomes non-smooth:

$$\Omega_{R,\infty} = \max\left(R^{-2}\mathrm{Y}, \mathrm{E}_{\mathrm{soft}}\right), \qquad 8.27$$

which results in considerable algorithmic complications due to nonexistence of derivatives at some points in the space of recombination matrices. The same problem will be manifest for large $B$ in the form of numerical instabilities.

So large values of $B$ are excluded by the requirement of algorithmic simplicity. The same requirement favors the linear choice $B=1$. However, most algorithmic efforts in a computer implementation of the corresponding minimum search algorithm [7] are spent on a proper handling of the recombination matrix $z_{aj}$ and the computation of $q_j$ etc., and only a fraction of the total code deals with the formulas such as 7.10 and 6.27, so that the values $B>1$ are not, strictly speaking, excluded.

The linear choice $B=1$ is also singled out by a similar requirement of analytical simplicity (needed to facilitate theoretical studies of e.g. power-suppressed effects using the Y–$\mathrm{E}_{\mathrm{soft}}$ plot).

Considering the alternative values for $B$, one might be tempted to add $R^{-2}\mathrm{Y}$ and $\mathrm{E}_{\mathrm{soft}}$ in quadrature ($B=2$). The corresponding region $\Omega < \omega_{\mathrm{cut}}$ would be a quarter of an ellipsoid (cf. the dotted boundary in the right figure of 8.20). As a further example, the rectangular region corresponds to $\Omega < \omega_{\mathrm{cut}}$ with $\Omega$ defined using $B=\infty$ (Eq. 8.27). A typical shape of the region $\Omega_R < \omega_{\mathrm{cut}}$ for the linear choice 6.27 is shown with the dashed straight line; larger $R$ correspond to steeper slopes.

The position of each event on the plane is determined by minimization of $\Omega$ and therefore depends on its specific form, so that a straightforward comparison of shapes of the regions $\Omega < \omega_{\mathrm{cut}}$ is in general not meaningful. However:

> For sufficiently small deviations of the fragmented event from the parent partonic event (the neighborhood of the origin of the Y–$\mathrm{E}_{\mathrm{soft}}$ plot, which corresponds to very small $\alpha_S$) the resulting values of Y and $\mathrm{E}_{\mathrm{soft}}$ will not depend on the specific form of $\Omega$.
> 8.28

This is because the minima of $\Omega$ tend to correspond to configurations with $z_{aj}=0$ or 1 (Sec. 9.1), which fact ensures some stability of resulting jet configurations with respect to small deformations (unless the event is such that $\Omega$ has a degenerate global minimum — a situation which occurs with probability zero). This phenomenon of "snapping" seems to persist for all $\Omega$ for which the corresponding function $W(\upsilon)$ (recall the reasoning in Sec. 6.22) is a convex function of the 2-dimensional vector $\upsilon$ (i.e. a norm in the mathematical sense). For instance, the already described (in Sec. 8.14) mechanism of balance between Y and $\mathrm{E}_{\mathrm{soft}}$ which makes a particle as a whole to either belong to a jet or be relegated to soft energy, remains operative irrespective of whether one compares Y and $\mathrm{E}_{\mathrm{soft}}$ or their positive powers as would be the case with the choice 6.24.

The proposition 8.28 means that in some neighborhood of the origin of the Y–$\mathrm{E}_{\mathrm{soft}}$ plot, the distribution of events is independent of the specific choice of $\Omega$ as long as the corresponding function $W(\upsilon)$ is a norm, so that all one has to take into consideration is the shape of the region $\Omega < \omega_{\mathrm{cut}}$.

Then from a purely geometrical point of view (justified by the tradition of using visual arguments in the construction of e.g. cone jet algorithms), one can reason as follows. There are two alternative ways to distort the parent parton event (the point (0) in Fig. 8.20): one is to add a non-negligible soft background to narrow jets (the arrow directed to the right from the origin in 8.20); the other is to make wider jets without much soft background (the arrow directed upwards). It is geometrically clear that the situations where one of these mechanisms dominates correspond better to the notion that the number of jets in the resulting event stayed the same as in the parent parton configuration, than a simultaneous effect of both mechanisms (the diagonal direction). In the latter case one would prefer to count the same number of jets only if both distortions are reduced. This seems to disfavor the shapes of the region $\Omega < \omega_{\mathrm{cut}}$ which are protruding along the diagonal (such as the rectangle in the right Fig. 8.20), and favor the more "flat" boundaries like the one corresponding to the linear choice 6.27.

In the final respect, the best argument for fine-tuning the form of $\Omega$ may be based on dynamical considerations such as suppressing sensitivity to higher-order and hadronization effects. The pattern exhibited by the right figure 8.20 and the additivity of small perturbative corrections (at least for small $\alpha_S$) seems to be rather universal and again favors the linear choice $B=1$ (which leads to Eq. 6.27).

To conclude:

> There seems to be no obvious general argument to counter the appeal of simplicity of the linear form of the criterion, Eq.6.27, which also retains the most important degree of freedom represented by the parameter $R$.
>
> The linear form is compatible with the additive nature of small perturbative corrections and seems to conform well to the intuitive notion of which deformations of the parton event preserve the "number of jets" best.
> 8.29

## Combining contributions from different jets 8.30

A similar ambiguity may be seen in the way contributions from different jets, $\mathrm{Y}_j$, are combined into a single expression Y (the transformation of the sum over $j$ in the transition from the second to third line in 6.5). Most arguments of Secs. 6.22–6.25 remain valid here too. However, in the case of combining Y and $\mathrm{E}_{\mathrm{soft}}$ in a factorized estimate the problem was due to two unknown coefficients $C_{f,i}$ (see 6.19) which vary independently under arbitrary changes of the observable $f$. In the present case there are no such unknown coefficients, and the inequality $\left|\sum \mathrm{Y}_j\right| \le \sum\left|\mathrm{Y}_j\right|$ becomes an exact equality for non-negative $\mathrm{Y}_j$. This seems to leave the linear form 7.10 as the only viable option here.

## Multiple jet configurations 9

If jet algorithms are supposed to invert hadronization then one should also take into account that there may be more than one (perhaps a continuum of) partonic configurations that could hadronize into a given hadronic state (Sec. 4.70). The problem is the more severe, the more pQCD corrections are taken into account. It is clear that at some level of precision, this effect must be taken into account in the theory of jet algorithms.

The multiplicity of parent partonic events for a given hadronic event **P** is reflected in a multiplicity of allowed jet configurations. In the context of OJD this is manifest in the fact that any jet configuration **Q** that satisfies 5.19 is a valid candidate, and the induced error can still be controlled via 5.24 and 5.25. The global minimum is the best choice from the viewpoint of minimizing the overall error but there are at least two cases when a unique choice may be hard to make. One case is the potential occurrence of multiple global minima (Sec. 9.1). Another is when equality is reached in 5.19 (Sec. 9.17). Both situations occur with theoretical probability zero but acquire significance in presence of detector errors.

The phrase "a unique choice may be hard to make" means that there is a discontinuity in the mapping $\mathbf{P}\to\mathbf{Q}$. As any discontinuity, this is manifest as an instability that causes a enhancement of errors for events near the discontinuity (Sec. 2.47 and [4]). Algorithmically, the handling of the problem of multiple jet configurations is a special case of the general method of regularization (Secs. 2.51, 2.52).

Two remarks:

(i) The options discussed here go beyond the conventional data processing scheme 4.38.

(ii) These options emerge naturally in the theory of OJD but they can be used in conjunction with conventional algorithms although in somewhat cruder forms because then one would not have the fine control of the weights offered by OJD (Sec. 9.16).

### Multiple minima 9.1

Numerical experiments (sufficiently extensive to accept these conclusions[bb]) show the following:

(i) MULTIPLICITY OF LOCAL MINIMA. Quite often (enough so that the issue may not be ignored, depending on the problem), there is more than one local minimum for the expression 6.27 as a function of **Q** (or, more precisely, $z_{aj}$) for fixed **P** and $R$. The simplest example is an event consisting of exactly three particles with equal energies and arranged symmetrically. Then among all possible 2-jet configurations, there are three isolated global minima with the same value of $\Omega[\mathbf{P},\mathbf{Q}]$. If one deforms the event slightly[cc] then the three minima remain local minima but in general only one will be the global minimum.

The number of local minima is not large ($O(1)$ on the average). It seems to correlate positively with the number of hard partons in the underlying partonic event.

(ii) AT THE POINTS OF MINIMA $z_{aj}$ ARE EQUAL TO EITHER 0 OR 1. In other words, particles tend to belong to a jet or the soft energy as a whole rather than are split between them. In this respect OJD is similar to the conventional algorithms. (However, first principles do allow solutions with fractional $z_{aj}$.)

(iii) MINIMA $z_{aj}$ ARE LOCALIZED AT ISOLATED POINTS. This directly follows from (ii).

The connection of multiple local minima with the multiplicity of jet configurations as produced by conventional algorithms is discussed in Section 10.

The occurrence of local minima in addition to the global one poses the following problem. No minimum search algorithm can absolutely guarantee that it has found the global minimum — especially for problems in $O(100)$ dimensions (recall that the dimensionality in our case is $N_{\text{particles}}\times N_{\text{jets}}$). The best one can hope to achieve is reduce the probability of missing the true global minimum e.g. by repeating searches from random initial configurations. Numerical experiments show that, given the efficiency of the minimum search algorithm described in [7], an exhaustive search of all local minima with a high confidence level does not constitute a practical difficulty.

The possible occurrence of several global minima poses the following problems.

On the one hand, probability of production of events **P** for which $\Omega[\mathbf{P},\mathbf{Q}]$ as a function of **Q** has a degenerate global minimum, is zero. Indeed, there is a finite probability to produce events with exactly $N$ particles. In the subset of such events, the events for which $\Omega[\mathbf{P},\mathbf{Q}]$ as a function of **Q** has a degenerate global minimum is a set of measure zero because minima are localized at isolated points. The probability density is a continuous function, whence follows the proposition.

Small deformations of such event (denote it as $\mathbf{P}_{\text{disc}}$) in general leave only one global minimum as shown in Fig. 9.2 where the curves describe trajectories of the minima, with solid parts corresponding to global minima and dashed parts, to local minima.

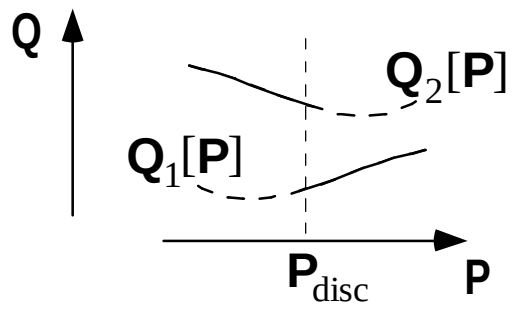

9.2

However, different deformations cause different global minima to survive. This means that with a non-zero probability detector errors may distort some events close to $\mathbf{P}_{\text{disc}}$ so that a local minimum will be seen as a global one.

Consider an observable defined via intermediacy of the mapping $\mathbf{P}\to\mathbf{Q}$:

$$\mathbf{P}\xrightarrow[\text{j.a.}]{}\mathbf{Q}\xrightarrow[\varphi]{}\varphi(\mathbf{Q}[\mathbf{P}])=f(\mathbf{P}). \qquad 9.3$$

This differs from the conventional scheme 4.38 in that now we do not assume any cut to be applied to the events. Any such cut is assumed to be incorporated into $\varphi$ as a $\theta$-factor (cf. 2.40).

Such an observable will in general be discontinuous near $\mathbf{P}_{\text{disc}}$ because the values $\varphi(\mathbf{Q}_k)$ where $\mathbf{Q}_k$ are different global minima for $\mathbf{P}_{\text{disc}}$, are in general all different. Then slight deformations of **P** would cause erratic jumps of $f(\mathbf{P})$ between all $\varphi(\mathbf{Q}_k)$, causing a non-optimal sensitivity of the observable to detector errors. One can suppress these fluctuations using the trick described below.

[bb] We used several hundred events generated by Jetset/Pythia [30] for typical processes studied at CERN and FNAL. Note that the mechanism of how $\Omega_R$ organizes particles into jets is essentially insensitive to the underlying physics. We have also used some simple events constructed manually to test the findings (iii) and (iv) in a more controlled manner.

[cc] A deformation may involve: a deformation of any particle's parameters; splitting particles into slightly acollinear fragments; adding soft arbitrarily directed particles to the event.

### The regularization trick 9.4

We are going to construct an observable $f_{\text{reg}}$ which would coincide with $f$ away from $\mathbf{P}_{\text{disc}}$. But near $\mathbf{P}_{\text{disc}}$, it would in general perform a continuous interpolation between different branches of $f$.

Let $\mathbf{Q}_k$ be the candidate local minima (they are actually functions of **P**; we will later discuss how $\mathbf{Q}_k$ can be selected). Suppose one can find weights[dd] $W_k$ normalized so that

$$\sum_k W_k = 1 . \qquad 9.5$$

Then one would define

$$f_{\text{reg}}(\mathbf{P}) = \sum_k W_k\, \varphi(\mathbf{Q}_k) . \qquad 9.6$$

If the weights $W_k$ depend on **P** continuously then so does the expression 9.6, and the discontinuity of the mapping $\mathbf{P} \to \mathbf{Q}$ is effectively masked.

One should also ensure that the expression 9.6 coincides with 9.3 for **P** outside some neighborhood $O$ of the point $\mathbf{P}_{\text{disc}}$. For that, it is sufficient that the weights vary in such a way that only one of them remains non-zero outside $O$ — the one which corresponds to the true global minimum $\mathbf{Q}_k$. Then outside $O$ only one term in the sum 9.6 survives with a unit weight, and $f_{\text{reg}}(\mathbf{P})$ coincides with $f(\mathbf{P})$ defined by 9.3.

The weights $W_k$ can be heuristically interpreted as probabilities that the event **P** resulted from hadronization of the partonic configuration $\mathbf{Q}_k$. See, however, the warning preceding 2.58.

In terms of collections of events, the described mechanism amounts to a replacement of the initial collection of events **P** with a collection of weighted (pseudo) events **Q**:

$$\{\mathbf{P}_i\}_i \to \{\mathbf{Q}_k, W_k\}_k , \qquad 9.7$$

where the r.h.s. comprises all jet configurations for all $\mathbf{P}_i$. Then

$$\frac{1}{N}\sum_i f(\mathbf{P}_i) \to \frac{1}{N}\sum_i f_{\text{reg}}(\mathbf{P}_i) = \frac{1}{N}\sum_k W_k\, \varphi(\mathbf{Q}_k) , \qquad 9.8$$

where

$$N = \sum_k W_k \qquad 9.9$$

in virtue of 9.5.

### A prescription for $W_k$ based on linear splines 9.10

Let us now present a simple prescription for constructing such weights. It should be remembered that there is no a priori recipe here (apart from the general desire to obtain a quasi-optimal observable; see Sec. 2.25). Also, it is not always necessary to eliminate all discontinuities: one may decide to patch some discontinuities and leave alone the rest (e.g. the kinds of discontinuities that occur seldom), depending on the problem. So the prescriptions described in what follows should be considered as merely examples.

Fix an event **P** and let $\mathbf{Q}_0$ be the point of global minimum of the function $\omega(\mathbf{Q}) = \Omega[\mathbf{P},\mathbf{Q}]$ with $\omega_0 = \Omega[\mathbf{P},\mathbf{Q}_0]$.

Choose the regularization parameter $r > 0$ so that

$$\omega_0 + r < \omega_{\text{cut}} . \qquad 9.11$$

Recall that $r$ should at least satisfy the restriction 2.59 where $\sigma_{\text{meas}}$ should now be taken to be a typical error induced in $\Omega_0$ by detector errors in **P**. Eq. 9.11 can be satisfied together with the mandatory restriction 2.59 provided

$$\omega_0 \ll \omega_{\text{cut}} \qquad 9.12$$

in the sense that $\sigma_{\text{meas}}$ is small enough that the condition $\omega_0 < \omega_{\text{cut}}$ is determined with a high reliability. The cases when 9.11 cannot be satisfied will be considered separately (Sec. 9.17).

Let $\mathbf{Q}_k$, $k = 1,\ldots$ be all local minima of $\omega(\mathbf{Q})$ (with the corresponding $\omega_k = \omega(\mathbf{Q}_k)$) which satisfy the restriction

$$\omega_k < \omega_0 + r . \qquad 9.13$$

Compute the weights $W_k$, $k = 0,1,\ldots$, from the conditions:

$$W_k \propto r^{-1}\left(r + \omega_0 - \omega_k\right) . \qquad 9.14$$

This together with the normalization 9.5 determines $W_k$.

The described trick eliminates $C$-discontinuities due to degenerate global minima at least for events which satisfy 9.12. This is because the values $\omega_k$ vary $C$-continuously with the event **P**¸ in general. However, this is not always the case, as discussed below.

### Cheshire local minima 9.15

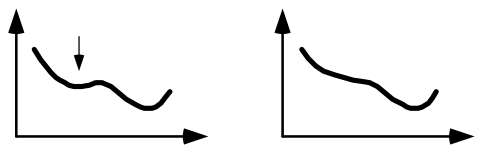

Indeed, any event can be $C$-continuously deformed into any other event, and the number of local minima of $\Omega$ in general differs for different events. This means that some local minima disappear under small deformations.[ee] This may somewhat spoil the regularization effect of the prescription if one of the local minima happens to be such a Cheshire minimum and vanishes while the corresponding $\omega_k$ is non-zero. This is more likely for larger values of the regularization parameter $r$ because the regularization procedure would then see more local minima.

It is possible to detect the Cheshire minima in the context of the minimum search algorithm described in [7] by looking at the values of gradient of $\Omega$. If these are smaller than some threshold then a corresponding factor should be introduced into 9.14 to effect a suppression. Then the weight $W_k$ would vanish in a continuous fashion as the corresponding $\Omega_k$ approaches the point where the local minimum disappears.

In any event, it seems that the effect of Cheshire minima could be dangerous only if detector errors are large enough that there is a sufficiently sizeable fraction of events whose local minima could be regarded as candidate global minima (and only a fraction of that fraction would exhibit the effect). In such a case it cannot be excluded that one might have to abandon the scheme 9.3 altogether in favor of the more complicated $C$-continuous observables, e.g. those constructed along the lines of [4].

### Comments for conventional jet algorithms 9.16

It is possible to use the described scheme with conventional algorithms. In the context of, say, cone algorithms, $\mathbf{Q}_k$ may be candidate jet configurations obtained e.g. for different initial configurations of cones (or other variations of the algorithm). In this cases one would have to take all weights $W_k$ equal.

---

[dd] We adopt the convention that $i$ labels events **P** and $k$ labels jet configurations **Q**. So $W_i$ and $W_k$ denote different arrays of numbers.

[ee] Remember that there always is at least one global minimum for any event.

Then as **P** varies, the weights will no longer vary continuously but since in general only one weight jumps at a time, the discontinuities would be mitigated.

A better option is to evaluate $W_k$ for each $\mathbf{Q}_k$ using $\omega_k = \Omega_R[\mathbf{P},\mathbf{Q}_k]$ with $\Omega_R$ borrowed from OJD (Eq. 6.27) even if $\mathbf{Q}_k$ are found using a conventional jet algorithm. This is possible because all one needs is the corresponding recombination matrices. These are easily restored from the output of any conventional jet algorithm.

## Regularizing the cut $\Omega[\mathbf{P},\mathbf{Q}] < \omega_{\text{cut}}$ 9.17

In the situation we have just considered all candidate jet configurations have the same number of jets. Next we suppose that this is no longer the case.

In the notations of 9.10, assume that $r$ cannot be chosen small enough to satisfy 9.11 for whatever reason (e.g. because the condition 9.12 is not satisfied). It is assumed that

$$\omega_0 < \omega_{\text{cut}} + \tfrac{1}{2} r\ . \qquad 9.18$$

$K$ is the minimal number of jets for which this condition is achieved.

Define $\alpha$ to be a function of $\omega_0$ that interpolates between the values 1 and 0 at the ends of regularization interval, e.g.:

$$\alpha = \begin{cases} 1 & \text{if} \quad \omega_0 < \omega_{\text{cut}} - \tfrac{1}{2} r, \\ 0 & \text{if} \quad \omega_0 > \omega_{\text{cut}} + \tfrac{1}{2} r, \\ r^{-1}\left(r + \omega_{\text{cut}} - \omega_0\right) & \text{otherwise}. \end{cases} \qquad 9.19$$

Also define $\beta = 1 - \alpha$.

Heuristically, $\alpha$ is interpreted as the probability that the event **P** has $K$ jets, then $\beta$ is the probability that the event has at least $K+1$ jets.

For simplicity we assume that $\min_{\mathbf{Q}} \Omega[\mathbf{P},\mathbf{Q}]$ on configurations with $K+1$ jets does not exceed $\omega_{\text{cut}} - \tfrac{1}{2} r$ . Otherwise the construction is to be iterated in the same spirit. (This is not likely to be needed often because $\min_{\mathbf{Q}} \Omega[\mathbf{P},\mathbf{Q}]$ as a function of $K$ decreases rather fast.)

Now, in the $K$-jet sector, define $\mathbf{Q}_k$ and the corresponding weights $W_k$ as in Sec. 9.10 except that the sum of $W_k$ is normalized to $\alpha$ rather than 1. Perform a similar procedure in the $(K+1)$-jet sector with the only modification that the sum of weights is normalized to $\beta$. Consider the collection of all jet configurations thus found together with their weights. The total sum of weights is equal to 1 by construction. The regularized $f$ is obtained according to 9.6 where now the summation runs over jet configurations with different numbers of jets.

If $\varphi$ in 9.3 incorporates a jet-number cut then one may choose to drop from the r.h.s. of 9.7 the jet configurations which do not satisfy the cut. The weights $W_k$ are evaluated prior to application of the cut, and the weights of the jet configurations retained in 9.7 are not affected thereby. The relation 9.9 is no longer valid, and the value of the normalizing factor, $N$, has to be remembered separately. (This is similar to how luminosity may have to be measured via special independent procedures rather than counting events for which a collision with a high transverse momentum occurred.)

## Regularization by variations of $R$ 9.20

An interesting variation is to evaluate jet configurations for each event for a sequence $R_n$ of values of the jet radius parameter $R$ — e.g. a few values around the standard value $R = 1$ (recall 8.17). For instance, $R_1 = 1-\varepsilon$, $R_2 = 1$, $R_3 = 1+\varepsilon$ with some $\varepsilon$.

This is motivated by the formal meaning of the parameter $R$ (see Eq. 6.28 and the discussion around it) which may motivate one to perform an averaging over $R$.

This option may be useful because events with clearly defined jets would tend to yield similar jet configurations for different values of $R$ whereas more fuzzy events would yield different jet configurations for different $n$.

So if one performs, say, histogramming of events in order to detect a peak, then the events which yielded several similar jet configurations would contribute in a more “focused” fashion.

On the other hand, the events which otherwise may have been entirely eliminated by selection procedures now have a chance to contribute their share of signal albeit with a weight $<1$.

9.21

Let $\omega_n$ be the value of $\Omega_{R_n}[\mathbf{P},\mathbf{Q}_{\text{opt}}]$ , with $\mathbf{Q}_{\text{opt}}$ found according to OJD with $R = R_n$. Let $\alpha_n = A(\omega_{\text{cut}} - \omega_n)$ where $A(x)$ is any monotonically increasing function, e.g. $A(x) = x$. (A function such as $A(x) = x^2$ would emphasize jet configurations which are farther from the cut.) Renormalize $\alpha_n$ so that their sum is equal to 1. The values thus found are larger for $n$ for which the optimal jet configuration effects a better approximation of the original event.

Then for each $n$, find jet configurations together with the corresponding weights normalized in such a way that the sum of weights for each event is equal to $\alpha_n$. The jet configurations for each $n$ can be found in arbitrarily sophisticated fashion. In the simplest case one takes one jet configuration found according to OJD without regularizations, and sets its weight to be $\alpha_n$. Alternatively, several jet configurations may be found using the regularization tricks described in Secs. 9.1 and 9.17.

In any case one ends up with a collection of jet configurations and weights whose total sum is equal to 1, and the regularized observables are found using 9.6.

This option could be used with conventional algorithms if one takes the resulting jet configurations with equal weights or evaluates the weights as described in Sec. 9.16.

## Discussion 9.22

(i) The described three regularization tricks regulate *any* observable, irrespective of its specific shape and meaning. For instance, $\varphi(\mathbf{Q})$ could be the (integer) number of dijets from **Q** whose mass belongs to some interval (bin) on the real axis. The corresponding regularized observable $f_{\text{reg}}$ takes non-integer values but its continuity is exactly what is needed to suppress fluctuations induced by detector errors.

(ii) One may consider replacing the prescription 9.6 by the following one. Let the recombination matrices $z_{aj}^{(i)}$ corresponding to the local minima $\mathbf{Q}_i$. Then define

$$z_{aj}^{(\text{reg})} = \sum\nolimits_{i=0,1\ldots} W_i\, z_{aj}^{(i)}\ . \qquad 9.23$$

If Eq. 9.5 holds, then this is a correct recombination matrix corresponding to some jet configuration $\mathbf{Q}_{\text{reg}}$ . One may be tempted to accept it as the resulting jet configuration. Since the corresponding recombination matrix 9.23 has fractional matrix elements, $\mathbf{Q}_{\text{reg}}$ can vary continuously in response to deformations of the original event. Then one would define the regularized observables $\varphi_{\text{reg}}(\mathbf{Q})$ to be simply $\varphi(\mathbf{Q}_{\text{reg}})$.

Unfortunately, such an interpretation is only valid if the condition 9.5 holds, and the important option of regularizing the cut involved in the definition of jets, Eq. 5.19, must still follow the scheme of Eq. 9.8.

Furthermore, the regularization effect for discontinuous $\varphi(\mathbf{Q})$ (cf. the example 4.42) is weaker here compared with the prescription 9.8. This is because the values $\varphi(\mathbf{Q}_{\text{reg}})$ still jump in response to variations in $\mathbf{P}$, although less erratically thanks to the more stable $\mathbf{Q}_{\text{reg}}$ as a function of $\mathbf{P}$.

(iii) The available experience seems to indicate that the values of $\Omega$ at different local minima (if there are any) for the same event may exhibit a significant spread. This means that local minima with values close to the value at the global minimum occur rarely.

(iv) The regularization tricks that yield a mixture of jet configurations with different number of jets may help to extract signal from events that would otherwise be dropped owing to the jet-number cut. For instance, suppose one looks at some process with 4 jets in the final state. Then events that would normally be counted as 3-jet events may, with regularization tricks, yield meaningful 4-jet configurations (with fractional weights $<1$). And vice versa: events that would normally be counted as having 4 jets but with some pairs of jets close, would "spill" some of their content into the 3-jet sector. The net effect here is equivalent to a relaxation of the rigid conventional jet-number cuts.

• All in all, the described regularization schemes are equivalent to a more sophisticated representation of the event's physical information — a representation in terms of several weighted jet configurations whose number may fluctuate depending on the event's fuzziness, etc. Such a representation preserves more information about the original event than any one jet configuration. This is especially true for the events with jets that are hard to resolve. The conventional jet-finding schemes correspond to enforcing a choice of one jet configuration even in situations when the choice is not clear-cut, and the jet configuration chosen may be a wrong one. On the other hand, with a regularized choice, the "correct" jet configuration will have chances to survive the jet-number cut, perhaps, with a fractional weight.

## Comparison with conventional algorithms 10

The proposition 8.16 establishes that OJD is essentially a cone algorithm with an inclusive treatment of soft energy (Sec. 6.9) rewritten in terms of thrust-like shape observables (Sec. 8.11). In this section we compare OJD with the conventional jet algorithms in a more systematic manner using the criterion of Sec. 5.26 for guidance.

There are two widely used classes of jet algorithms developed by trial and error: cone and recombination algorithms.[ff] We will consider them in turn (Sec. 10.1 and Sec. 10.9, respectively).

### Comparison with cone algorithms 10.1

Cone algorithms were introduced in [8] and define jets in a purely geometric fashion using cones of a fixed shape and angular radius $R$, so that the finding of jets reduces to finding the number and positions of the corresponding cones.

Cone positions are found via some kind of iterative procedure. We note that such an iterative search procedure can always be interpreted as a search of a minimum of some implicitly defined function on jet configurations; the function is parametrized by the event. It is clear that in general such a function may have many local minima, similarly to what was observed for OJD in Section 9.

The choice of initial configuration to start iterations from is not fixed by scientific considerations. Depending on how one makes this choice, one ends up with different jet configurations in the end. It is not difficult to realize that:

> The problem of choosing the initial configuration — which has as a consequence non-uniqueness of the resulting jet configuration — represents a **vicious circle** in the definition of cone algorithms. To break it one needs an extraneous principle, which for conventional algorithms is usually replaced by a convention.
> 10.2

In the case of OJD one simply opts for the global minimum of a well-defined shape observable, which corresponds to minimization of the information loss incurred in the transition from events to jets. It should be emphasized that the candidate jet configurations of the cone algorithms correspond to the local minima of OJD — not the degenerate global minima which occur much less often and to handle which our theory provides simple options (Section 9).

The termination condition for the cone algorithm is usually ad hoc too. For instance, the algorithm may seek to make the cone axes coincide with the corresponding jets' 3-momenta [29].

The original proposition of [8] was to minimize the energy left outside all jet cones, which is similar to the mechanism of OJD (8.18; cf. Eq. 10.7). However, the algorithm of [8] is algorithmically inconvenient, so the currently used variations [2] abandon the theoretically preferred inclusive treatment of soft energy (Sec. 6.9) in favor of lower cuts on energies of candidate jets (the so-called 'f'-cuts).

Note also how a scientific consideration is sacrificed here in favor of convenience of implementation of an ad hoc scheme, whereas with OJD, the theoretically preferred treatment of soft energy (Sec. 6.9) also leads to a simpler, faster and more robust computer code [7].

A murky problem specific to cone algorithms is how to treat cone overlaps. It remains essentially unsolved because of a lack of a guiding principle beyond the basic boundary condition 4.33. For this reason one usually recurs here to ad hoc conventions.[gg]

The mechanism represented by the parameter $R$ in OJD indicates its similarity to the conventional cone algorithms. The similarity is further exhibited by the algorithmic implementation of OJD described in [7].

We conclude:

[ff] See also sections 5.2.1–5.2.2 of [1] where they are called cluster and combination algorithms, respectively.

[gg] Perhaps, after intense discussions in a working group ;)

OJD is a cone algorithm in disguise with jet shapes determined (and jet overlaps handled) dynamically by means of a shape observable taking into account the distribution of energy in jets.

10.3

(See also 8.16 and 8.17.)

One can obtain a less optimal (in the sense of Sec. 5.26) jet definition via a cruder estimate for $\Omega$, but such as would be closer to the cone algorithms. It is easy to obtain the following simple upper bound for the fuzziness of the $j$-th jet:

$$Y_j[\mathbf{P},\mathbf{Q}] \le \mathcal{E}_j R_j^2, \quad R_j = \max_{a\in j} \theta_{aj}, \qquad 10.4$$

where the maximum is evaluated over all particles contributing to the jet, so that $R_j$ is interpreted as the jet's radius. The resulting less optimal variant of the criterion would be

$$\tilde{\Omega}_R = \sum_j \mathcal{E}_j \left(R_j/R\right)^2 + E_{\text{soft}}. \qquad 10.5$$

The mechanism of minimization is more transparent here than in the non-simplified case: take a particle from one jet and move it to another or to soft energy. Then the criterion 10.5 would decrease or increase depending on the induced changes in the two jets' radii $R_j$.

An even cruder version is obtained via the following upper bound for Eq. 10.5:

$$\tilde{\Omega}_R \le \left(\sum_j \mathcal{E}_j\right) \max_j \left(R_j/R\right)^2 + E_{\text{soft}}. \qquad 10.6$$

So one could define a jet finding scheme similarly to OJD but based on the following shape observable which is the same as the r.h.s. of 10.6:

$$\approx{\Omega}_R^{\text{S-W}} = \left(E_{\text{tot}} - E_{\text{soft}}\right) \max_j \left(R_j/R\right)^2 + E_{\text{soft}}. \qquad 10.7$$

In this variant jets' radii ignore the details of the energy distribution between particles — as in the conventional cone algorithms.

Verbally: the criterion 10.7 would attempt to include as much energy as possible into as few jets as possible with the jets' radii as narrow as possible but not exceeding $R$ (as with OJD, a particle farther than $R$ from any jet's axis is relegated to soft energy). If the event consists of non-overlapping sprays of particles with angular radii (measured from the spray's 3-momentum) not exceeding $R$, the criterion 10.7 will find jets in one-to-one correspondence with the sprays.

In the absence of jet overlaps, the mechanism of the criterion 10.7 is essentially equivalent to the original cone algorithm of Sterman and Weinberg [8], and similarly to the latter, it handles the soft energy in a theoretically correct fully inclusive fashion.

Unlike the algorithm of [8], the criterion 10.7 does not require additional prescriptions to handle jets' overlaps.

10.8

The criterion 10.7 may be implemented similarly to the simplex method [36]. Unfortunately, the analytical structure of 10.5 and 10.7 does not seem to allow the tricks which contributed to the efficiency of the implementation of OJD described in [7].

A general conclusion is that the conventional cone algorithms are non-optimal (in the sense of Sec. 5.26) whenever jet energies exhibit a significant variation.

## Comparison with recombination algorithms 10.9

Recombination algorithms emerged in the context of Monte Carlo hadronization models (the Luclus algorithm [30]) with inversion of hadronization as a primary motivation, apparently. The recombination scheme was popularized by the JADE algorithm [31], and subsequently improved by the $k_T$/Durham [32] and Geneva [33] variations.

### General discussion 10.10

A recombination algorithm iteratively replaces a pair of particles by one (pseudo) particle using some criterion to decide whether a given pair is to be recombined or left as is.

There are three problems here — all similar to what one encounters with cone algorithms.

One problem is the treatment of soft energy, and everything said about it in the context of cone algorithms is applicable here (the theoretically preferred inclusive treatment is abandoned owing to a conflict with an ad hoc algorithmic scheme).

Another problem is the lack of any firm principle to determine the order of recombinations. Intuition suggests that closest neighbors should be recombined first but with $O(100)$ particles in the event, there is still much choice. Similarly, one may start recombinations with the most energetic particles. This prescription is actually born out by the analogy with OJD: as is seen from the expression 6.20, starting to collect particles into jets from the most energetic ones allows one to focus from the very beginning on jet configurations in which largest contributions — those from the most energetic particles — are suppressed.

However, selecting the most energetic particles is a non-IR safe procedure, so some preclustering is needed, which cannot be too coarse. This introduces an undesirable non-inclusivity which may result in an enhancement of power corrections.

These ambiguities take the place of the problem of choosing initial conditions for the cone algorithms, so that the jet definitions based on recombination algorithms also contains a vicious circle similar to the one pointed out for cone algorithms (10.2).

The third problem concerns the choice of the recombination criterion used to decide when two particles are to be recombined into one (this has a parallel in the problem of handling overlapping cones in the case of cone algorithms). There seems to be a consensus emerging about a preferential status of the $k_T$ criterion [32] as enabling better theoretical calculations.

### $2 \to 1$ recombinations 10.11

Let us now take a closer look at recombination criteria.

To begin with, we note that a series of recombinations

$$\mathbf{P} \xrightarrow{2\to1} \mathbf{P}' \xrightarrow{2\to1} \mathbf{P}'' \xrightarrow{2\to1} \ldots \xrightarrow{2\to1} \mathbf{Q}, \qquad 10.12$$

is naturally interpreted in the framework of the definition 5.9 as a series of approximations

$$f(\mathbf{P}) \approx f(\mathbf{P}') \approx f(\mathbf{P}'') \approx \ldots \approx f(\mathbf{Q}), \qquad 10.13$$

so that each recombination can be analyzed within the framework of the developed theory.

It is perfectly obvious that even if one performs each $2 \to 1$ recombination in an optimal way, the scheme 10.12–10.13 in general causes an accumulation of additional errors (instabilities) compared with a global optimization such as done in OJD.

10.14

To appreciate this, recall how economically cancellations were arranged in our derivation of the corresponding error estimates; cf. e.g. Eq. 6.5.

Let us now obtain the $2 \to 1$ recombination criterion which corresponds to OJD. Consider how it treats a narrow spray consisting of two particles $a$ and $b$. In this case one sees from 6.20 that if one combines them into one jet $j$ then

$$\mathrm{Y}_j^{\mathrm{opt}} \approx E_a E_b \left(E_a + E_b\right)^{-1} \theta_{ab}^2 . \qquad 10.15$$

This is exactly the geometric mean of the JADE criterion [31],

$$\mathrm{Y}_j^{\mathrm{JADE}} \approx E_a E_b \, \theta_{ab}^2 , \qquad 10.16$$

and the Geneva criterion [33],

$$\mathrm{Y}_j^{\mathrm{Geneva}} \approx E_a E_b \left(E_a + E_b\right)^{-2} \theta_{ab}^2 . \qquad 10.17$$

(Remember that in our case all energies are fractions of the total energy of the event.)

Furthermore, when recombining pairs of soft particles, the JADE criterion underestimates 10.15 and thus would tend to combine them into spurious jets — exactly the problem which gave rise to the Geneva [33] and $k_\mathrm{T}$/Durham [32] variations. The Geneva and $k_\mathrm{T}$ criteria (as well as the earlier Luclus criterion [30]) overestimate 10.15 in such cases. From the viewpoint of the developed theory, this is indicative of their non-optimality but is otherwise safe (overestimating induced errors is not dangerous).

To further compare Eq. 10.15 with the $k_\mathrm{T}$ criterion, rewrite the latter by normalizing to the total energy and taking square root to achieve first order homogeneity in energies:

$$\mathrm{Y}_j^{k_\mathrm{T}} \to \left(E_a + E_b\right) \min(x, 1-x)\, \theta_{ab} , \qquad 10.18$$

where $x = E_a \left(E_a + E_b\right)^{-1}$. Eq. 10.15 in similar notations becomes

$$\mathrm{Y}_j^{\mathrm{opt}} \approx \left(E_a + E_b\right) x(1-x)\, \theta_{ab}^2 . \qquad 10.19$$

The difference in the $x$-dependence is inessential (cf. Fig. 10.20; one can bound one function by the other times a coefficient) unlike the angular dependence which is qualitatively different. A tentative conclusion from 10.18 and 10.19 would be that the $k_\mathrm{T}$ criterion would tend to yield more jets at smaller angular separations than the variant 10.19. It is thus less optimal than 10.19 in the sense that in general it requires more jets to ensure that the same amount of information from the event is preserved in the resulting jet configuration.

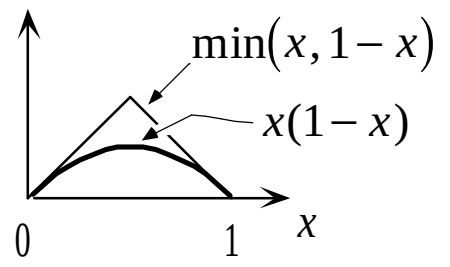

10.20

Of course, the latter property need not necessarily be a drawback because it ensures that the shape of the $K$-jet sectors in the space of events (4.35) is qualitatively different here compared with OJD, and this may be useful in practice (see the discussion in Sec. 5.1).

- On the other hand, the advantage of the $k_\mathrm{T}$ criterion over other conventional schemes (better theoretical predictions) seems to be overshadowed by an ultimate amenability to theoretical analyses of the shape observables in terms of which OJD is formulated.

Recall also the remarks in Sec. 5.31 concerning how dynamical information could be incorporated into OJD.

The last remark concerning the $k_\mathrm{T}$ algorithm is as follows. It is an attempt to make use of the theoretical pQCD results to improve upon the recombination scheme and is obviously motivated by the fact that the kinematics of $2 \to 1$ recombinations makes irresistible an inclusion into the picture of theoretical results such as the Sudakov formfactor. However, this per se can hardly be regarded as a justification for the recombination scheme as such and does not correct its fundamental deficiency — the ambiguity of the order of recombinations (see Sec. 10.9).

> The point here is not that QCD should be ignored in the construction of jet algorithms but that the recombination scheme may not be the best receptacle to pour dynamical QCD wisdom into. 10.21

## Non-uniqueness of jet configurations and the meaning of $\omega_\mathrm{cut}$ 10.22

The above analysis indicates that the conventional algorithms behave as imperfect heuristics for the minimization problem in OJD. This observation reveals an interesting point, namely, existence of a source of errors entirely specific to conventional algorithms and uncorrelated for different algorithms.

Indeed, consider possible existence of several local minima of $\Omega_R$ (when the event does not appear to have well-defined jets at a given resolution $\omega_\mathrm{cut}$). The optimal algorithm simply repeats the search from different initial configurations (e.g. randomly generated), and if it finds more than one local minimum then the global minimum is selected simply by comparison of the corresponding values of $\Omega_R$.

It is not difficult to realize that situations with several local minima seen by OJD have an immediate analog in the situations where the conventional algorithms find different configurations depending on minor algorithmic variations such as the choice of initial configuration or the order of recombinations.

The conventional algorithms, however, provide no criterion to select the best configuration from several such candidates: Any ad hoc prescription amounts to a more or less random choice — and from the viewpoint of OJD, a random choice of the local minimum results in a jet configuration which may inherit less information from the initial event than is actually possible. In other words, the use of conventional algorithms implies a systematically larger loss of information compared with OJD.

The instability of the found configuration of jets which results from random choice of a local minimum is due to the stochastic nature of hadronization and is manifest on a per event basis.

OJD would similarly fail in situations with several global minima but such situations occur with theoretical probability zero, and if they do become important due to detector errors, there are specific prescriptions to regularize the corresponding instabilities (Section 9).

To summarize:

(i) Ambiguities of conventional algorithms are an additional source of errors in physical results — additional compared with the theoretically optimal behavior of OJD.

(ii) Then OJD is preferable over conventional schemes in proportion to how the number of events with more than one local minimum of $\Omega_R$ exceeds those with several global minima (taking into account various experimental and theoretical uncertainties). Events with several local minima seem to prolifer-ate in proportion to importance of higher-order and power-suppressed corrections.

(iii) It should be possible to quantify these effects by determining the fraction of events with several local minima (checking along the way how occurrence of several local minima is reflected in ambiguities of results of conventional algorithms), and the fraction of events with several global minima (modulo various sources of uncertainties taken into account via [simple] models).

(iv) We are also compelled to conclude that working with not too small values of the parameters such as $\omega_{\text{cut}}$ and $y_{\text{cut}}$ imposes a fundamental limit on the potentially attainable precision of interpreted physical information such as parameters of the Standard Model, obtained via intermediacy of jet algorithms, although the potential numerical magnitude of the effect (or rather defect) remains unclear. This has a simple explanation:

> The parameter $\omega_{\text{cut}}$ and the similar parameters of the conventional algorithms actually describe the errors induced in the physical information by the approximate description of events in terms of jets.
>
> 10.23

Cf. the estimate 5.18 from which OJD 5.20 is derived.

- The conclusion 10.23 has to be taken into account when comparing results of different jet algorithms. This may help to explain the finding that the prospective dominant error in the planned top mass measurements at the LHC is due to the ambiguities of jet definition [34]. It may be possible to reduce such an error using the methods described in Section 9.
- The conclusion 10.23 is also to be kept in view when comparing results obtained using the same jet algorithm but different event samples (e.g. CDF and D0).

## Conclusions 11

The discovered optimal jet definition (OJD; it is summarized in Sec. 7.16) is essentially a cone algorithm (Sec. 8.14) entirely reformulated in terms of shape observables (the fuzziness; Sec. 8.1) which generalize the well-known thrust to the case of any number of thrust semi-axes (Sec. 8.11). The cone shapes and positions are determined dynamically via minimization of the fuzziness. The soft energy is treated inclusively via a cumulative cut on the soft energy (Sec. 6.9), which is similar to the original prescription of Sterman and Weinberg [8] but differs from the currently preferred 'f'-cuts [2].

The criterion is controlled by two parameters: $R$ and $\omega_{\text{cut}}$. The parameter $R$ sets an upper limit on the maximal angular radius of jets (Sec. 8.14). The parameter $\omega_{\text{cut}}$ effectively sets an upper bound on the soft energy allowed to be left out of jet formation, but its primary role is to control the loss of information entailed by the transition from the event to jets (Sec. 5.17).

### The synthesis of OJD 11.1

It is rather remarkable that OJD turns out to be a smooth blend of many things and tricks tried in the practice of jet algorithms.

We have already noted that it is essentially a cone algorithm rewritten in terms of thrust-like shape observables. It even allowed us to obtain a shape-observables-based analog of the original cone algorithm of [8] (Eq. 10.7).

Neither is new the idea of jet finding via a global optimization — such a version of recombination algorithms was earlier explored in [35].

Curiously, OJD yields for each jet what we called the physical 4-momentum $q_j$ with $q_j^2 > 0$ (Eq. 7.1), and simultaneously a light-like 4-vector $\tilde{q}_j$ (Eqs. 7.9 and 7.15), both closely related and playing an important role in the definition; cf. 7.10. Note in this respect that the variants of cone algorithm used by D0 and CDF yield, respectively, massive and massless jets [2], which can be associated with $q_j$ and $\mathcal{E}_j \tilde{q}_j$ .

The options for inclusion of dynamical QCD information are also available although in a different form than with the $k_T$ algorithm — via dependence of $\omega_{\text{cut}}$ on events (Sec. 5.31) and via theoretical analyses of the shape observables $\Omega_R$ (Sec. 8.1).

Even the recombination scheme — although it did not find a visible place in OJD — can still be regarded as a heuristic for minimum search (Sec. 10.11).

The only important element of the conventional schemes not incorporated in OJD is the lower ('f'-) cuts on jets' energies. Stray soft particles are now handled via an inclusive energy cut (8.18).

### What OJD derives from 11.2

A widely held opinion (cf. [2]) is that the definition of jets is subjective in nature. The developed theory shows that it is not quite so.

The important ingredient which has been missing from the conventional discussions of jet definition (it would be misleading to use the word theory here) is the information analysis of the problem of jet definition. Our analysis is based on an earlier groundwork [4] which emphasized a purely kinematical viewpoint on jet algorithms as approximation tricks rooted in — but not identical with — the dynamics of QCD.

The most important clarification of the theory of [4] obtained in the present paper is the notion of optimal observables for measurements of fundamental parameters (Secs. 2.7 and 4.19). The notion (together with the resulting practical prescriptions, Sec. 2.25) provides a guidance for a systematic improvement upon the conventional scheme of measurements based on the notion of jets (Sec. 4.28; cf. the new options described in Section 9).

The notion of optimal observables allows one to interpret the event's information content (which is the basis of OJD, Section 5) in the light of the fundamental Rao-Cramer inequality of mathematical statistics (Sections 2, 4.19 and 5.26).

The general considerations which went into the derivation of OJD are as follows:

1. A systematic reliance on first principles of physical measurement, quantum field theory and QCD.
2. Avoidance of ad hoc choices not fortified by strictly analytical arguments.
3. The requirement that the jet configuration must inherit maximum information from the event.
4. Conformance to the Snowmass conventions in regard of kinematics of hadron collisions.
5. Maximal computational simplicity.

A remarkable fact is that other properties usually postulated for jet algorithms emerge as mere consequences of the re-quirement of computational simplicity, which fact provides their ultimate justification thus replacing the usual aesthetic arguments:

6. Energy-momentum conservation in the formation of jets from particles (Sec.6.15, 7.3).
7. Conformance to relativistic kinematics (Sec.7.7).
8. Maximal inclusiveness of the criterion (needed to reduce sensitivity to hadronization effects which correspond to higher order logarithmic and power corrections).

Lastly — and most surprisingly — the found criterion possesses a property which is naturally interpreted as

9. An optimal inversion of hadronization (Sec.5.10).

## Mutable and immutable elements of OJD 11.3

One has to distinguish between the handling of a single event and the construction of observables for collections of events.

At the level of an individual event, all the arbitrariness is in the form of Ω. There is not much room left for modifications of the Ω as given by 6.27 ∪ 7.10 ∪ 6.8. The internal structure of Y and $E_{soft}$ does not seem to allow meaningful modifications. So the only reasonable option might have been in how Y and $E_{soft}$ are combined into a single quantity; it is represented by Eq. 8.26. It results in only marginal computational complications but seems to make theoretical analyses more difficult without offering clear advantages (see the discussion in Sec. 8.25).

At the level of collections of events things are more interesting. In particular, making $\omega_{cut}$ depend on the event **P** is the way to include dynamical QCD information into the picture (Sec. 5.17). However, lifting the restrictions of the standard scheme 4.38 (cf. Section 9) may be more important in the end (cf. the remarks after Eq. 5.33).

To summarize: the form of $\Omega_R$ (6.27 ∪ 7.10 ∪ 6.8) is the least mutable element of the described scheme, so that all variations would use as the main building block a minimization procedure for Ω (such a procedure is provided in the code developed in [7]).

The most interesting variations (i.e. those which allow improvements of the conventional scheme 4.38 in the direction of constructing better approximations of the ideal optimal observables 4.20) concern the definition of observables on collection of events (see Section 9 for tricks to start from).

The simplest universal (i.e. dynamics-agnostic) optimal jet definition is based on the linear choice 6.27 ($B = 1$ in 8.26) and an event-independent $\omega_{cut}$. This is closely related to the way the conventional cone algorithms are defined and may be accepted, in the context of the developed theory, as a default definition for all comparisons.

In short, the theory of OJD only deals with the function to be minimized in order to find jets (the fuzziness Ω, 6.27 ∪ 7.10 ∪ 6.8) — but it only provides guiding principles (the method of quasi-optimal observables; Sec. 2.25) for how observables are to be constructed. It is up to the user to decide whether to stick to the conventional scheme 4.38 or go beyond its limitations using e.g. the tricks of Section 9.

## Remarks on implementation 11.4

That OJD (summarized in Sec. 7.16) is fully constructive is in itself rather wonderful given that it was derived in a straightforward fashion from the seemingly innocent (to the point of appearing meaningless) criterion 5.9 which, however, only accurately expresses a fundamental idea implicit in the jet paradigm — that the configuration of jets inherits the essential physical information of the corresponding event (5.8).

Computationally, the problem of finding jets in our formulation reduces to finding the recombination matrix $z_{aj}$ which minimizes $\Omega[\mathbf{P}, \mathbf{Q}]$ given by 6.27. For an event which lit up 150 detector cells and contains 4 jets, $z_{aj}$ has 600 independent components, so that one has an optimization problem in a domain of a very large dimensionality. Such problems are notoriously difficult. Fortunately, the analytical simplicity of both the function to be minimized and the regularity of the domain in which the minimum is to be found (a direct product of standard simplices, one simplex per particle; cf. 6.3, 6.4) can be effectively employed to design an efficient algorithm [7].

Although the minimization algorithm of [7] was obtained from purely analytical considerations (a variant of the gradient search which makes a heavy use of the analytical specifics of the problem) plus some experimentation, a posteriori it is naturally interpreted as follows:

— the algorithm starts with some (perhaps randomly generated) distribution of particles between jets;

— the jets perform iterative “negotiations” by considering particles one by one and deciding if and how their energy should be redistributed between the jets and the soft energy in order to improve upon the current configuration;

— the algorithm stops when no particle can be further redistributed to decrease Ω.

This is reminiscent of the iterative adjustment of jets’ positions in the cone algorithms. However, the jets’ axes and shapes are specified in the conventional algorithms directly, and in the optimal criterion, indirectly via the recombination matrix.

Feasibility of implementation of OJD is thus not an issue.

• A liberating consequence of the jet definition via minimization of a simple function is that a specific implementation of the minimum-finding algorithm is of no consequence whatever (physical or other) provided it yields the optimal jet configuration with required precision. Thus different groups of physicists are free to explore their favorite algorithms — from simplest low-overhead methods for theoretical computations with a few partons, to neural, genetic, Danzig’s [36], equidomoidal [37], … algorithms for experimental data processing — as long as they minimize the same criterion and control approximation errors sufficiently well in doing so. This would be a truly satisfactory way to resolve the difficulties encountered in comparison of physical results from groups which use different variants of jet algorithms [2].

The criterion 6.27 tends to prefer configurations with $z_{aj}$ equal to exactly 0 or 1 (remark (ii) at the beginning of Sec. 9.1). This makes the problem very similar to that of linear programming for which a vast theory exists (see e.g. [36]) where one can borrow ideas for more efficient or fancy implementations.

Note that allowing fractional values for $z_{aj}$ proves to be extremely convenient algorithmically: the domain in which the minimum is to be found is then a convex region, so one chooses an internal point (which corresponds to some fractional values) as a starting point for minimum search and then descends into a minimum via a most direct route.

### New options for jet-based data processing and ancillary results 11.5

The theory of OJD offers new options for improvements upon the conventional scheme 4.38. Some such options are described in Section 9. Also the additional information about events contained in the parameters $Y_j$ and $E_{soft}$ can be used to expand the phase space of jet configurations in order to enhance informativeness of the resulting observables and thus approach the theoretical Rao-Cramer limit of the optimal observables 4.20.

Also the following results deserve to be mentioned here:

(i) The usefulness of the method of quasi-optimal observables (Sec. 2.25) goes beyond jet-related measurements.

(ii) The $Y$–$E_{soft}$ distribution (Sec. 8.19) offers a new model-independent window on the dynamics of hadronization thus allowing a new class of tests of pQCD as well as theoretical descriptions of hadronization models.

### Advantages of OJD 11.6

OJD — even interpreted narrowly in the context of the conventional scheme 4.38 — has the following advantages over the conventional algorithms:

(i) OJD solves the problem of non-uniqueness of jet configurations which is insurmountable in the context of conventional schemes. It thus eliminates a source of errors entirely due to the structure of jet algorithms (Sec. 10.22).

(ii) OJD extirpates the difficulties of conventional algorithms usually "solved" via ad hoc prescriptions (the handling of cone overlaps, the choice of order of recombinations, etc.).

(iii) The shape observables on which OJD is based generalize the well-known thrust and are therefore superbly amenable to theoretical studies — evidently more so than any imaginable modification of the conventional schemes (cf. the pQCD calculations for the thrust reviewed in [20]).

(iv) OJD allows independent implementations so that different experimental and theoretical groups only have to agree upon the function to be minimized (Sec. 11.4).

Furthermore, OJD offers new options for improving upon the conventional jet-based data-processing scheme 4.38 as described in Section 9 in the direction of approaching the theoretical Rao-Cramer limit on precision of extracted fundamental parameters (see Sec. 2.19).

The simplest dynamics-agnostic OJD allows modifications to incorporate dynamical QCD information (Sec. 5.31).

A fast and robust implementation of OJD is available as a Fortran code [7].

In conclusion, the most important result of this paper is a systematic analytical theory of jet definition based on first principles, with explicitly formulated assumptions, and with the logic of jet definition elucidated in a formulaic fashion. If one were to construct as precise as possible an approximation to the optimal observable in a specific application then it is a theorem that OJD is a better tool for that than any conventional jet algorithm. However, it is not clear whether the cost of such an ideal solution would be justified by the resulting increase in precision of results.

In any event, the developed systematic framework reveals some new options (e.g. the regularization via multiple jet configurations) which may be useful even within the conventional approach.

ACKNOWLEDGMENTS. A NORDITA secretary who failed to submit an earlier text [6] to a journal thus gave me an opportunity to revisit the subject and discover the trick with the recombination matrix. Ya.I. Azimov, E.E. Boos, and I.F. Ginzburg offered their insights on the issue of quark-hadron duality. Walter Giele informed me of the jet definition activities at FNAL and the related Web sites. The collaboration with Dima Grigoriev on the minimum search algorithm [7] led to a revision of an imprecise treatment of soft energy in the first posting of this paper. Pablo Achard ran important large scale tests of the algorithm on real data and provided valuable comments. Monique Werlen offered useful criticisms of a draft of this paper. Denis Perret-Gallix provided an encouragement and supplied information about the on-going discussions of jet definition. Dima Bardin helped to clarify the bibliographic status of the concept of optimal observables. I thank all the above people for their valuable inputs.

This work was supported in part by the Russian Foundation for Basic Research under grant 99-02-18365.

■